\documentclass{LMCS}

\usepackage{amssymb,algorithm,algorithmic,graphicx,hyperref,enumerate}

\def\doi{4 (4:15) 2008}
\lmcsheading%
{\doi}
{1--44}
{}
{}
{Nov.~\phantom{0}2, 2006}
{Dec.~23, 2008}
{}   

\begin{document}

\title{The Wadge Hierarchy of Deterministic Tree Languages\rsuper*}

\author[F.~Murlak]{Filip Murlak}

\address{Institute of Informatics, University of Warsaw, ul.~Banacha
  2, 02--097 Warszawa, Poland}

\email{fmurlak@mimuw.edu.pl}

\thanks{Supported by KBN Grant~4~T11C~042~25.}



\titlecomment{{\lsuper*}An extended abstract of this paper was
  presented at ICALP'06 in Venice, Italy.}

\begin{abstract} 
We provide a~complete description of the Wadge hierarchy for
deterministically recognisable sets of infinite trees. In particular
we give an elementary procedure to decide if one deterministic tree
language is continuously reducible to another. This extends Wagner's
results on the hierarchy of $\omega$-regular languages of words to the
case of trees. 
\end{abstract}

\maketitle

\section{Introduction}

Two measures of complexity of recognisable languages of infinite words
or trees have been considered in literature: the index hierarchy,
which reflects the combinatorial complexity of the recognising
automaton and is closely related to $\mu$-calculus, and the Wadge
hierarchy, which is the refinement of the Borel/projective hierarchy
that gives the deepest insight into the topological complexity of
languages. Klaus Wagner was the first to discover remarkable relations
between the two hierarchies for finite-state recognisable
($\omega$-regular) sets of infinite
words~\cite{wagner0}. Subsequently, decision procedures determining an
$\omega$-regular language's position in both hierarchies were given
\cite{kupferman,kwiatek,wagner}.

For tree automata the index problem is only solved when the input is
a~deterministic automaton \cite{hie,urban}. As for topological
complexity of recognisable tree languages, it goes much higher than
that of $\omega$-regular languages, which are all
$\Delta^0_3$. Indeed, co-B\"uchi automata over trees may recognise
$\Pi^1_1$-complete languages \cite{gap}, and Skurczy\'nski
\cite{skurcz} proved that there are even weakly recognisable tree
languages in every finite level of the Borel hierarchy. This may
suggest that in the tree case the topological and combinatorial
complexities diverge. On the other hand, the investigations of the
Borel/projective hierarchy of deterministic languages \cite{split,gap}
reveal some interesting connections with the index hierarchy.

Wagner's results \cite{wagner0,wagner}, giving rise to what is now
called the Wagner hierarchy (see~\cite{perrin}), inspire the search
for a~complete picture of the two hierarchies and the relations
between them for recognisable tree languages. In this paper we
concentrate on the Wadge hierarchy of deterministic tree languages: we give a
full description of the Wadge-equivalence classes forming the
hierarchy, together with a procedure calculating the equivalence class
of a given deterministic language. In particular, we show that the hierarchy has the height $\omega^{\omega \cdot 3}+3$, which should be compared with $\omega^\omega$ for regular $\omega$-languages \cite{wagner}, $\omega^{\omega^2}$ for deterministic context-free $\omega$-languages \cite{contextfree}, $(\omega_1^{CK})^\omega$ for $\omega$-languages recognised by deterministic Turing machines \cite{selivanov}, or an unknown ordinal $\xi>(\omega_1^{CK})^\omega$ for $\omega$-languages recognised by nondeterministic Turing machines, and the same ordinal $\xi$ for nondeterministic context-free languages \cite{finkel}.

The key notion of our argument is an adaptation of the Wadge game to
tree languages, redefined entirely in terms of automata. Using this
tool we construct a~collection of canonical automata representing the
Wadge degrees of all deterministic tree languages. Then we provide a~procedure calculating the canonical form of a~given deterministic automaton, which runs within the time of finding the productive states of the automaton (the exact complexity of this problem is unknown, but not worse than exponential).

\section{Automata}\label{sect:automata}

We use the symbol $\omega$ to denote the set of natural numbers $\{0, 1, 2, \ldots \}$. For an alphabet $\Sigma$, $\Sigma^*$  is the set of finite words over $\Sigma$ and $\Sigma^\omega$ is the set of infinite words over $\Sigma$. The {\em concatenation}  of words $u\in \Sigma^*$ and $v \in \Sigma^* \cup \Sigma^\omega$ will be denoted by $uv$, and the empty word by $\varepsilon$. The concatenation is naturally generalised for infinite sequences of finite words $v_1v_2v_3\dots$. The concatenation of sets $A \subseteq \Sigma^*$, $B \subseteq \Sigma^* \cup \Sigma^\omega$ is $AB=\{uv: u\in A, v \in B \}$.

A {\em tree}  is any subset of $\omega^*$ closed under the prefix relation. An element of a~tree is usually called a~{\em node}.  A {\em leaf}  is any node of a~tree which is not a~(strict) prefix of some other node. A~{\em $\Sigma$-labelled tree}  (or a~tree over $\Sigma$) is a~function $t: {\rm dom}\,  t \to \Sigma$ such that ${\rm dom}\,  t$ is a~tree. For $v \in {\rm dom}\,  t$ we define $t.v$  as a~subtree of $t$ rooted in $v$, i.~e., ${\rm dom}\,  (t.v) = \{ u: vu \in {\rm dom}\,  t\}$, $t.v(u) = t(vu)$.

A {\em full $n$-ary $\Sigma$-labeled}  tree is a~function $t: \{0,1, \ldots, n-1\}^* \to \Sigma$. The symbol $T_{\Sigma}$ will denote the set of full binary trees over $\Sigma$. From now on, if not stated otherwise, a~``tree'' will mean a~full binary tree over some alphabet.

Out of a~variety of acceptance conditions for automata on infinite
structures, we choose the parity condition. A~{\em nondeterministic
  parity automaton on words}  can be presented as a~tuple $A = \langle
\Sigma, Q, \delta, q_0, {\rm rank}\, \!\rangle$, where $\Sigma$ is
a~finite input alphabet, $Q$ is a finite set of states, $\delta
\subseteq Q \times \Sigma \times Q$ is the transition relation, and $q_0 \in Q$ is the initial state. The meaning of the function ${\rm rank}\,\! : Q \to \omega$ will be explained later. Instead of $(q,\sigma, q_1) \in \delta$ one usually writes $q \stackrel{\sigma}{\longrightarrow} q_1$. A {\em run} of an automaton $A$ on a~word $w\in \Sigma^\omega$ is a~word $\rho_w\in Q^\omega$ such that $\rho_w(0) = q_0$ and if $\rho_w(n) = q$, $\rho_w(n+1) = q_1$, and $w(n) = \sigma$, then $q \stackrel{\sigma}{\longrightarrow} q_1$. A~run $\rho_w$ is {\em accepting} if the highest rank repeating infinitely often in $\rho_w$ is even; otherwise $\rho_w$ is {\em rejecting}. A~word is {\em accepted} by $A$ if there exists an accepting run on it. The language recognised by $A$, denoted $L(A)$ is the set of words accepted by $A$. An automaton is {\em deterministic} if its relation of transition is a~{\em total} function $Q \times \Sigma \to Q$. Note that a deterministic automaton has a~unique run (accepting or not) on every word. We call a~language deterministic if it is recognised by a~deterministic automaton.

A {\em nondeterministic automaton on trees}  is a~tuple $A = \langle \Sigma, Q, \delta, q_0, {\rm rank}\, \!\rangle$, the only difference being that $\delta \subseteq Q \times \Sigma \times Q \times Q$. Like before, $q \stackrel{\sigma}{\longrightarrow} q_1, q_2$ means $(q,\sigma, q_1, q_2) \in \delta$. We write  $q \stackrel{\sigma, 0}{\longrightarrow} q_1$ if there exists a~state $q_2$ such that $q \stackrel{\sigma}{\longrightarrow} q_1, q_2$. Similarly for  $q \stackrel{\sigma, 1}{\longrightarrow} q_2$. A~{\em run} of $A$ on a~tree $t\in T_\Sigma$ is a~tree $\rho_t\in T_Q$ such that $\rho_t(\varepsilon) = q_0$ and if $\rho_t(v) = q$, $\rho_t(v0) = q_1$, $\rho_t(v1) = q_2$ and $t(v) = \sigma$, then $q \stackrel{\sigma}{\longrightarrow} q_1, q_2$. A~path $\pi$ of the run $\rho_t$ is {\em accepting} if the highest rank repeating infinitely often in $\pi$ is even; otherwise $\pi$ is {\em rejecting}. A~run is called accepting if all its paths are accepting. If at least one of them is rejecting, so is the whole run. An automaton is called deterministic if its transition relation is a~total function $Q \times \Sigma \to Q \times Q$.

By $A_q$ we denote the automaton $A$ with the initial state set to $q$. A~state $q$ is {\em all-accepting}  if $A_q$ accepts all trees, and {\em all-rejecting} if $A_q$ rejects all trees. A~state (a transition) is called {\em productive} if it is used in some accepting run. Observe that being productive is more than just not being all-rejecting. A~state $q$ is productive if and only if it is not all-rejecting and there is a~path $q_0 \stackrel{\sigma_0,d_0}{\longrightarrow} q_1  \stackrel{\sigma_1,d_1}{\longrightarrow} \ldots \stackrel{\sigma_n,d_n} {\longrightarrow}q$ such that $q_i \stackrel{\sigma_i,\bar d_i}{\longrightarrow} q'_i$, $\bar d_i \neq d_i$, and $q'_i$ is  not all-rejecting for $i=0,1,\ldots,n$.

Without loss of generality we may assume that all states in $A$ are productive save for one all-rejecting state $\bot$ and that all transitions are either productive or are of the form $q\stackrel{\sigma}{\longrightarrow} \bot,\bot$. The reader should keep in mind that this assumption has influence on the complexity of our algorithms. Transforming a~given automaton into such a~form of course needs calculating the productive states, which is equivalent to deciding a~language's emptiness. The latter problem is known to be in $\textrm{NP}\cap \textrm{co-NP}$ and has no polynomial time solutions yet. Therefore, we can only claim that our algorithms are polynomial for the automata that underwent the above preprocessing. We will try to mention it whenever it is particularly important. \label{explicitproductive}

\begin{figure}
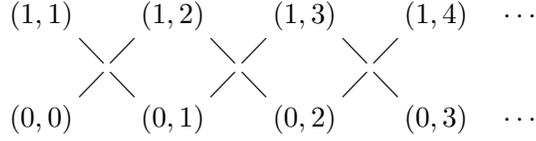

\centering
{\setlength\arraycolsep{1pt}
$\begin{array}{ccccccccccc}
(1,1) &&& (1,2) &&& (1,3) &&& (1,4)  & \quad \cdots\\
& \diagdown & \diagup && \diagdown & \diagup && \diagdown & \diagup \\ 
& \diagup & \diagdown && \diagup & \diagdown && \diagup & \diagdown \\  
(0,0) &&& (0,1) &&& (0,2) &&& (0,3) & \quad \cdots 
\end{array}$}
\caption{The Mostowski--Rabin index hierarchy.}
\label{fig:indexhierarchy}
\end{figure}

The {\em Mostowski--Rabin index of an automaton} $A$ is a~pair \[(\min {\rm rank}\, Q,\max {\rm rank}\, Q )\,.\] An automaton with index $(\iota, \kappa)$ is often called a~$(\iota,\kappa)$-automaton. Scaling down the $\mathrm{rank}$ function if necessary, one may assume that $\min {\rm rank}\, Q$ is either 0 or 1. Thus, the indices are elements of $\{0,1\}\times\omega \,\setminus\, \{(1,0)\}$. For an index $(\iota,\kappa)$ we shall denote by $\overline {(\iota,\kappa)}$ {\em the dual index}, i.~e., $\overline{(0,\kappa)} = (1,\kappa+1)$, $\overline{(1,\kappa)} = (0,\kappa-1)$.
Let us define an ordering of indices with the following formula
\[(\iota,\kappa)<(\iota',\kappa') \;\; \textrm{if and only if} \;\; \kappa-\iota < \kappa' - \iota' \,.\]
In other words, one index is smaller than another if and only if it uses less ranks. This means that dual indices are not comparable. 
 {\em The Mostowski--Rabin index hierarchy} for a~certain class of automata consists of ascending sets (levels) of languages recognised by $(\iota,\kappa)$-automata (see Fig.~\ref{fig:indexhierarchy}). 

The fundamental question about the hierarchy is the strictness, i.~e., the existence of languages recognised by a~$(\iota, \kappa)$-automaton, but not by a $\overline{(\iota, \kappa)}$-automaton. The strictness of the hierarchy for deterministic automata follows easily from the strictness of the hierarchy for deterministic word automata \cite{wagner}: if a~word language $L$ needs at least the index $(\iota, \kappa)$, so does the language of trees that have a word from $L$ on the leftmost branch. The index hierarchy for nondeterministic automata is also strict \cite{klony}. In fact, the languages showing the strictness may be chosen deterministic: one example is the family of the languages of trees over the alphabet $\{\iota, \iota+1, \ldots, \kappa\}$ satisfying the parity condition on each path.

The second important question one may ask about the index hierarchy is how to determine the exact position of a~given language. This is known as the {\em index problem}.

Given a~deterministic language, one may ask about its {\em deterministic index}, i.~e., the exact position in the index hierarchy of deterministic automata (deterministic index hierarchy). This question can be answered effectively. Here we follow the method introduced by Niwi{\'n}ski and Walu\-kiewicz \cite{kwiatek}. 

\begin{figure}
\centering
\includegraphics[width=0.6\textwidth]{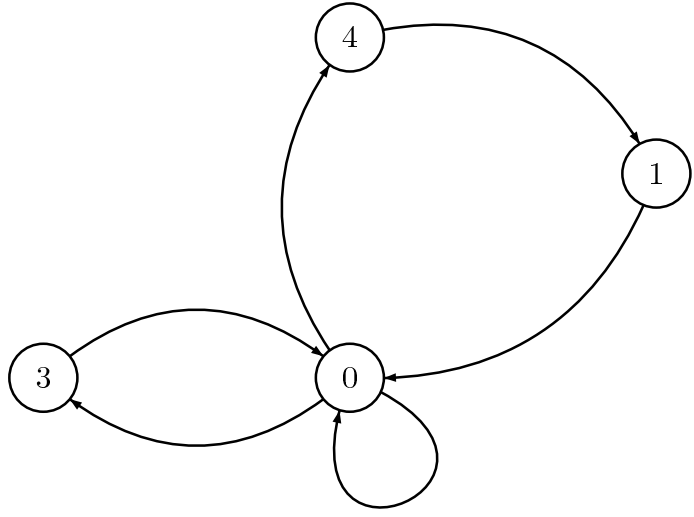}
\caption{A~$(0,2)$-flower.}
\label{fig:02flower}
\end{figure}

A path in an automaton is a~sequence of states and transitions: \[p_0 \stackrel{\sigma_1,d_1}{\longrightarrow} p_1 \stackrel{\sigma_2,d_2}{\longrightarrow} \ldots \stackrel{\sigma_{n-1},d_{n-1}}{\longrightarrow} p_n\,.\]
A {\em loop}  is a~path starting and ending in the same state, $p_0 {\longrightarrow} p_1 {\longrightarrow} \ldots {\longrightarrow} p_0$. A~loop  is called {\em accepting}  if $\max_i \mathrm{rank}\,(p_i)$ is even. Otherwise it is {\em rejecting}. A~$j$-loop is a~loop with the highest rank on it equal to $j$. A~sequence of loops $\lambda_\iota, \lambda_{\iota+1}, \ldots , \lambda_\kappa$ in an automaton is called {\em an alternating chain} if the highest rank appearing on $\lambda_i$ has the same parity as $i$ and it is higher then the highest rank on $\lambda_{i-1}$ for $i=\iota, \iota+1, \ldots, \kappa$. A~{\em $(\iota,\kappa)$-flower} \index{flower} is an alternating chain $\lambda_\iota, \lambda_{\iota+1}, \ldots, \lambda_\kappa$ such that all loops have a~common state $q$ (see Fig. \ref{fig:02flower}). \footnote{This is a slight modification of the original definition from \cite{kwiatek}.} 

Niwi{\'n}ski and Walukiewicz use flowers in their solution of the index problem for deterministic word automata. 

\begin{thm}[Niwi{\'n}ski, Walukiewicz \cite{kwiatek}] \label{omegaindch}
A deterministic automaton on words is equivalent to a~deterministic $(\iota,\kappa)$-automaton iff it does not contain a~$\overline{(\iota,\kappa)}$-flower. \qed
\end{thm}

\noindent For a~tree language $L$ over $\Sigma$, let $\mathrm{Paths}(L) \subseteq (\Sigma \times \{0,1\})^\omega$ denote the language of generalised paths of $L$, \[\mathrm{Paths}(L) = \left \{ \langle (\sigma_1, d_1), (\sigma_2, d_2), \ldots \rangle \colon \exists_{t \in L}\; \forall_i \; t(d_1 d_2 \ldots d_{i-1}) = \sigma_i\right \} \, .\]
A deterministic tree automaton $A$, can be treated as a~deterministic
word automaton recognising $\mathrm{Paths}(L(A))$. Simply for $A = \langle Q, \Sigma,  q_0, \delta, \mathrm{rank} \rangle$, take $\langle Q, \Sigma \times \{0,1\},  q_0, \delta', \mathrm{rank} \rangle$, where $(p,(\sigma, d),q) \in \delta' \iff (p,\sigma, d,q) \in \delta$. Conversely, given a~deterministic word automaton recognising $\mathrm{Paths}(L(A))$, one may interpret it as a~tree automaton, obtaining thus a~deterministic automaton recognising $L(A)$. Hence, applying Theorem \ref{omegaindch} one gets the following result.

\begin{prop} \label{indch}
For a~deterministic tree automaton $A$ the language $L(A)$ is recognised by a deterministic $(\iota,\kappa)$-automaton iff $A$ does not contain a $\overline{(\iota,\kappa)}$-flower. \qed
\end{prop}

\begin{figure}
\centering
\includegraphics[width=0.8\textwidth]{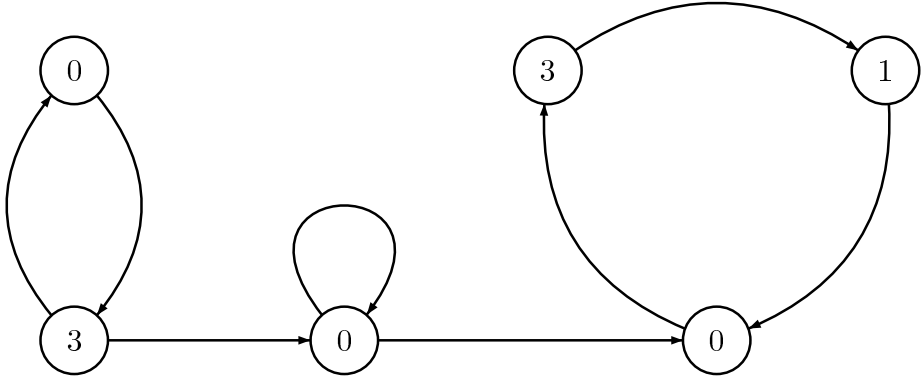}
\caption{A weak $(1,3)$-flower.}
\label{fig:weak02flower}
\end{figure}

In \cite{split} it is shown how to compute the {\em weak deterministic index} of a~given deterministic language. An automaton is called {\em weak} if the ranks may only decrease during the run, i.~e.,  if $p \longrightarrow q$, then $\mathrm{rank}(p) \geq \mathrm{rank}(q)$. The weak deterministic index problem is to compute a~weak deterministic automaton with minimal index recognising a given language. The procedure in \cite{split} is again based on the method of difficult patterns used in Theorem \ref{omegaindch} and Proposition \ref{indch}. We need the simplest pattern exceeding the capability of weak deterministic $(\iota, \kappa)$-automata. Just like in the case of the deterministic index, it seems natural to look for a~generic pattern capturing all the power of  $\overline{(\iota,\kappa)}$. Intuitively, we need to enforce the alternation of ranks provided by $\overline {(\iota,\kappa)}$. Let a~{\em weak $(\iota,\kappa)$-flower} \label{weakflowers} be a~sequence of loops $\lambda_\iota, \lambda_{\iota+1} \ldots , \lambda_\kappa$ such that $\lambda_{j+1}$ is reachable from $\lambda_j$, and $\lambda_j$ is accepting iff $j$ is even (see Fig. \ref{fig:weak02flower}). 

\begin{prop}[\cite{split}] \label{windch}
A deterministic automaton $A$ is equivalent to a~weak deterministic $(\iota,\kappa)$-automaton iff it does not contain a~weak $\overline{(\iota,\kappa)}$-flower. \qed
\end{prop}

For a~deterministic language one may also want to calculate its {\em nondeterministic index},  i.~e., the position in the hierarchy of nondeterministic automata. This may be lower than the deterministic index, due to greater expressive power of nondeterministic automata. Consider for example the language $L_M$ consisting of trees whose leftmost paths are in a~regular word language $M$. It can be recognised by a~nondeterministic $(1,2)$-automaton, but its deterministic index is equal to the deterministic index of $M$, which can be arbitrarily high.

The problem transpired to be rather difficult and has only just been solved in \cite{hie}. Decidability of the general index problem for nondeterministic automata is one of the most important open questions in the field.

\section{Topology}\label{sect:topology}

We start with a~short recollection of elementary notions of descriptive set theory. For further information see \cite{kechris}.

Let $2^{\omega}$ be the set of infinite binary sequences with a~metric given by the formula 
\[d(u,v) =  \left \{ 
  \begin{array}{l l}
     2^{-\min\{i \in \omega \;:\;\; u_i \neq v_i\}} & \textrm{iff } u \neq v \\
     0 & \textrm{iff } u=v \\
  \end{array}
\right .
\] and~$T_{\Sigma}$ be the set of infinite binary trees over $\Sigma$ with a metric
\[ d(s,t) = \left \{ 
  \begin{array} {l l}
    2 ^{-\min \{|x|\;: \;\; x \in \{0,1\}^*, \; s(x) \neq t(x)\}} & \textrm{iff } s \neq t \\
    0 & \textrm{iff } s=t
  \end{array}
\right . 
.\]
Both $2^{\omega}$ and $T_{\Sigma}$, with the topologies induced by the above metrics, are Polish spaces  (complete  metric spaces with countable dense subsets). In fact, both of them are homeomorphic to the Cantor discontinuum. 

The class of Borel sets of a~topological space $X$ is the closure of the class of open sets of $X$ by complementation and countable sums. Within this class one builds so called {\em Borel hierarchy}. The initial (finite) levels of the Borel hierarchy are defined as follows:
\begin{enumerate}[$\bullet$]
  \item $\Sigma^0_1(X)$ --  open subsets of $X$,
  \item $\Pi^0_k(X)$ --  complements of sets from $\Sigma^0_k(X)$, 
  \item $\Sigma^0_{k+1}(X)$ --  countable unions of sets from $\Pi^0_k(X)$.
\end{enumerate}
For example, $\Pi^0_1(X)$ are closed sets, $\Sigma^0_2(X)$ are $F_\sigma$ sets, and $\Pi^0_2(X)$ are $G_\delta$ sets. By convention, $\Pi^0_0(X) = \{X\}$ and $\Sigma^0_0(X) = \{ \emptyset \}$.

Even more general classes of sets from the {\em projective hierarchy}. \index{projective hierarchy} We will not go beyond its lowest level:
\begin{enumerate}[$\bullet$]
  \item $\Sigma^1_1(X)$ --  {\em analytical} subsets of $X$, i.~e., projections of Borel subsets of $X^2$ with product topology,
  \item $\Pi^1_1(X)$ --  complements of sets from $\Sigma^1_1(X)$.
\end{enumerate}

Whenever the space $X$ is determined by the context, we omit it in the notation above and write simply $\Sigma^0_1$, $\Pi^0_1$, and so on. 

Let $\varphi : X \to Y$ be a~continuous map of topological spaces. One
says that $\varphi$ is a~{\em reduction} of $A \subseteq X$ to $B
\subseteq Y$, if $\forall_{x \in X}\; x \in A \leftrightarrow
\varphi(x) \in B$. Note that if $B$ is in a~certain class of the above
hierarchies, so is $A$. For any class ${\mathcal C}$ a~set $B$ is
${\mathcal C}$-{\em hard}, if for any set $A \in {\mathcal C}$ there
exists a~reduction of $A$ to $B$. The topological hierarchy is strict
for Polish spaces, so if a~set is ${\mathcal C}$-hard, it cannot be in
any lower class. If a~${\mathcal C}$-hard set $B$ is also an element
of ${\mathcal C}$, then it is ${\mathcal C}$-{\em complete}.

\begin{figure}
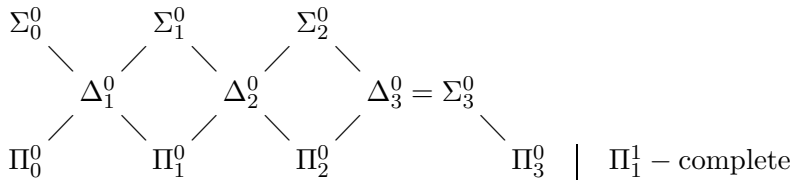

\centering
{\setlength\arraycolsep{1pt}
$\begin{array}{ccccccccccccc|c}
\Sigma^0_0 &&&& \Sigma^0_1 &&&& \Sigma^0_2 \\
& \diagdown && \diagup && \diagdown && \diagup && \diagdown &&\\ 
& & \Delta^0_1 & & & & \Delta^0_2 & & & & \Delta^0_3 = \Sigma^0_3\\
& \diagup &&  \diagdown && \diagup && \diagdown && \diagup & &\diagdown \\  
\Pi^0_0 &&&& \Pi^0_1 &&&& \Pi^0_2 &&&& \Pi^0_3 \quad & \quad \Pi^1_1-\textrm{complete}
\end{array}$}
\caption{The Borel hierarchy for deterministic tree languages.}
\label{fig:borelhierarchy}
\index{Borel hierarchy!of deterministic languages}
\end{figure}

In 2002 Niwi\'nski and Walukiewicz discovered a~surprising dichotomy
in the topological complexity of deterministic tree languages:
a~deterministic tree language has either a~very low Borel rank or it
is not Borel at all (see Fig.~\ref{fig:borelhierarchy}). We say that
an automaton $A$ admits a~{\em split} if there are two loops $p
\stackrel{\sigma, 0}{\longrightarrow} p_0 \longrightarrow \ldots
\longrightarrow p$ and $p \stackrel{\sigma, 1}{\longrightarrow} p_1
\longrightarrow \ldots \longrightarrow p$ such that the highest ranks
occurring on them are of different parity and the higher one is odd.

\begin{thm}[Niwi{\'n}ski, Walukiewicz \cite{gap}] \label{topgap} \index{Gap Theorem}
For a~deterministic automaton $A$, $L(A)$ is on the level $\Pi^0_3$ of the Borel hierarchy iff $A$ does not admit split; otherwise $L(A)$ is $\Pi^1_1$-complete (hence non-Borel). \qed
\end{thm}

An important tool used in the proof of the Gap Theorem is the
technique of difficult patterns. In the topological setting the
general recipe goes like this: for a~given class identify a~pattern
that can be ``unravelled'' to a language complete for this class; if
an automaton does not contain the pattern, then $L(A)$ should be in
the dual class. In the proof of the Gap Theorem, the split pattern is
``unravelled'' into the language of trees having only finitely many
1's on each path. This language is $\Pi^1_1$-complete (via a~reduction
of the set of well-founded trees).

In \cite{split} a~similar characterisation was obtained for the
remaining classes from the above hierarchy. Before we formulate these
result, let us introduce one of the most important technical notions
of this study. A~state $p$ is {\em replicated} \label{replicated}
\index{state!replicated} \index{replication} by a~loop $q_1 \stackrel
      {\sigma, d_0}\longrightarrow q_2 \longrightarrow \ldots
      \longrightarrow q_1 $ if there exist a~path $q_1
      \stackrel{\sigma, d_1} \longrightarrow q_2' \longrightarrow
      \ldots \longrightarrow p$ such that $d_0 \neq d_1$. We will say
      that a~flower is replicated by a~loop $\lambda$ if it contains
      a~state replicated by $\lambda$. The phenomenon of replication
      is the main difference between trees and words. We will use it
      constantly to construct hard languages that have no counterparts
      among word languages. Some of them occur in the proposition
      below.

\begin{thm}[Murlak \cite{split}] \label{borelch}
Let $A$ be a~deterministic automaton.
\begin{enumerate}[\em(1)]
\item $L(A) \in \Sigma^0_1$ iff $A$ does not contain a~weak $(0,1)$-flower.
\item $L(A) \in \Pi^0_1$ iff $A$ does not contain a~weak $(1,2)$-flower.
\item $L(A) \in \Sigma^0_2$ iff $A$ does not contain a~$(1,2)$-flower nor a weak $(1,2)$-flower replicated by an accepting loop.
\item $L(A) \in \Pi^0_2$ iff $A$ does not contain a~$(0,1)$-flower.
\item $L(A) \in \Sigma^0_3$ iff $A$ does not contain a~$(0,1)$-flower replicated by an accepting loop. \qed
\end{enumerate}
\end{thm}

\section{The Main Result}

The notion of continuous reduction defined in
Sect. \ref{sect:topology} yields a preordering on sets. Let $X$ and
$Y$ be topological spaces, and let $A\subseteq X$, $B \subseteq Y$.
We write $A \leq_W B$ (to be read ``$A$ is {\em Wadge reducible} to
$B$''), if there exists a~continuous reduction of $A$ to $B$, i.~e.,
a~continuous function $\varphi \colon X \to Y$ such that
$A=\varphi^{-1}(B)$. We say that $A$ is {\em Wadge equivalent} to $B$,
in symbols $A\equiv_W B$, if $A \leq_W B$ and $A \leq_W B$. Similarly
we write $A <_W B$ if $A \leq_W B$ and $B \not\leq_W A$. The {\em
  Wadge ordering} is the ordering induced by $\leq_W$ on the
$\equiv_W$-classes of subsets of Polish spaces. The Wadge ordering
restricted to Borel sets is called the {\em Wadge hierarchy}.

In this study we only work with the spaces $T_\Sigma$ and
$\Sigma^\omega$. Since we only consider finite $\Sigma$, these spaces
are homeomorphic with the Cantor discontinuum $\{0,1\}^\omega$ as long
as $|\Sigma|\geq 2$. In particular, all the languages we consider are
Wadge equivalent to subsets of $\{0,1\}^\omega$. Note however that the
homeomorphism need not preserve recognisability. In fact, no
homeomorphism from $T_{\Sigma}$ to $\{0,1\}^\omega$ does: the Borel
hierarchy for regular tree languages is infinite, but for words it
collapses on $\Delta^0_3$. In other words, there are regular tree
languages (even weak, or deterministic), which are not Wadge
equivalent to regular word languages. Conversely, each regular word
language $L$ is Wadge equivalent to a~deterministic tree language $L'$
consisting of trees which have a~word from $L$ on the leftmost branch.
As a~consequence, the height of the Wadge ordering of regular word
languages gives us a~lower bound for the case of deterministic tree languages, and this is essentially everything we can conclude from the word case. 

The starting point of this study is the  Wadge reducibility problem. 

\begin{center}
\fbox{
\begin{tabular}{rl}
{\sc Problem}: & Wadge reducibility \\
{\sc Input}: & Deterministic tree automata $A$ and $B$\\
{\sc Question}: & $L(A) \leq_W L(B)$?
\end{tabular}
}
\end{center}

\noindent An analogous problem for word automata can be solved fairly
easy by constructing a tree automaton
recognising Duplicator's winning strategies (to be defined in the next
section). This method however
does not carry over to trees. One might still try
to solve the Wadge reducibility problem directly by comparing carefully the
structure of two given automata, but we have chosen a~different
approach. We will provide a family of {\em canonical} deterministic
tree automata $ {\mathcal  A} = \{A_i \colon i \in I\}$ such that 
\begin{enumerate}[(1)]
\item given $i,j \in I$, it is decidable if $L(A_i) \leq_W L(A_j)$,
\item for each deterministic tree automaton there exists exactly
  one $i \in I$ such that $L(A) \equiv_W L(A_i)$, and this $i$ can be
  computed effectively for a given $A$.
\end{enumerate}
\noindent The decidability of the Wadge reducibility problem follows
easily from the existence of such a family: given two
deterministic automata $A$ and $B$, we compute $i$ and $j$ such that
$L(A) \equiv_W L(A_i)$ and $L(B) \equiv_W L(A_j)$, and check if
$L(A_i) \leq _W L(A_j)$. 

More precisely, we prove the following theorem. 
\begin{thm} \label{main}
There exists a family of deterministic tree automata \[{\mathcal C}' = \{C_\alpha \colon \alpha \in
I\} \cup \{D_\alpha, E_\alpha \colon \alpha \in J\}\] with $I= \{\alpha \colon 0<\alpha \leq \omega^{\omega\cdot3} +2\}$,
$J= \{n \colon  0<n< \omega\} \cup \{ \omega^{\omega\cdot2}
\alpha_2 + \omega^\omega \alpha_1 + n \colon  \alpha_2 <
\omega^\omega, \; 0 < \alpha_1 < \omega^\omega, \; n<\omega \}$
such that  
\begin{enumerate}[\em(1)]
\item for $0 < \alpha < \beta \leq \omega^{\omega\cdot3} +2$, whenever
  the respective automata are defined, we have 
\[\begin{array}{ccccccc}
 L(C_\alpha) &          &              &          & L(C_\beta) &          &             \\
             & \searrow &              & \nearrow &             & \searrow &             \\
             &          & L(E_\alpha) &          &             &          &  L(E_\beta) \\ 
             & \nearrow &              & \searrow &             & \nearrow &             \\
L( D_\alpha) &          &              &          & L(D_\beta) &          &             
\end{array}\]
where $\to$ means $<_W$, and $L(C_\alpha)$ and
$L(D_\alpha)$ are incomparable,

\item for each deterministic tree automaton $A$ there exists exactly
  one automaton $A' \in {\mathcal C}'$ such that $L(A') \equiv_W L(A)$
  and it is computable, i.e., there exists an algorithm
  computing for a given $A$ a pair  $(\Xi, \alpha) \in \{C\}\times I
  \cup \{D,E\}\times J$ such that $L(A) \equiv_W L(\Xi_\alpha)$. 
\end{enumerate}
\end{thm}
\noindent The family ${\mathcal C}'$ satisfies the conditions
postulated for the family of canonical automata ${\mathcal A}$: for ordinals presented as
arithmetical expressions over $\omega$ in
Cantor normal form the ordinal order is decidable, so we can take  $\{C\}\times I
  \cup \{D,E\}\times J$ as the indexing set of ${\mathcal A}$. 

Observe that the pair $(\Xi, \alpha)$ computed for a given $A$ can be seen
as a name of the $\equiv_W$-class of $L(A)$. Hence, the set 
$\{C\}\times I \cup \{D,E\}\times J$ together with the order defined in the
statement of theorem provides a complete effective description of the Wadge hierarchy
restricted to deterministic tree languages. One thing that follows is that the height of the hierarchy is $\omega^{\omega
  \cdot 3}+3$.

The remaining part of the paper is in fact a single long proof.
We start by reformulating the classical criterion of reducibility via
Wadge games in terms of automata (Sect. \ref{gamesandautomata}). This
will be the main tool of the whole argument. Then we define four ways
of composing automata: sequential composition $\oplus$, replication
$\to$, parallel composition $\land$, and alternative $\lor$
(Sect. \ref{operations}). Using the first three operations we
construct the canonical automata, all but top three ones  
(Sect. \ref{canonicalautomata}). Next, to rehearse our proof method,
we reformulate and prove
Wagner's results in terms of canonical automata (Sect.
\ref{withoutbranching}). Finally, after some preparatory remarks
(Sect. \ref{theuseofreplication}), we prove the first part of Theorem
\ref{main}, modulo three missing canonical automata.

Next, we need to show that our family our family contains all deterministic
tree automata up to Wadge equivalence of the recognised languages. Once again we turn to the
methodology of patterns used in Sect. \ref{sect:automata} and
Sect. \ref{sect:topology}. We introduce a fundamental notion of
admittance, which formalises what it means to contain an automaton as
a~pattern (Sect. \ref{patternsinautomata}). Then we generalise $\to$
to $(\iota, \kappa)$-replication
$\stackrel{(\iota,\kappa)}{\longrightarrow}$ in order to define the remaining
three canonical automata, and
rephrase the results on the Borel hierarchy and the Wagner hierarchy
in terms of admittance of canonical automata
(Sect. \ref{wagnerhierarchyandbeyond}). Basing on these results, we
show that the family of canonical automata is closed by the
composition operations (Sect. \ref{closureproperties}), and prove the
Completeness Theorem asserting that (up to Wadge equivalence) each
deterministic automaton may be obtained as an iterated composition of $C_1$ and $D_1$ (Sect. \ref{sect:completeness}). As
a~consequence, each deterministic automaton is equivalent to
a~canonical one. From the proof of the Completeness Theorem
we extract an algorithm calculating the equivalent canonical automata, which concludes the proof of Theorem~\ref{main}.

\section{Games and Automata} \label{gamesandautomata}

A classical criterion for reducibility is based on the notion of {\em Wadge games}. Let us introduce a~tree version of Wadge games (see \cite{perrin} for word version). By  {\em the $n$th level} of a tree we understand the set of nodes $\{0,1\}^{n-1}$. The 1st level consists of the root, the 2nd level consists of all the children of the root, etc. For any pair of tree languages $L\subseteq T_{\Sigma_1}, M \subseteq T_{\Sigma_2}$ the game $G_W(L,M)$ is played by Spoiler and Duplicator. Each player builds a tree, $t_S \in T_{\Sigma_1}$ and $t_D \in T_{\Sigma_2}$ respectively. In every round, first Spoiler adds some levels to $t_S$ and then Duplicator can either add some levels to $t_D$ or skip a~round (not forever). The result of the play is a pair of full binary trees. Duplicator wins the play if $t_S\in L \iff t_D \in M$. We say that Spoiler {\em is in charge of } $L$, and Duplicator {\em is in charge of} $M$. 

Just like for the classical Wadge games, a~winning strategy for Duplicator can be easily transformed into a~continuous reduction, and vice versa. 

\begin{lem} \label{wadgegame}
Duplicator has a~winning strategy in $G_W(L,M)$ iff $L\leq_W M$. 
\end{lem}

\proof A~strategy for Duplicator defines a~reduction in an obvious
way. Conversely, suppose there exist a~reduction $t \mapsto
\varphi(t)$. It follows that there exist a~sequence $n_k$ (without
loss of generality, strictly increasing) such that the level $k$ of
$\varphi(t)$ depends only on the levels $1,2, \ldots, n_k$ of
$t$. Then the strategy for Duplicator is the following: if the number
of the round is $n_k$, play the $k$th level of $t_D$ according to
$\varphi$; otherwise skip. \qed

\vspace{5pt}

We would like to point out that Wadge games are much less interactive
than classical games. The move made by one player has no influence on
the possible moves of the other. Of course, if one wants to win, one
has to react to the opponent's actions, but the responses need not be
immediate. As long as the player keeps putting some new letters, he
may postpone the real reaction until he knows more about the
opponent's plans. Because of that, we will often speak about
strategies for some language without considering the opponent and even
without saying if the player in charge of the language is Spoiler or
Duplicator.

Since we only want to work with deterministically recognisable
languages, let us redefine the games in terms of automata. Let $A$,
$B$ be deterministic tree automata. The {\em automata game} $G(A,B)$
starts with one token put in the initial state of each automaton. In
every round players perform a~finite number of the actions described
below.
\begin{enumerate}[\hbox to6 pt{\hfill}]
\item\noindent{\hskip-11 pt\bf Fire a~transition:}\ for a~token placed
  in a~state $q$ choose a transition $q
  \stackrel{\sigma}{\longrightarrow}q_1, q_2$, take the old token away
  from $q$ and put new tokens in $q_1$ and $q_2$.
\item\noindent{\hskip-11 pt\bf Remove:}\ remove a~token placed in
  a~state different from $\bot$.
\end{enumerate}\smallskip
Spoiler plays on $A$ and must perform one of these actions at least
for all the tokens produced in the previous round. Duplicator plays on
$B$ and is allowed to postpone performing an action for a~token, but
not forever. Let us first consider plays in which the players never
remove tokens. The paths visited by the tokens of each player define
a~run of the respective automaton. We say that Duplicator wins a~play
if both runs are accepting or both are rejecting. Now, removing
a~token from a~state $p$ is interpreted as plugging in an accepting
subrun in the corresponding node of the constructed run. So,
Duplicator wins if the runs obtained by plugging in an accepting
subrun for every removed token are both accepting or both rejecting.

Observe that removing tokens in fact does not give any extra power to
the players: instead of actually removing a~token, a~player may easily
pick an accepting subrun, and in future keep realising it level by
level in the constructed run. The only reason for adding this feature
in the game is that it simplifies the strategies. In a~typical
strategy, while some tokens have a significant role to play, most are
just moved along a~trivially accepting path. It is convenient to
remove them right off and keep concentrated on the real actors of the
play.

We will write $A\leq B$ if Duplicator has a~winning strategy in $G(A,B)$. Like for languages, define $A \equiv B$ iff $A\leq B$ and  $A \geq B$. Finally, let $A<B$ iff $A\leq B$ and $A\not \geq B$.

\begin{lem} \label{automatagame} 
For all deterministic tree automata $A$ and $B$, \[A \leq B \iff L(A) \leq_W L(B)\,.\]
\end{lem}

\proof First consider a modified Wadge game $G'_W(L,M)$, where players are allowed to build their trees in an arbitrary way provided that the nodes played always form one connected tree, and in every round Spoiler must provide both children for all the nodes that were leaves in the previous round. It is very easy to see that Duplicator has a winning strategy in $G'_W(L,M)$ iff he has a winning strategy in $G_W(L,M)$.

Suppose that Duplicator has a~winning strategy in $G(A,B)$. We will show that Duplicator has a~winning strategy in $G'_W(L(A), L(B))$, and hence $L(A) \leq_W L(B)$. What Duplicator should do is to simulate a~play of $G(A,B)$ in which an imaginary Spoiler keeps constructing the run of $A$ on the tree $t_S$ constructed by the real Spoiler in $G'_W(L(A), L(B))$, and Duplicator replies according to his winning strategy that exists by hypothesis. In $G'_W(L(A), L(B))$ Duplicator should simply construct a~tree such that $B$'s run on it is exactly Duplicator's tree from $G(A, B)$. 

Let us move to the converse implication. Now, Duplicator should simulate a~play in the game $G'_W(L(A), L(B))$ in which Spoiler keeps constructing a~tree such that $A$'s run on it is exactly the tree constructed by the real Spoiler in $G(A,B)$, and Duplicator replies according to his winning strategy. In $G(A,B)$ Duplicator should keep constructing the run of $B$ on $t_D$ constructed in the simulated play.  \qed

\vspace{5pt}

As a~corollary we have that all automata recognising a~given language have the same ``game power''. 

\begin{cor} 
For deterministic tree automata $A$ and $B$, if $L(A) = L(B)$, then $A \equiv B$. \qed
\end{cor}

Classically, in automata theory we are interested in the language recognised by an automaton. One language may be recognised by many automata and we usually pick the automaton that fits best our purposes. Here, the approach is entirely different. We are not interested in the language itself, but in its Wadge equivalence class. This, as it turns out, is reflected in the general structure of the automaton. Hence, our main point of interest will be that structure. 

We will frequently modify an automaton in a~way that does change the
recognised language, but not its $\equiv_W$-class. One typical thing we need to do with an automaton, is to treat it as an automaton over an extended alphabet in such a~way, that the new recognised language is Wadge equivalent to the original one. This has to be done with some care, since the automaton is required to have transitions by each letter from every state.  Suppose we want to extend the input alphabet \label{extalphabet} by a~fresh letter $\tau$. Let us construct an automaton $A_\tau$. First, if $A$ has the all-rejecting state $\bot$, add a~transition $\bot\stackrel{\tau}{\longrightarrow}\bot,\bot$. Then add an all-accepting state $\top$ with transitions $\top\stackrel{\sigma}{\longrightarrow}\top,\top$ for each $\sigma \in \Sigma \cup \{\tau\}$ (if $A$ already has the state $\top$, just add a~transition $\top\stackrel{\tau}{\longrightarrow}\top,\top$). Then for each $p \notin \{\bot,\top\}$, add a~transition $p\stackrel{\tau}{\longrightarrow}\top,\top$. 

\begin{lem} \label{extended}
For every deterministic tree automaton $A$ over $\Sigma$ and a~letter $\tau\not\in\Sigma$, $A \equiv A_\tau$.
\end{lem}

\proof It is obvious that $A \leq A_\tau$: since $A_\tau$ contains all transitions of $A$, a~trivial winning strategy for Duplicator in $G(A,A_\tau)$ is to copy Spoiler's actions. Let us see that new transitions do not give any real power. Consider $G(A_\tau, A)$. While Spoiler uses old transitions, Duplicator may again copy his actions. The only difficulty lies in responding to a~move that uses a~new transition. Suppose Spoiler does use a~new transition. If Spoiler fires a~transition $p\stackrel{\tau}{\longrightarrow}\top, \top$ for a~token $x$ in a~state $p \neq \bot$, Duplicator simply removes the corresponding token in $p$, and ignores the further behaviour of $x$ and all his descendants. The only other possibility is that Spoiler fires $\bot\stackrel{\tau}{\longrightarrow}\bot, \bot$.  Then for the corresponding token Duplicator should fire $\bot\stackrel{\sigma}{\longrightarrow}\bot, \bot$ for some $\sigma\in\Sigma$. The described strategy is clearly winning for Duplicator. \qed

\vspace{5pt}

An automaton for us is not as much a~recognising device, as a~device to carry out strategies. Therefore even two automata with substantially different structure may be equivalent, as long as they enable us to use the same set of strategies. A~typical thing we will be doing, is to replace a~part of an automaton with a~different part that gives the same strategical possibilities. Recall that by $A_q$ we denote the automaton $A$ with the initial state changed to $q$. For $q \in Q^A$ let $A_{q:=B}$ denote the automaton obtained from a copy of $A$ and a~copy of $B$ by replacing each $A$'s transition of the form $p\stackrel{\sigma, d}{\longrightarrow} q$ with $p\stackrel{\sigma, d}{\longrightarrow} q_0^B$. Note that $A_{q:=A_q}$ is equivalent to $A$. 

\begin{lem}[Substitution Lemma] \label{substitution} 
Let $A$, $B$, $C$ be deterministic automata with pairwise disjoint sets of states, and let $p$ be a~state of $C$. If $A \leq B$, then $C_{p:=A} \leq C_{p:=B}$. 
\end{lem}

\proof Consider the game $G(C_{p:=A}, C_{p:=B})$ and the following strategy for Duplicator. In $C$ Duplicator copies Spoiler's actions. If some Spoiler's token $x$ enters the automaton $A$, Duplicator should put its counterpart $y$ in the initial state of $B$, and then $y$ and its descendants should use Duplicator's winning strategy from $G(A,B)$ against $x$ and its descendants. 

Let us see that this strategy is winning. Suppose first that Spoiler's run is rejecting. Then there is a~rejecting path, say $\pi$. If on $\pi$ the computation stays in $C$, in Duplicator's run $\pi$ is also rejecting. Suppose $\pi$ enters $A$. Let $v$ be the first node of $\pi$ in which the computation is in $A$. The subrun of Spoiler's run rooted in $v$ is a~rejecting run of $A$. Since Duplicator was applying a~winning strategy form $G(A,B)$, the subrun of Duplicator's run rooted in $v$ is also rejecting. In either case, Duplicator's run is rejecting. 

Now assume that Spoiler's run is accepting, and let us see that so is Duplicator's. All paths staying in $C$ are accepting, because they are identical to the paths in Spoiler's run. For every $v$ in which the computation enters $B$, the subrun rooted in $v$ is accepting thanks to the winning strategy form $G(A,B)$ used to construct it. \qed

\section{Operations} \label{operations}

It this section we introduce four operations that will be used to
construct canonical automata representing Wadge degrees of
deterministic tree languages.

\begin{figure}
\centering
\includegraphics[width=\textwidth]{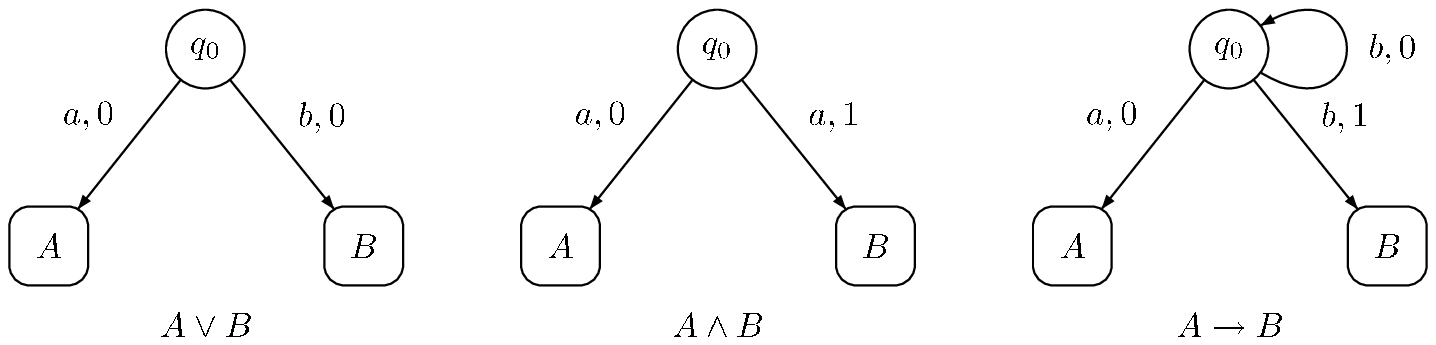}
\caption{The alternative $A \lor B$, the parallel composition $A \land
  B$, and the replication $A\to B$ (transitions to $\bot$ and $\top$ are omitted).}
\label{fig:alternative}
\end{figure}

The first operation yields an automaton that lets a~player choose between $A$ and $B$. For two deterministic tree automata $A$ and $B$ over $\Sigma$, the {\em alternative}  $A \lor B$ (see Fig. \ref{fig:alternative}) is an automaton with the input alphabet $\Sigma \cup \{a,b\}$ consisting of disjoint copies of $A$ and $B$ over the extended alphabet $\Sigma \cup \{a,b\}$, $A_{a,b}$ and $B_{a,b}$, and a~fresh initial state $q_0$ with transitions \[q_0 \stackrel{a}{\longrightarrow} q_0^{A_{a,b}},\top\,,\quad q_0 \stackrel{b}{\longrightarrow} q_0^{B_{a,b}}, \top\,, \quad \textrm{and} \;\; q_0 \stackrel{\sigma}{\longrightarrow} \top,\top \;\; \textrm{for}  \;\; \sigma\notin\{a,b\}\] (only if $L(A) = L(B) = \emptyset$ put $q_0 \stackrel{\sigma}{\longrightarrow} \bot,\bot$). By Lemma \ref{substitution}, $\equiv$ is a~congruence with respect to $\lor$. Furthermore, $\lor$ is associative and commutative up to $\equiv$.  Multiple alternatives are performed from left to right:  \[A_1 \lor A_2 \lor A_3 \lor A_4= ((A_1 \lor A_2) \lor A_3)\lor A_4\,.\]

The {\em parallel composition}  $A \land B$ is defined analogously, only now we extend the alphabet only by $a$ and add transitions \[q_0 \stackrel{a}{\longrightarrow} q_0^{A}, q_0^{B}\,, \quad \textrm{and} \;\;  q_0 \stackrel{\sigma}{\longrightarrow} \top, \top \;\; \textrm{for} \;\; \sigma \neq a\] (only if $L(A)= \emptyset$ or $L(B) = \emptyset$, put $q_0 \stackrel{\sigma}{\longrightarrow} \bot,\bot$).  Note that, while in $A \lor B$ the computation must choose between $A$ and $B$, here it continues in both. Again, $\equiv$ is a~congruence with respect to $\land$. The language $L(A \land B)$ is Wadge equivalent to $L(A) \times L(B)$ and $\land$ is associative and commutative up to $\equiv$. Multiple parallel compositions are performed from left to right, and for $n>0$ the symbol $(A)^n$ denotes $\underbrace{A \land \ldots \land A}_n$.\label{automatapower}

To obtain the {\em  replication} $A \to B$, extend the alphabet again by
$\{a,b\}$, set $\mathrm{rank}(q_0)=1$, and add and transitions 
\[q_0\stackrel{a}{\longrightarrow} q_0^{A}, \top \,,
\quad q_0 \stackrel{b}{\longrightarrow} q_0,
q_0^{B}\,, \quad \textrm{ and }\; q_0 \stackrel{\sigma}{\longrightarrow}
\bot,\bot \;\; \textrm{for}  \;\; \sigma\notin\{a,b\}\,.\] Like for two previous operations, $\equiv$ is a congruence with respect to $\to$.

\begin{figure}
\centering
% 6ta
\includegraphics[width=0.7\textwidth]{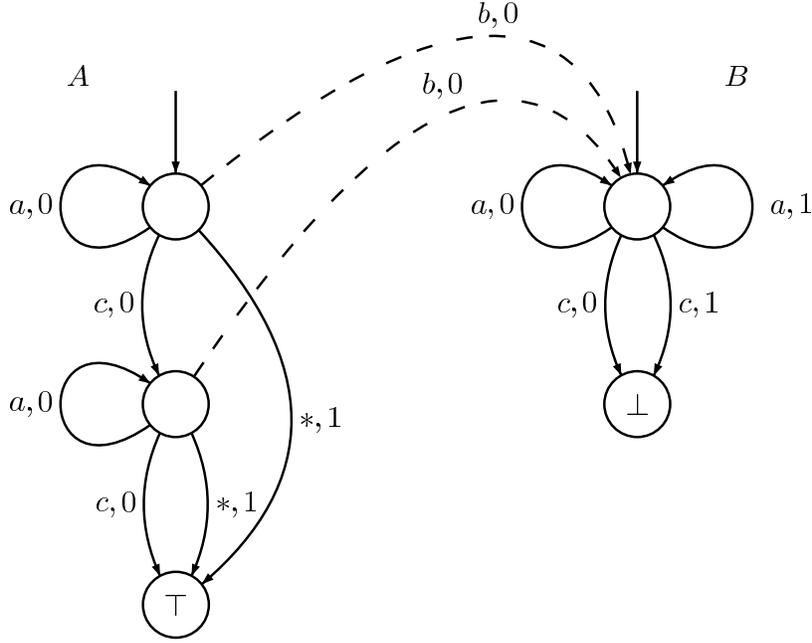}
\caption{The sequential composition $A\oplus B$.}
\label{fig:sequential}
\end{figure}

The last operation we define produces out of $A$ and $B$ an automaton that behaves as $A$, but in at most one point (on the leftmost path) may switch to $B$. A state $p$ is {\em leftmost}  if no path connecting the initial state with $p$ uses a~right transition. In other words, leftmost states are those which can only occur in the leftmost path of a run. Note that an automaton may have no leftmost states. Furthermore, a leftmost state cannot be reachable from a non-leftmost state. In particular, if an automaton has any leftmost states at all, the initial state has to be leftmost. For deterministic tree automata $A$ and $B$ over $\Sigma$, the {\em sequential composition} $A \oplus B$ (see Fig. \ref{fig:sequential})  
is an automaton with the input alphabet $\Sigma \cup \{ b\}$, where $b$ is a~fresh letter. It is constructed by taking copies of $A$ and $B$ over the extended alphabet $\Sigma \cup \{ b\}$  and replacing the transition $p\stackrel{b, 0}{\longrightarrow} r$ with $p \stackrel{b, 0}{\longrightarrow} q_0^{B_b}$ for each leftmost state $p$ and $r \in \{\bot,\top\}$.  Like for $\land$ and $\lor$, we perform the multiple sequential compositions from left to right. For $n>0$ we often use an abbreviation $n A = \underbrace{A \oplus \ldots \oplus A}_n$.  Observe that if $A$ has a leftmost state, then a state in $A \oplus B$ is leftmost iff it is a leftmost state of $A$ or a leftmost state of $B$. It follows that the $\equiv$-class of a~multiple sequential composition does not depend on the way we put parentheses. An analog of $\oplus$ for word automata defines an operation on $\equiv$-classes, but for tree automata this is no longer true. We will also see later that $\oplus$ is not commutative even up to $\equiv$.

The priority of the operations is $\oplus, \land, \lor, \to$. For instance $A_1 \to A_2 \oplus A_3 \land A_4 \lor A_5 = A_1 \to ((( A_2 \oplus A_3) \land A_4) \lor A_5)$. Nevertheless, we usually use parentheses to make the expressions easier to read. 

\begin{figure}
\centering
% 6ta
\includegraphics[width=0.6\textwidth]{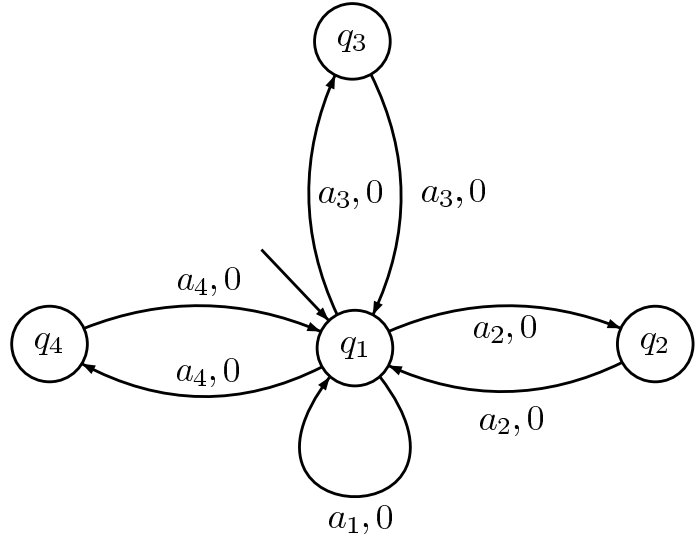}
\caption{The canonical $(1,4)$-flower $F_{(1,4)}$.}
\label{fig:flower2}
\end{figure}

Finally, let us define the basic building blocks, to which we will
apply the operations defined above. The {\em canonical $(\iota, \kappa)$-flower} $F_{(\iota, \kappa)}$ (see Fig. \ref{fig:flower2})
is an automaton with the input alphabet $\{a_\iota, a_{\iota+1} \ldots, a_\kappa\}$,  the states $q_\iota, q_{\iota+1},\ldots, q_\kappa$ where the initial state is $q_\iota$ and $\mathrm{rank}(q_i)=i$, and transitions  \[q_\iota \stackrel{a_\iota}{\longrightarrow} q_\iota,\top\,, \;\; q_\iota \stackrel{a_j}{\longrightarrow} q_j,\top\,, \;\;   q_j\stackrel{a_j}{\longrightarrow} q_0, \top\,, \;\; \textrm{and} \;\;  q_j\stackrel{a_k}{\longrightarrow} \top,\top \;\;\]
for $j=\iota+1, \iota+2, \ldots, \kappa$ and $k \neq j$. A~flower $F_{(\iota, \kappa)}$ is {\em nontrivial} if  $\iota <\kappa$.

In the definitions above we often use an all-accepting state $\top$. This is in fact a~way of saying that a~transition is of no importance when it comes to possible strategies: a~token moved to $\top$ has no use later in the play. Therefore, we may assume that players remove their tokens instead of putting them to $\top$. In particular, when a~transition is of the form $p \stackrel{\sigma}{\longrightarrow} q, \top$, it is convenient to treat it as a ``left only'' transition in which no new token is created, only the old token is moved from $p$ to $q$. Consequently, when analysing games on automata, we will ignore the transitions to $\top$.

\section{Canonical Automata} \label{canonicalautomata}

For convenience, in this section we put together the definitions of all
canonical automata (save for three which will be defined much
later) together with some very simple properties. More explanations and
intuitions come along with the proofs in the next three 
sections. 

 For each $\alpha < \omega^{\omega\cdot3}$ we define the {\em canonical automaton} $C_\alpha$. The automata $D_\alpha$ and $E_\alpha$ will only be defined for $0 < \alpha < \omega$ and $\alpha = \omega^{\omega\cdot2} \alpha_2 + \omega^\omega \alpha_1 + n$ with $0 < \alpha_1 < \omega^\omega$, $\alpha_2 < \omega^\omega$,  $n<\omega$. All the defined automata have at least one leftmost state, so the operation $\oplus$ is always non-trivial. 

Let $C_1 = F_{(0,0)}$, $D_1 =  F_{(1,1)}$, and $E_1 = F_{(0,0)} \lor F_{(1,1)}$. For $1 < \alpha< \omega$ define 
\begin{eqnarray}
C_\alpha & = & C_1 \oplus (\alpha-1) E_1\,, \nonumber \\
D_\alpha & = & D_1 \oplus (\alpha-1) E_1\,,\nonumber \\
E_\alpha & = & \alpha E_1\,. \nonumber
\end{eqnarray}

For $\omega \leq \alpha < \omega^\omega$ we only define $C_\alpha$. Let $C_{\omega} = C_1 \to C_3$ and $C_{\omega^{k+1}} = C_1 \to (C_1 \oplus C_{\omega^k})$ for $1 \leq k < \omega$. For every $\alpha$ from the considered range we have a~unique presentation $\alpha = \omega^{l_k}n_k + \omega^{l_{k-1}}n_{k-1} + \ldots + \omega^{l_0} n_0$, with $\omega> l_k >0$, $l_k > l_{k-1} > \ldots > l_0$ and $0<n_i<\omega$. For $l_0=0$ define
\begin{eqnarray}
C_\alpha &=& C_{n_0} \oplus n_1 C_{\omega^{l_1}} \oplus \ldots \oplus n_k C_{\omega^{l_k}} \quad \textrm{for odd } n_0\,, \nonumber \\
C_\alpha &=& D_{n_0} \oplus n_1 C_{\omega^{l_1}} \oplus \ldots \oplus n_k C_{\omega^{l_k}} \quad \textrm{for even } n_0\,, \nonumber
\end{eqnarray}
and for $l_0>0$ set
\begin{eqnarray}
C_\alpha &=&  n_0 C_{\omega^{l_0}} \oplus n_1 C_{\omega^{l_1}} \oplus \ldots \oplus n_k C_{\omega^{l_k}} \,. \nonumber
\end{eqnarray}

Now consider $\omega^\omega \leq \alpha < \omega^{\omega\cdot2}$. For $k<\omega$ let $C_{\omega^{\omega+k}} = F_{(0,k+1)}$, $D_{\omega^{\omega+k}} = F_{(1,k+2)}$ and $E_{\omega^{\omega+k}} = F_{(0,k+1)} \lor F_{(1,k+2)}$.  For every $\alpha$ from the considered range we have a~unique presentation $\alpha=\omega^\omega \alpha_1 + \alpha_0$ with $\alpha_0, \alpha_1<\omega^\omega$ and $\alpha_1>0$. Let $\alpha_1 = \omega^{l_k}n_k + \omega^{l_{k-1}}n_{k-1} + \ldots + \omega^{l_0} n_0$, with $\omega> l_k > l_{k-1} > \ldots > l_0$ and $0<n_i<\omega$.  For $\alpha_0=0$ and $l_0 = 1$ let 
\begin{eqnarray}
C_\alpha &=& C_{\omega^{\omega+l_0}} \oplus n_1E_{\omega^{\omega + l_1}} \oplus \ldots \oplus n_kE_{\omega^{\omega+l_k}} \,, \nonumber \\
D_\alpha &=& D_{\omega^{\omega+l_0}} \oplus n_1E_{\omega^{\omega + l_1}} \oplus \ldots \oplus n_kE_{\omega^{\omega+l_k}} \,, \nonumber \\
E_\alpha &=& E_{\omega^{\omega + l_0}} \oplus n_1E_{\omega^{\omega + l_1}} \oplus \ldots \oplus n_kE_{\omega^{\omega+l_k}} \,, \nonumber
\end{eqnarray}
for  $\alpha_0=0$ and $l_0 > 1$ let
\begin{eqnarray}
C_\alpha &=& C_{\omega^{\omega+l_0}} \oplus (n_0 - 1)E_{\omega^{\omega + l_0}} \oplus n_1E_{\omega^{\omega + l_1}} \oplus \ldots \oplus n_kE_{\omega^{\omega+l_k}} \,, \nonumber \\
D_\alpha &=& D_{\omega^{\omega+l_0}} \oplus (n_0 - 1)E_{\omega^{\omega + l_0}} \oplus n_1E_{\omega^{\omega + l_1}} \oplus \ldots \oplus n_kE_{\omega^{\omega+l_k}} \,, \nonumber \\
E_\alpha &=& n_0E_{\omega^{\omega + l_0}} \oplus n_1E_{\omega^{\omega + l_1}} \oplus \ldots \oplus n_kE_{\omega^{\omega+l_k}} \,, \nonumber
\end{eqnarray}
for $\omega>\alpha_0>0$ let 
\begin{eqnarray}
C_\alpha &=& C_{\alpha_0} \oplus E_{\omega^\omega\alpha_1} \,, \nonumber \\
D_\alpha &=& D_{\alpha_0} \oplus E_{\omega^\omega\alpha_1} \,, \nonumber \\
E_\alpha &=& E_{\alpha_0} \oplus E_{\omega^\omega\alpha_1}, \nonumber
\end{eqnarray}
and for $\alpha_0>\omega$ let 
\begin{eqnarray}
C_\alpha &=& C_{\alpha_0} \oplus E_{\omega^\omega\alpha_1} \,. \nonumber
\end{eqnarray}

Finally consider $\omega^{\omega\cdot2} \leq \alpha < \omega^{\omega\cdot3}$. Let $C_{\omega^{\omega\cdot2}} = C_1 \to F_{(0,2)}$, and for $k<\omega$ let $C_{\omega^{\omega\cdot2 +k +1}} = C_1 \to ( C_1 \oplus C_{\omega^{\omega\cdot2 +k}})$. We have a~unique presentation $\alpha = \omega^{\omega\cdot2}\alpha_2 + \omega^\omega \alpha_1 + \alpha_0$ with $\alpha_0, \alpha_1, \alpha_2 < \omega^\omega$ and $\alpha_2>0$. Let $\alpha_2 = \omega^{l_k}n_k + \omega^{l_{k-1}}n_{k-1} + \ldots + \omega^{l_0} n_0$, with $\omega> l_k > l_{k-1} > \ldots > l_0$ and $0<n_i<\omega$. For $\alpha_0=\alpha_1=0$ let 
\begin{eqnarray}
C_\alpha &=& n_0 C_{\omega^{\omega\cdot2+l_0}} \oplus n_1C_{\omega^{\omega\cdot2+l_1}} \oplus \ldots \oplus n_k C_{\omega^{\omega\cdot2+l_k}}\,, \nonumber
\end{eqnarray}
for $\alpha_0=0$ and $\alpha_1>0$ let 
\begin{eqnarray}
C_\alpha & = & C_{\omega^\omega\alpha_1} \oplus C_{\omega^{\omega\cdot2}\alpha_2}\,, \nonumber \\
D_\alpha & = & D_{\omega^\omega\alpha_1} \oplus C_{\omega^{\omega\cdot2}\alpha_2}\,, \nonumber \\
E_\alpha & = & E_{\omega^\omega\alpha_1} \oplus C_{\omega^{\omega\cdot2}\alpha_2}\,, \nonumber 
\end{eqnarray}
for $\omega>\alpha_0>0$ and $\alpha_1=0$ let
\begin{eqnarray}
C_\alpha &=& C_{\alpha_0} \oplus C_{\omega^{\omega\cdot2}\alpha_2}  \quad \textrm{for odd}\;\; \alpha_0\,, \nonumber \\
C_\alpha &=& D_{\alpha_0} \oplus C_{\omega^{\omega\cdot2}\alpha_2}  \quad \textrm{for even}\;\; \alpha_0\,, \nonumber
\end{eqnarray}
and in the remaining case ($\alpha_0>\omega$ or $\alpha_1>0$) let
\begin{eqnarray}
C_\alpha &=& C_{\omega^\omega \alpha_1 + \alpha_0} \oplus C_{\omega^{\omega\cdot2}\alpha_2}\,. \nonumber
\end{eqnarray}

Let $\mathcal C$ denote the family of the canonical automata, i.~e.,
\begin{eqnarray}
{\mathcal C} &=& \{C_\alpha: \alpha<\omega^{\omega\cdot3}\} \cup \{D_n, E_n \colon n<\omega\} \cup \nonumber\\
         & & \cup \; \{D_{\omega^{\omega\cdot2} \alpha_2 + \omega^\omega \alpha_1 + n}, E_{\omega^{\omega\cdot2} \alpha_2 + \omega^\omega \alpha_1 + n} \colon 0 < \alpha_1 < \omega^\omega\,,\; \alpha_2 < \omega^\omega\,, \; n<\omega\} \,. \nonumber
\end{eqnarray} 
In the next three sections we will investigate the order induced on ${\mathcal C}$ by the Wadge ordering of the recognised languages. 

Now, let us discuss briefly the anatomy and taxonomy of the canonical automata. {\em Simple}  automata are those canonical automata that cannot be decomposed with respect to~$\oplus$,  i.~e., the automata on the levels $\omega^{k}$, $\omega^{\omega+k}$, and $\omega^{\omega\cdot 2+k}$ for $k<\omega$. {\em Complex}  automata are those obtained from simple ones by means of $\oplus$. If for some automata $A_1, A_2, \ldots, A_n$ we have  $A=A_1 \oplus A_2 \oplus \ldots \oplus A_n$, we call $A_i$ {\em components} of $A$. If $A_i$ are simple, they are called {\em simple components} of $A$, $A_1$ is the {\em head component}, and $A_n$ is the {\em tail component}.  {\em Non-branching} canonical automata are those constructed from flowers without the use of $\to$, i.~e., $C_{\omega^\omega \alpha + n}, D_{\omega^\omega \alpha + n}, E_{\omega^\omega \alpha + n}$ for $\alpha<\omega^\omega$ and $n<\omega$. The remaining automata are called {\em branching}. The term {\em head loop} refers to any minimal-length loop around the initial state. If the head component of a canonical automaton is branching, then the automaton has only one head loop. Similarly, if the head component is $C_1$ or $D_1$.

According to the definition of the automata game, in a~branching
transition a token is split in two. However in branching canonical
automata, the role to be played by two new tokens is very
different. Therefore, we prefer to see the process of splitting
a~token as producing a~new token that moves along the right branch of
the transition, while the original one moves left. Thus each token
moves along the leftmost path from the node it was born in, bubbling
out new tokens to the right. Let us prove the following simple yet 
useful property of those paths. 

\begin{prop}\label{rejectingpath}
If a~run constructed by a~player in charge of a canonical automaton is
rejecting, one of the tokens has visited a~rejecting path.  
\end{prop}

\proof  Observe that in a~canonical automaton the only loop using
right transitions is the loop around $\top$. In other words, each path
of the constructed computation that does not reach $\top$ goes right
only a~bounded number of times (depending on the automaton). Now,
consider a~rejecting run constructed during a~play. It must
contain a~rejecting path $\pi$. The token created during the last
right transition on $\pi$ visits a~suffix of $\pi$, which of course is
a~rejecting path.\qed 

\vspace{5pt}

Recall that we have defined the operation $\oplus$ in such a~way, that
the second automaton can only be reached via a~leftmost path. This
means that the only token that can actually move from one simple
automaton to another is the initial token. Since passing between the
simple automata forming a~canonical automaton is usually the key
strategic decision, we call the initial token {\em critical}, and the
path it moves along, the {\em critical path}.

Since we can remove the tokens from $\top$ with no impact on the
outcome of the game, we can assume that in transitions of the form $p
\stackrel{\sigma}{\longrightarrow} q, \top$ or 
$p \stackrel{\sigma}{\longrightarrow} \top, q$  no new tokens are
produced, only the old token moves from $p$ to $q$. The following fact
relies on this convention. 

\begin{prop}\label{finitelymanytokens} 
If a~player in charge of a canonical automaton produces infinitely may
tokens, the resulting run is rejecting.  
\end{prop}

\proof We will proceed by structural induction. The claim holds
trivially for non-branching automata. Suppose now that $A = C_1 \to
A'$. If the constructed run is to be accepting, the player can only
loop a finite number of times in the head loop of
$A$, thus producing only a~finite number of new tokens. 
By the induction hypothesis for $A'$, those tokens can only have
finitely many descendants. Hence, in the whole play there can be only
finitely many tokens.

Now, take $A=A'\oplus A''$. Suppose there were infinitely many tokens
in some play on $A$. Observe that all the tokens in $A''$ are
descendants of $A''$'s critical token. Hence, if there were infinitely
many tokens in $A''$, by the induction hypothesis for $A''$ the whole
run is rejecting. Suppose there were infinitely many tokens in
$A'$. Consider a~play in which the critical token instead of moving to
$A''$ stays in the last accepting loop of $A'$ (it exists by the
definition of canonical automata). In such a~play a~run of $A'$ is
build. Since there are infinitely many tokens used, the run is
rejecting by the induction hypothesis for $A'$. Consequently, the run
of $A$ constructed in the original play must have been rejecting as
well. \qed

\section{Without Branching} \label{withoutbranching}

In this section we briefly reformulate Wagner's results on regular word languages \cite{wagner} in terms of canonical automata. For the sake of completeness, we reprove them in our present framework. 

The scenario is just like for tree languages: define a~collection of canonical automata, prove that they form a~strict hierarchy with respect to the Wadge reducibility, check some closure properties, and provide an algorithm calculating the equivalent canonical automaton for a~given deterministic automaton, thus proving that the hierarchy is complete for regular languages.

Since the non-branching canonical automata have only left transitions, they only check a~regular word property on the leftmost path. It is easy to see that for each word language $K$, the language of trees whose leftmost branch is in $K$ is Wadge equivalent to $K$. Based on this observation, we will treat the non-branching canonical automata as automata on words. 

Let $L_{(\iota, \kappa)}$ denote the language of infinite words over $\{\iota, \iota+1, \ldots, \kappa\}$ that satisfy the parity condition, i.~e., the highest number occurring infinitely often is even.

\begin{lem} \label{eliotakappa}
For every index $(\iota, \kappa)$ and every deterministic tree automaton $A$ of index at most $(\iota, \kappa)$, 
\begin{enumerate}[\em(1)]
\item $L(A) \leq L_{(\iota, \kappa)}$,
\item $L(F_{(\iota, \kappa)}) \equiv_W L_{(\iota, \kappa)}$, 
\item $L_{(\iota,\kappa)} \leq L_{(\iota,\kappa')}$ iff $(\iota,\kappa) \leq (\iota',\kappa')$.
\end{enumerate}
\end{lem}

\proof A reduction showing (1) is given by $w \mapsto
\mathrm{rank}(q_0)\mathrm{rank}(q_1)\mathrm{rank}(q_2) \ldots $, where
$q_0q_1q_2\ldots$ is the run of $A$ on the word $w$.

For (2) the remaining reduction is obtained by assigning to a~sequence
$n_1n_2n_3\ldots$ the tree with the word
$a_{n_1}a_{n_1}a_{n_2}a_{n_2}a_{n_3}a_{n_3}\ldots$ on the leftmost
branch, and a $a_{\iota}$ elsewhere.

Since $L_{(\iota,\kappa)}$ can be recognised by a~$(\iota, \kappa)$
automaton, one implication in (3) follows from (1). To prove the
remaining one, it is enough to show that $L_{(\iota,\kappa)} \not \leq
L_{\overline{(\iota,\kappa)}}$. Let us fix $\iota$ and proceed by
induction on $\kappa$. For $\iota=\kappa$ the claim holds trivially:
$\emptyset \subseteq T_{\{1\}}$ and $T_{\{1\}}$ are not reducible to
each other. Take $\iota<\kappa$ and let $(\iota',\kappa') =
\overline{(\iota, \kappa)}$. Consider the game $G_\kappa =
G_W(L_{(\iota, \kappa)}, L_{(\iota', \kappa')})$. As long as
Duplicator does not play $\kappa'$, Spoiler can follow the strategy
from $G_{\kappa-1} = G_W(L_{(\iota,\kappa-1)},
L_{\overline{(\iota,\kappa-1)}})$. If Duplicator never plays
$\kappa'$, he loses. When Duplicator plays $\kappa'$, Spoiler should
play $\kappa$, and then again follow the strategy from $G_{\kappa-1}$,
and so on. Each time, Duplicator has to play $\kappa'$ finally,
otherwise he loses. But then he must play $\kappa'$ infinitely many
times, and he loses to, since $\kappa'$ and $\kappa$ have different
parity. \qed

\vspace{5pt}

For the sake of convenience let us renumber the non-branching automata. For $\eta<\omega^\omega$ let \[\hat C_{\omega \eta  + n} = C_{\omega^\omega \eta + n}\,, \qquad \hat D_{\omega \eta  + n} = D_{\omega^\omega \eta  + n}\,, \qquad \hat E_{\omega \eta  + n} = E_{\omega^\omega \eta  + n}\,. \]
Let  $\hat {\mathcal C} = \{\hat C_\alpha, \hat D_\alpha, \hat E_\alpha: 1<\alpha<\omega^\omega\}$.\label{hat}

\begin{prop} \label{omegaorder}
For $0 < \alpha < \beta < \omega^\omega$ we have
\[\begin{array}{ccccccc}
\hat C_\alpha &          &              &          & \hat C_\beta &          &             \\
             & \searrow &              & \nearrow &             & \searrow &             \\
             &          & \hat E_\alpha &          &             &          & \hat E_\beta \\ 
             & \nearrow &              & \searrow &             & \nearrow &             \\
\hat D_\alpha &          &              &          & \hat D_\beta &          &             
\end{array}\]
where $\to$ means $<$. Furthermore, $\hat C_\alpha \not \leq \hat D_\alpha$ and $\hat D_\alpha \not \leq \hat C_\alpha$.
\end{prop}

\proof First, observe that $\hat C_\alpha \leq \hat E_\alpha$: a~winning strategy for Duplicator in $G(\hat C_\alpha, \hat E_\alpha)$ is to move the initial token to $F_{(0,\kappa)}$, and then simply copy Spoiler's actions. Analogously, $\hat D_\alpha \leq \hat E_\alpha$.

Let us now suppose that $\beta=\omega^k$ for some $k<\omega$. Then $\alpha=\omega^{k-1}n_{k-1} + \ldots + n_0$. 
By definition,  $\hat E_\alpha$, has index at most  $(0, k)$. Hence, by Lemma \ref{eliotakappa}, $\hat E_\alpha \leq F_{(0,k)} = \hat C_\beta$. If we increase the ranks in each $F_{(0,l)}$ in $\hat E_\alpha$ by 2, we obtain an automaton with index at most $(1,k+1)$ recognising the same language. Hence, we also have $\hat E_\alpha \leq F_{(1,k+1)}=\hat D_\alpha$.
 
Now, consider the general case.  We have a~unique pair of presentations $\alpha=\omega^km_k + \ldots + m_0$ and $\beta=\omega^kn_k + \ldots + n_0$ with $n_k>0$. Let $i$ be the largest number satisfying $m_i \neq n_i$. Since $\alpha<\beta$, $m_{i_0}<n_{i_0}$. Thus we have $\hat E_\alpha \equiv \hat E_{\alpha_0} \oplus \hat E_{\gamma}$, $\hat C_\beta \equiv \hat E_{\beta_0}  \oplus \hat C_{\gamma}$, where $\gamma = \omega^k m_k+\ldots+\omega^im_i$, $\alpha_0 = \omega^{i-1}m_{i-1} + \ldots + m_0$, $\beta_0 = \omega^i (n_i-m_i) + \omega^{i-1}m_{i-1} + \ldots + m_0$.  Consider the game $G(\hat E_{\alpha_0} \oplus \hat E_{\gamma}, \hat E_{\beta_0} \oplus \hat C_{\gamma})$. The strategy for Duplicator is as follows. First move the token to the last $F_{(0,i)}$ in $\hat C_{\beta_0}$. Then follow the strategy given by the inequality $\hat E_{\alpha_0} \leq F_{(0,i)}$, as long as Spoiler stays in $\hat E_{\alpha_0}$. If he stays there forever, Duplicator wins. If Spoiler moves to $\hat E_\gamma$, Duplicator should do the same and keep copying Spoiler's move from that moment on. This also guarantees winning. The proof for $\hat D_\beta$ is entirely analogous.

In order to prove that the inequalities are strict it is enough to show that $\hat C_\alpha \not \leq \hat D_\alpha$ and $\hat D_\alpha \not \leq \hat C_\alpha$. We only prove that $\hat C_\alpha \not \leq \hat D_\alpha$; the proof for $\hat D_\alpha \not \leq \hat C_\alpha$ is entirely analogous. Let us proceed by induction. The assertion holds for $\alpha=1$: the whole space is not reducible to the empty set. Let us take $\alpha>1$. By the definition, $\hat C_{\alpha} = F_{(0,k)} \oplus \hat E_{\gamma}$, $\hat D_\alpha = F_{(1,k+1)} \oplus \hat E_\gamma$, where $\alpha = \omega^k + \gamma$.  Consider the game $G(F_{(0,k)} \oplus \hat E_{\gamma}, F_{(1,k+1)} \oplus \hat E_\gamma)$. We have to find a~winning strategy for Spoiler. If Duplicator never leaves $F_{(1,k+1)}$ Spoiler can stay in $F_{(0,k)}$ and win using the strategy given by the Lemma \ref{eliotakappa} (3). Otherwise, after Duplicator enters $\hat E_\gamma$, he must make choice between $\hat C_\gamma$ and $\hat D_\gamma$. Spoiler should loop in any loop of $F_{(0,k)}$ waiting for Duplicator's choice. When Duplicator chooses one of  $\hat C_\gamma$, $\hat D_\gamma$, Spoiler should choose the other one and use the strategy given by the induction hypothesis. \qed

\vspace{5pt}

The third step is proving closure by natural operations. For word automata only the operations $\oplus$ and $\lor$ make sense. The operation $\lor$ is defined just like for trees. To define $\oplus$, simply assume that all states are leftmost. It is easy to see that $\equiv$ is a congruence with respect to $\oplus$ and $\land$. Both operations are associative up to $\equiv$.

\begin{prop} \label{omegaclosure}
For each $A_1,A_2 \in \hat {\mathcal C}$, one can find in polynomial time automata $A_\lor, A_\oplus \in \hat {\mathcal C}$ such that $A_1 \lor A_2 \equiv A_\lor$ and $A_1 \oplus A_2 \equiv A_\oplus$.  
\end{prop}

\proof Closure by $\lor$ is easy. For $A_1 \geq A_2$ it holds that $A_1 \lor A_2 \equiv A_1$. Indeed,  $A_1 \lor A_2 \geq A_1$, as $A_1 \lor A_2$ contains a copy of $A_1$. For the converse inequality consider $G(A_1 \lor A_2, A_1)$. In the first move, Spoiler moves his initial token either to $A_1$ or to $A_2$. If Spoiler chooses $A_1$, Duplicator may simply mimic Spoiler's actions in his copy of $A_1$. If Spoiler chooses $A_2$, Duplicator wins by applying the strategy from $G(A_2, A_1)$, guaranteed by the inequality $A_1 \geq A_2$. 

In the remaining case $A_1$ and $A_2$ are incomparable. But then $A_1 = \hat C_\alpha$,  $A_2 = \hat D_\alpha$ for some $\alpha<\omega^\omega$  (or symmetrically). It is very easy to see that $\hat C_\alpha \lor \hat D_\alpha \equiv \hat E_\alpha$. 

Let us now consider $A_1\oplus A_2$.  Since $\oplus$ is associative up to $\equiv$ and only depends on the $\equiv$-classes of the input automata, it is enough to prove the claim for simple $A_1$; in order to obtain a~canonical automaton for $(A_1^{(1)} \oplus \ldots \oplus A_1^{(n)}) \oplus A_2$, take $A_1^{(1)} \oplus (A_1^{(2)} \oplus \ldots (A_1^{(n)} \oplus A_2) \ldots )$. Let us first consider $A_1=\hat C_{\omega^k}$.  Observe that if $\hat C_{\omega^k} \geq B$, $\hat C_{\omega^k} \oplus B \equiv \hat C_{\omega^k}$. It is enough to give a strategy for Duplicator in $G(\hat C_{\omega^k} \oplus B, \hat C_{\omega^k})$, since the other inequality is obvious. To win, Duplicator should first copy Spoiler's actions, as long as Spoiler stays in $\hat C_{\omega^k}$. When Spoiler moves to $B$, Duplicator should simply switch to the strategy from $G(B, \hat C_{\omega^k})$.

Using the property above, we easily reduce the general situation to one of the following cases: $\hat C_{\omega^k} \oplus \hat C_{\eta \omega^{k+1}}$, $\hat C_{\omega^k} \oplus \hat D_{\eta \omega^{k}}$, or $\hat C_{\omega^k} \oplus \hat E_{\eta \omega^{k}}$.  In the third case, the automaton is already canonical. Let us calculate the result in the first two cases. 

In the first case we have $\hat C_{\omega^k} \oplus \hat C_{\eta \omega^{k+1}} \equiv \hat C_{\eta \omega^{k+1}}$. Consider the game $G(\hat C_{\omega^k} \oplus \hat C_{\eta \omega^{k+1}},  \hat C_{\eta \omega^{k+1}})$. Let $\hat C_{\omega^l}$ be the head component of $\hat C_{\eta \omega^{k+1}}$. It holds that $l>k$. In order to win the game, while Spoiler stays inside $\hat C_{\omega^k}$, Duplicator should stay in $\hat C_{ \omega^{l}}$ and use the strategy from $G(\hat C_{\omega^k},\hat C_{\omega^{l}})$. When  Spoiler enters $\hat C_{\eta \omega^{k+1}}$, Duplicator may simply copy his actions. The converse inequality is trivial.

In the second case  there are two possibilities. If the head component of $\hat D_{\eta \omega^{k}}$ is $\hat D_{\omega^{l}}$ with $l>k$, proceeding as before one proves $\hat C_{\omega^k} \oplus \hat D_{\eta \omega^{k+1}} \equiv \hat D_{\eta \omega^{k+1}}$. But if $l=k$, we have $\hat C_{\omega^k} \oplus \hat D_{\eta \omega^{k}} \equiv \hat C_{\eta\omega^{k} + \omega^k}$. Consider the game $G( \hat C_{\eta\omega^{k} + \omega^k}, \hat C_{\omega^k} \oplus \hat D_{\eta \omega^{k}})$. While Spoiler stays in $\hat C_{\omega^k}$, Duplicator should copy his actions. When Spoiler leaves $\hat C_{\omega^k}$, he has to choose between $\hat D_{\omega^k}$ and the next copy of $\hat C_{\omega^k}$. If he chooses $\hat D_{\omega^k}$, Duplicator also moves to his copy of $\hat D_{\omega^k}$, and mimics Spoiler actions. Suppose Spoiler chooses $\hat C_{\omega^k}$. Then Duplicator stays in his head component, and mimics  Spoiler's actions, as long as he stays in $\hat C_{\omega^k}$. When Spoiler leaves $\hat C_{\omega^k}$, he enters the initial state of $\hat E_{\eta' \omega^k}$, where $\eta'+1 = \eta$. Duplicator should exit $\hat C_{\omega^k}$, go past $\hat D_{\omega^k}$,  and enter his copy of $\hat E_{\eta' \omega^k}$. From now on, he can copy Spoiler's actions. 

For $A_1 = \hat D_{\omega^k}$, simply dualise the claims and the proofs. For  $A_1 = \hat E_{\omega^k}$, note that $\hat E_{\omega^k} \oplus A_2 \equiv \hat C_{\omega^k} \oplus A_2 \,\lor\, \hat D_{\omega^k} \oplus A_2$, and the equivalent canonical automaton can be obtained by previous cases. \qed

\vspace{5pt} 

Let us now see that the hierarchy is complete for word languages.

\begin{thm} \label{wagner}
For each word automaton $A$ one can find in polynomial time a~canonical non-branching automaton $B$ such that $L(A) \equiv_W L(B)$.
\end{thm}

\proof We will proceed by induction on the height of the DAG of
strongly connected components of $A$. Without loss of generality we
may assume that all states of $A$ are reachable from the initial
state. In such case, the DAG of SCCs is connected and has exactly one {\em root}
component, the one containing the initial state of the automaton. 

 Suppose that the automaton is just one strongly connected component. Let $(\iota, \kappa)$ be the highest index for which $A$ contains  a~$(\iota, \kappa)$-flower. It is well defined, because if $A$ contains a~$(0,k)$-flower and a~$(1,k+1)$-flower, it must also contain a~$(0, k+1)$-flower. By Theorem \ref{indch}, $A$ is equivalent to a~$(\iota, \kappa)$-automaton and so, by Lemma \ref{eliotakappa}, $A \leq F_{(\iota,\kappa)}$. On the other hand it is easy to see, that in $G(F_{(\iota, \kappa)}, A)$, Duplicator may easily use the $(\iota, \kappa)$-flower in $A$ to mimic Spoiler's actions in $F_{(\iota, \kappa)}$. Hence, $A \equiv F_{(\iota, \kappa)}$.

Now, suppose that the DAG of SCCs of $A$ has at least two nodes. Let
$X$ be the root SCC. Like before, let $(\iota,\kappa)$ be the maximal index such that $X$ contains a~$(\iota,\kappa)$-flower. Let $q_1, \ldots, q_m$ be all the states reached by the transitions exiting $X$ (the ``initial'' states of the SCCs that are children of $X$). Recall that $A_q$ is the automaton $A$ with the initial state set to $q$. Let $B_i$ be the canonical non-branching automaton equivalent to $A_{q_i}$. It is easy to see that $A \equiv F_{(\iota,\kappa)} \oplus (B_1 \lor B_2 \lor \ldots \lor B_m)$. \qed

\begin{figure}
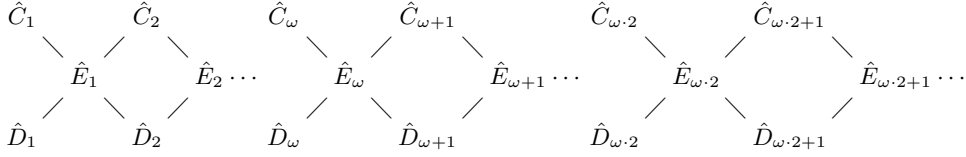

\centering
{\footnotesize {\setlength\arraycolsep{1pt}
$\begin{array}{ccccccccccccccccccccccccccccccc}
\hat C_1      &         &   &         &\hat C_2        &         &   &      &
\hat C_{\omega}&         &   &         &\hat C_{\omega+1}&         &   &      &
\hat C_{\omega\cdot2}&         &   &         &\hat C_{\omega\cdot2+1}&         &   &      &
        &   &         \\

   &\diagdown&   &\diagup  &   &\diagdown&   &      & 
   &\diagdown&   &\diagup  &   &\diagdown&   &      & 
   &\diagdown&   &\diagup  &   &\diagdown&   &      & 
   &   & \\

   &         &\hat E_1&         &   &         &\hat E_2&\cdots&
   &         &\hat E_{\omega}&         &   &         &\hat E_{\omega+1}&\cdots&
   &         &\hat E_{\omega\cdot2}&         &   &      
   &\hat E_{\omega\cdot2+1}&\cdots&
        & \\

   &\diagup  &   &\diagdown&   &\diagup  &   &      & 
   &\diagup  &   &\diagdown&   &\diagup  &   &      & 
   &\diagup  &   &\diagdown&   &\diagup  &   &      & 
   &   & \\

\hat D_1&         &   &         &\hat D_2&  &   &      &
\hat D_{\omega}&         &   &         &\hat D_{\omega+1}  &         &   &      &
\hat D_{\omega\cdot2}&         &   &         &\hat D_{\omega\cdot2+1}  &         &   &      &
        &   &        \\
\end{array}$}}

\caption{An initial segment of the Wagner hierarchy}
\label{fig:wagnerhierarchy}
\index{Wagner hierarchy}
\end{figure}

\section{The Use of Replication} \label{theuseofreplication}

Branching automata are defined by iterating $\to$. The significance of $\to$ lies in the fact that closing the family of non-branching automata by this operation gives, up to Wadge equivalence, almost all deterministic tree languages (only $C_{\omega^{\omega\cdot3}}$, $C_{\omega^{\omega\cdot3}+1}$, and $C_{\omega^{\omega\cdot3}+2}$ will be defined by means of a~stronger replication). In particular, we will show that the operation $\land$ is not needed. In other words, $\to$ is everything that deterministic tree automata have, which word automata have not. Let us see then what the use of the operation $\to$ is.

There are two kinds of simple branching automata. The first one is obtained by iterating $\to$ on $C_3$, and generalises $C_n$. Intuitively, $C_n=C_1\oplus(n-1)E_1$ lets a~player in an automata game change his mind $n-1$ times in the following sense. First, the player moves his (only) token along the head loop. The head loop is accepting, so if he keeps looping there forever, the resulting run will be accepting. But after some time he may decide that producing an accepting run is not a~good idea. In such a~case he can move to the rejecting loop in the first copy of $E_1$. Later he may want to change his mind again, and again, until he reaches the last copy of $E_1$. Now, when the player is in charge of $C_\omega = C_1 \to C_3$ he can choose a~number $n<\omega$, and looping in the head loop of $C_\omega$ produce $n$ tokens in the head loop of his copy of $C_3$. We will see that with those tokens it is possible to simulate any strategy designed for $C_{n+2}$. In other words, $C_\omega$ offers the choice between $C_n$ for arbitrarily high $n \geq 3$. The automaton $C_{\omega^2} = C_1 \to (C_1\oplus (C_1\to C_3))$ lets you choose the number of times you will be allowed to choose some $C_n$, and so on. 

The second kind of simple branching automata, obtained by iterating $\to$ on $C_{\omega^{\omega+1}}$, does the same with $C_{\omega^{\omega+n}}$ instead of $C_n$. For instance, $C_{\omega^{\omega\cdot2}} = C_1 \to C_{\omega^{\omega+1}}$ lets the player choose any $C_{\omega^{\omega+n}} = \hat C_{\omega^n}$ (see page \pageref{hat}), and in consequence $L(C_{\omega^{\omega\cdot2}})$ is hard for the class of regular languages of words.

Let us now see the proofs. The first lemma justifies the name replication. 

\begin{lem} \label{replication} For all automata $A, B$ and all $0< k <\omega$, 
\begin{enumerate}[\em(1)]
\item $A\to B\; \geq \; (A\to B) \land (B)^k$,
\item $C_1\to B\; \geq \; (B)^k$.
\end{enumerate} 
\end{lem} 

\proof To see that $(1)$ holds,  consider $G((A\to B)\land (B)^k, A\to B)$. Spoiler's initial moves produce a~token $x$ in the head loop of $A\to B$, and
tokens $x_1, \ldots, x_k$, each in a~different copy of $B$. Duplicator should
\begin{figure}
\centering
\includegraphics[width=\textwidth]{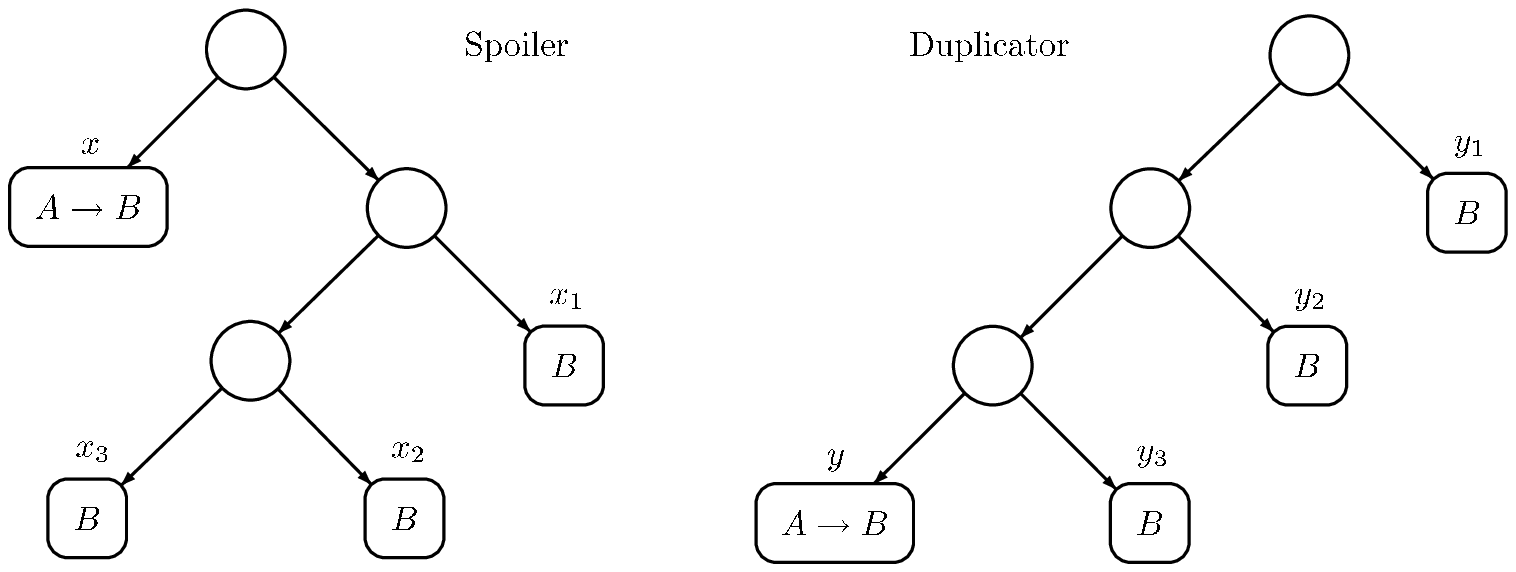}
\caption{An initial part of the play in $G((A\to B)\land (B)^3, A\to B)$.}
\label{fig:replication}
\end{figure}
loop his starting token $y$ around the head loop of $A\to B$ exactly $k$ times producing for each $x_i$ a~{\em doppelg\"anger} $y_i$ and move them all to the initial state of $B$ (see Fig.~\ref{fig:replication}). From now on $y$ mimics $x$, and $y_i$ mimics $x_i$ for $i=1, \ldots, k$. 

For the proof of $(2)$ it is enough to check that $(C_1 \to B) \land (B)^k \geq (B)^k$.  Clearly $C_1 \to B \geq C_1$. By Lemma \ref{substitution}, $(C_1 \to B) \land (B)^k \geq C_1 \land (B)^k$, and the claim follows. \qed

\vspace{5pt}

Next we need to calculate the value of $(C_3)^n$ and $(C_{\omega^{\omega+1}})^n$. Apart from canonical $(\iota,\kappa)$-flowers $F_{(\iota, \kappa)}$, we consider the following automata containing weak $(\iota,\kappa)$-flowers (see page \pageref{weakflowers}): \[WF_{(0,n)} = \underbrace{C_1 \oplus D_1 \oplus C_1 \oplus D_1 \oplus \ldots}_{n+1}\,,\;\; WF_{(1,n+1)} = \underbrace{D_1 \oplus C_1 \oplus D_1 \oplus C_1 \oplus \ldots}_{n+1}\,.\]  We will refer to these automata as weak $(\iota,\kappa)$-flowers too. In fact, $WF_{(0,n)} \equiv C_{n+1}$, $WF_{(1,n+1)} \equiv D_{n+1}$, but we find the notation convenient.\label{canonicalweakflowers} 

A pair $(i_1,i_2) \in \omega\times\omega$ is called {\em even} if both $i_1$ and $i_2$ are even. Otherwise $(i_1,i_2)$ is {\em odd}. Let $[\iota,\kappa]$ denote the set $\{\iota, \iota+1, \ldots, \kappa\} \subseteq \omega$ with the natural order. Consider the set $[\iota,\kappa] \times [\iota',\kappa']$ with the product order: $(x_1, y_1) \leq (x_2, y_2)$ if $x_1 \leq x_2$ and $y_1 \leq y_2$. For $m=0,1$ and $n \geq m$ define an {\em alternating chain of type $(m,n)$},  or $(m,n)$-chain, as a~sequence $(x_m,y_m) < (x_{m+1}, y_{m+1}) < \ldots < (x_n, y_n)$, such that $(x_i, y_i)$ is even iff $i$ is even. Suppose we have a~$(m,n)$-chain of maximal length in $[\iota,\kappa] \times [\iota',\kappa']$. The parity of $n$ is equal to the parity of $(\kappa, \kappa')$, as defined above, for otherwise we could extend the alternating chain with $(\kappa, \kappa')$ and get a~$(m,n+1)$-chain. Consequently, the following operation is well-defined:\[(\iota,\kappa) \land (\iota',\kappa') = \textrm{ the type of the longest alternating chain in } [\iota,\kappa] \times [\iota',\kappa']\,.\] 

\begin{lem} \label{flowers}
For all indices $(\iota_1,\kappa_1)$ and $(\iota_2, \kappa_2)$ it holds that 
\begin{eqnarray} 
F_{(\iota_1,\kappa_1)} \land F_{(\iota_2,\kappa_2)} &\equiv& F_{(\iota_1,\kappa_1) \land (\iota_2,\kappa_2)}\,, \nonumber\\
WF_{(\iota_1,\kappa_1)} \land WF_{(\iota_2,\kappa_2)} &\equiv& WF_{(\iota_1,\kappa_1) \land (\iota_2,\kappa_2)}\,. \nonumber
\end{eqnarray}
In particular, $(F_{(0,2)})^k \equiv F_{(0,2k)}$ and $(WF_{(0,2)})^k \equiv WF_{(0,2k)}$. Equivalently,  $(C_{\omega^{\omega+1}})^k \equiv C_{\omega^{\omega+1 + 2k}}$ and  $(C_3)^k = C_{2k+1}$.
\end{lem}

\proof By Lemma \ref{eliotakappa}, $L(F_{(i, j)}) \equiv_W L_{(i, j)}$, so  $L(F_{(\iota_1,\kappa_1)} \land F_{(\iota_2,\kappa_2)})\equiv_W L_{(\iota_1,\kappa_1)} \times L_{(\iota_2,\kappa_2)}$, where $L \times M = \{(x_1, y_1)(x_2, y_2)\ldots \colon x_1x_2\ldots \in L,\; y_1y_2\ldots \in M \}$. We will show that $L_{(\iota_1,\kappa_1)} \times L_{(\iota_2,\kappa_2)} \equiv_W L_{(\iota, \kappa)}$, where ${(\iota,\kappa) = (\iota_1,\kappa_1)\land(\iota_2,\kappa_2)}$.

Consider the following automaton $A$. The state space is the set \[[\iota_1, \kappa_1]\times[\iota_2,\kappa_2] \to \{0, 1, 2\}\] and the initial state is the function constantly equal $0$. The transition relation $\delta$ is defined as $(f,\sigma, g) \in \delta$ iff for all $i$ and $j$,  $(f(i,j), \sigma, g(i,j)) \in \delta_{(i,j)}$, where $\delta_{(i,j)}$ is defined as 
\begin{eqnarray}
&&0 \stackrel{(i,*)}{\longrightarrow} 1\,, \qquad 0 \stackrel{(k,*)}{\longrightarrow} 0 \textrm{ for all } k \neq i\,, \nonumber \\
&&1 \stackrel{(*,j)}{\longrightarrow} 2\,, \qquad 1 \stackrel{(*, k)}{\longrightarrow} 1 \textrm{ for all } k \neq j\,,  \nonumber \\
&&2 \stackrel{(*,*)}{\longrightarrow} 1\,, \nonumber 
\end{eqnarray}
with $*$ denoting any letter.

Let us now define the rank function. For $i\in[\iota_1,\kappa_1]$ and  $j\in[\iota_2,\kappa_2]$, let $(\iota',\kappa') = (\iota_1, i) \land (\iota_2,j)$ and $\mathrm{rank}(i,j) = \kappa'$. Observe that $\iota'=\iota$, so $\iota \leq \kappa'\leq \kappa$. Set the rank of the states that never take the value $2$ to $\iota$. For the remaining states set the rank to $\mathrm{rank}(\max_k i_k, \max_k j_k)$, where  $(i_1, j_1), (i_2, j_2), \ldots, (i_r, j_r)$ are the arguments for which the value $2$ is taken.

Let us check that the automaton recognises $L_{(\iota_1, \kappa_1)} \times L_{(\iota_2,\kappa_2)}$. Take a~word $w=(x_1,y_1)(x_2,y_2)\ldots\,$. Let $x=\max_k x_k$ and $y=\max_k y_k$. In the run of $A$ on $w$, the states $f$ satisfying $f(x,y)=2$ will occur infinitely often. Furthermore, from some moment on there only appear states $f$ satisfying $\forall_{(x',y')} \; f(x',y') = 2 \implies (x',y') \leq (x,y)$. Since $(x,y)\leq(x',y') \implies \mathrm{rank}(x,y) \leq \mathrm{rank}(x',y')$, the highest rank used infinitely often in the run on $w$ is $\mathrm{rank}(x,y)$. Finally, $\mathrm{rank}(x,y)$ is even iff $x$ and $y$ are even, so the run on $w$ is accepting iff $w\in L_{(\iota_1,\kappa_1)} \times L_{(\iota_2,\kappa_2)}$.

Since $A$ has the index $(\iota, \kappa)$, the automaton itself provides a reduction of $L_{(\iota_1, \kappa_1)} \times L_{(\iota_2,\kappa_2)}$ to $L_{(\iota,\kappa)}$. 

By definition of $(\iota, \kappa)$, there exists a~sequence of pairs \[(x_\iota,y_\iota) < (x_{\iota+1}, y_{\iota+1}) < \ldots < (x_{\kappa}, y_{\kappa})\] such that for all $i$ it holds that $\iota_1 \leq x_i \leq \kappa_1$, $\iota_2\leq y_i \leq \kappa_2$, and $x_i$ and $y_i$ are even iff $i$ is even. The reduction is given by the function \[\varphi(i_1i_2i_3\ldots) = (x_{i_1},y_{i_1})(x_{i_2},y_{i_2})(x_{i_3},y_{i_3})\ldots\,.\] 

The proof for weak flowers is entirely analogous. \qed

\begin{lem}\label{replicators} 
For all $0< k, l<\omega$ and all $m <\omega$
\begin{eqnarray} C_1 \oplus C_{\omega^{m}k}\, \land\, C_1 \oplus C_{\omega^{m}l} &\equiv& C_1 \oplus C_{\omega^{m}(k+l)}\,, \nonumber \\
C_1 \oplus C_{\omega^{\omega\cdot2+m}k} \land C_1 \oplus C_{\omega^{\omega\cdot2+m}l}  &\equiv &C_1 \oplus C_{\omega^{\omega\cdot2+m}(k+l)}\,. \nonumber 
\end{eqnarray}
In particular, $(C_1 \oplus C_{\omega^{m}})^k \equiv C_1 \oplus C_{\omega^{m}k}$ and $(C_1 \oplus C_{\omega^{\omega\cdot2+m}})^k \equiv C_1 \oplus C_{\omega^{\omega\cdot2+m}k}$.
\end{lem}

\begin{figure}
\centering
% 6ta
\includegraphics[width=\textwidth]{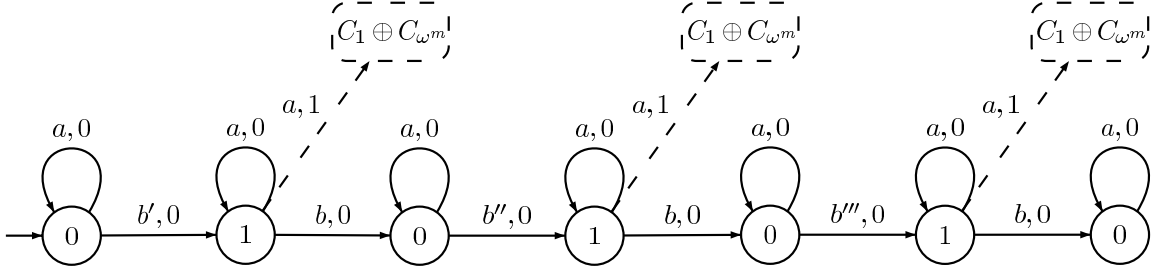}
\caption{The weak flower $WF_{(0,6)}$ formed by the leftmost states of $C_1 \oplus C_{\omega^{m+1}\cdot3}$.}
\label{fig:replicators}
\end{figure}

\proof Consider $G(C_1 \oplus C_{\omega^{m}k} \land C_1 \oplus C_{\omega^{m}l},C_1 \oplus C_{\omega^{m}k} \oplus C_{\omega^{m}l})$. Observe that Duplicator's critical token will move along a copy of $WF_{(0,2k+2l)}$ formed by the leftmost states of consecutive copies of $C_{\omega^{m}}$ (see Fig. \ref{fig:replicators}). 
Spoiler's initial token splits in the first move in two tokens which continue moving along $WF_{(0,2k)}$ and $WF_{(0,2l)}$. For the purpose of this proof, call them both critical. 

The strategy for Duplicator is based on the fact that $WF_{(0,2k)} \land WF_{(0,2l)} \equiv WF_{(0,2k+2l)}$ (Lemma \ref{flowers}). Duplicator can loop his critical token inside an accepting loop as long as both Spoiler's critical tokens loop inside accepting loops. When Spoiler changes his mind and moves one of them to a rejecting loop, Duplicator should move to a~rejecting loop too, and keep looping there until both Spoiler's tokens are again in accepting loops. This can only repeat $k+l$ times, so Duplicator is able to realise this strategy. 

This way, whenever Spoiler produces a~new token $x$ using one of the critical tokens, Duplicator can produce its {\em doppelg\"anger $y$}. The role of the {\em doppelg\"anger} is to mimic the original. \index{doppelg\"anger} The mimicking is in fact passed from generation to generation: if the original token bubbles a~new token $x'$, $y$ should bubble a~new {\em doppelg\"anger} $y'$ which is to mimic $x'$, and so on. 

In order to see that the strategy is winning it is enough to observe two facts: Duplicator's critical token stays in a~rejecting loop forever iff one of Spoiler's critical tokens does, and the sequence of ranks seen by any of Spoiler's non-critical tokens is equal to the one seen by its {\em doppelg\"anger}. Hence, $C_1 \oplus C_{\omega^{m}k} \land C_1 \oplus C_{\omega^{m}l} \leq C_1 \oplus C_{\omega^{m}(k+l)}$. 

The converse inequality is proved in a~similar way and for the second equivalence the same proof works.\qed

\vspace{5pt}

\begin{cor} \label{hardautomata} For all $l, \iota, \kappa <\omega$  and all $0<n<\omega$
\begin{enumerate}[\em(1)] 
\item $C_{\omega} > WF_{(\iota,\kappa)}$,  $C_{\omega^{l + 1 }} \geq C_{\omega^{l}n}$, 
\item $C_{\omega^{\omega\cdot2}} > F_{(\iota,\kappa)}$, $C_{\omega^{\omega\cdot2 + l + 1 }} \geq C_{\omega^{\omega\cdot2 + l}n}$. \end{enumerate}
\end{cor}

\proof Since $C_{\omega} = C_1 \to C_3 \equiv C_1 \to WF_{(0,2)}$,  by Lemma \ref{replication} and  Lemma \ref{flowers} we get  $C_{\omega} \geq (WF_{(0,2)})^m \equiv WF_{(0,2m)}$ and by the strictness of the hierarchy for word languages $C_{\omega} > WF_{(\iota,\kappa)}$. Similarly, using Lemma \ref{replication} and Lemma \ref{replicators} we get $C_{\omega^{l + 1 }} \geq (C_1 \oplus C_{\omega^{l}})^n \equiv C_1 \oplus C_{\omega^{l}n} \geq C_{\omega^{l}n}$. The remaining two inequalities are analogous. \qed

\section{Automata in Order} \label{automatainorder}

Let us start examining the order on canonical automata with the following simple observation.

\begin{lem} \label{CD1} 
For all $0 < \alpha < \omega^\omega$ \[C_\alpha \leq C_{\omega^\omega}\,,\qquad C_\alpha \leq D_{\omega^\omega}\,.\]
\end{lem}

\proof We give a~proof for the first inequality; the second one is proved analogously. Consider the following strategy for Duplicator in $G(C_\alpha, C_{\omega^\omega})$. In every move, if any of Spoiler's tokens is inside a rejecting loop, Duplicator should move his critical token around a~$1$-loop, otherwise he should loop around the $0$-loop. Let us see that the strategy is winning. 

By Proposition \ref{finitelymanytokens} if Spoiler's run is to be accepting, he must produce only finitely many tokens. All of those tokens must finally get to some $0$-loop, and stay there forever. This means that after some number of moves, all Spoiler's tokens are in $0$-loops which they will never leave later. But from this moment on Duplicator's critical token will keep looping around the $0$-loop, so Duplicator's run will also be accepting. 

By Proposition \ref{rejectingpath}, if Spoiler's run is to be rejecting, there must be a~token that from some moment on stays forever in a~$1$-loop. Then Duplicator's token will also get trapped in the rejecting loop in $C_{\omega^\omega}$, and Duplicator's run will be rejecting too. \qed

\vspace{5pt}

Let us now see that we can restrict the way the players use non-critical tokens. For a~simple automaton $A$ and a~canonical automaton $B=B_1\oplus \ldots \oplus B_n$ with $B_i$ simple, we say that $B$ {\em dominates}\index{domination}\label{domination} $A$ if one of the following conditions holds
\begin{enumerate}[$\bullet$]
\item $A$ is non-branching
\item $A=C_1 \to C_\alpha$,  $B_1 = C_1 \to C_\beta$, and $\beta\geq\alpha$,
\item $A=C_{\omega^m}$ and $B_1=F_{(\iota,\kappa)}$ or $B_1=F_{(\iota,\kappa)}\lor F_{\overline{(\iota,\kappa)}}$ for $\iota<\kappa$.
\end{enumerate}

\begin{lem} \label{resetrule}
Let $A_1,A_2, \ldots, A_n$ be simple and let $B$ be a~canonical automaton dominating all $A_i$. For every deterministic automaton $C$, if Spoiler has a winning strategy in $G(A_1 \oplus \ldots \oplus A_n \oplus B, C)$, then he also has a strategy in which he removes all non-critical tokens before entering $B$. Similarly for Duplicator in $G(C, A_1 \oplus \ldots \oplus A_n \oplus B)$. 
\end{lem} 

\proof Let $B=B_1\oplus \ldots \oplus B_n$ with $B_i$ simple. Suppose that at some moment the strategy tells Spoiler to enter $B$ (if this never happens, the claim is obvious). If there are no non-critical tokens left in $A_1,A_2, \ldots, A_n$, then we are done. However if there are, we have to take extra care of them. Suppose Spoiler has produced non-critical tokens $x_1, \ldots, x_r$, and $x_i$ is in $A_{m_i}$. Since $x_i$ is not on a~critical path of $A_{m_i}$, by the definitions of canonical automata, it will stay within a~copy of $C_{\alpha_i}$ over the alphabet extended to the alphabet of $B$. 

Suppose $B_1 = C_1 \to C_\beta$. Since $B$ dominates $A_i$, $\beta\geq\alpha_i$ for all $i$. Spoiler should replace the token $x_i$ with $x_i'$ and let $x_i'$ take over the duties of $x_i$. To produce $x'_i$, Spoiler should loop once in the head loop of $B_1$. If $B_1 = C_{\omega^k}$, or $A_i=C_{\omega^{\omega\cdot2 + k'}}$, Spoiler may simply move $x_i'$ to  a~copy of $C_{\alpha_i}$ and let it perform exactly the actions $x_i$ would take. If $\beta= C_{\omega^{\omega\cdot2 + k}}$, $\alpha_i =  C_{\omega^{k'}}$, Spoiler should move $x_i'$ to the copy of $F_{(0,2)}$ contained in $C_\beta$, and let it apply the strategy guaranteed by Lemma \ref{CD1}. To see that the strategy is applicable, it is enough to note that it does not require any waiting, and that $F_{(0,2)}$ contains a~copy of $F_{(0,1)}$.

Suppose now that $B_1$ is non-branching. Then, $\alpha_i <\omega^\omega$ for all $i$. In this case Spoiler cannot produce a~token to take over $x_i$'s duties. Instead, he has to modify the actions of the critical token. He should move the critical token according to his original strategy moving from flower to flower, only when one of his non-critical tokens would be in a~rejecting loop, he should choose a~$1$-loop in his current flower (instead of the loop suggested by the old strategy). Just like in the proof of Lemma \ref{CD1}, if in a~play according to the original strategy one of the non-critical tokens stays forever in a~rejecting loop, then in the game according to the new strategy the critical token finally also gets trapped in a~$1$-loop. Otherwise, there are only finitely many non-critical tokens, and all of them finally stabilise in an accepting loop. From that moment on, the critical token will see exactly the same ranks as it would see if Spoiler was playing with the original strategy. Hence, the modified strategy is also winning. 

If the original strategy brings Spoiler to a~branching automaton, he should produce counterparts of his non-critical tokens just like above. \qed

\begin{cor} \label{resetrule2}
For every canonical automaton of the form  $A \oplus B$ and every deterministic tree automaton $C$, if a~Spoiler has a~winning strategy in $G(A \oplus B, C)$, than he has also a~winning strategy which removes all non-critical tokens before entering $B$. Similarly for Duplicator in $G(C, A~\oplus B)$.
\end{cor}

\proof Let $A=A_1 \oplus A_2 \oplus \ldots \oplus A_n$ with $A_i$ simple. From the structure of canonical automata it follows that if $A \oplus B$ is canonical, $B$ dominates $A_i$ for $i=1,2,\ldots,n$. \qed

\vspace{5pt}

Now we are ready to get back to the order on ${\mathcal C}$. 

\begin{lem} \label{order}
If $0< \alpha \leq \beta < \omega^{\omega\cdot3}$ then 
$C_\alpha \leq C_\beta$ and whenever $D_\alpha$ and $E_\beta$ are defined,
$D_\alpha \leq E_\beta, \; C_\alpha \leq E_\beta$. If $\beta<\alpha$, then 
$E_\beta \leq D_\alpha, \; E_\beta \leq C_\alpha$.
\end{lem}

\proof As an auxiliary claim let us see that if $A \oplus B$ is canonical and $A'\geq A$, $A\oplus B \leq A' \oplus B$. Indeed, the following is a~winning strategy for Duplicator in $G(A\oplus B, A' \oplus B)$. While Spoiler keeps inside $A$, apply the strategy from $G(A,A')$. If Spoiler enters $B$, by Corollary~\ref{resetrule2} we may assume he removes all non-critical tokens. Hence, Duplicator may remove non-critical tokens, move the critical token to $B$ and copy Spoiler's actions. 

Let us now see that $C_\alpha \leq C_\beta$ for $\alpha < \beta < \omega^{\omega\cdot3}$; the other inequalities may be proved in an analogous way. We will proceed by induction on $(\alpha,\beta)$ with lexicographic order. If $\beta < \omega$, the result follows by the word languages case. Suppose that $\omega \leq \beta < \omega^\omega$. Let $\alpha=\omega^km_k + \ldots + m_0$ and $\beta = \omega^k n_k + \ldots + n_0$, $n_k > 0$.  First, assume that $m_k=0$. Obviously $C_{\omega^k}\leq C_\beta$, simply because $C_\beta$ contains a~copy of $C_{\omega^k}$. If $k=1$ the claim follows directly from Corollary~\ref{hardautomata}. For $k>1$, using the induction hypothesis and Corollary~\ref{hardautomata}, we get $C_\alpha \leq C_{\omega^{k-1}(m_{k-1}+1)} \leq C_{\omega^k}$. Now, assume that $m_k >0$. Then $\alpha = \omega^k + \alpha'$, $\beta = \omega^k + \beta'$ for some ordinals $\alpha' < \beta'$. By definition $C_\alpha = C_{\alpha'} \oplus C_{\omega^k}$, $C_\beta = C_{\beta'}\oplus C_{\omega^k}$, and by induction hypothesis, $C_{\alpha'} \leq C_{\beta'}$. Hence, by the auxiliary claim above, $C_\alpha\leq C_\beta$.

Now, suppose that $\omega^\omega \leq \beta < \omega^{\omega \cdot 2}$. Let $\alpha = \omega^\omega \alpha_1  + \alpha_0$, $\beta = \omega^\omega\beta_1 + \beta_0$ for  $\alpha_0, \alpha_1, \beta_0, \beta_1  <\omega^\omega$. If  $\alpha_1=\beta_1$, then by induction hypothesis $C_{\alpha_0}\leq C_{\beta_0}$, and  $C_\alpha\leq C_\beta$ follows by the auxiliary claim above.  Assume that $\alpha_1<\beta_1$. By Lemma \ref{CD1}, $C_{\alpha_0} \leq C_{\omega^\omega}$. Replacing $G(C_{\alpha_0}, C_{\beta_0})$ with $G(C_{\alpha_0}, C_{\omega^\omega})$ in the above strategy, we get $C_\alpha = C_{\alpha_0} \oplus E_{\omega^\omega \alpha_1} \leq C_{\omega^\omega} \oplus E_{\omega^\omega \alpha_1} = C_{\omega^\omega (\alpha_1 + 1)}$. By Proposition \ref{omegaorder}, $C_{\omega^\omega (\alpha_1 + 1)} \leq C_{\omega^\omega \beta_1}$ and  since $C_{\omega^\omega \beta_1}$ is contained in $C_{\omega^\omega \beta_1 + \beta_0}$, we get $C_{\omega^\omega \alpha_1 + \alpha_0} \leq C_{\omega^\omega \beta_1 + \beta_0}$. Observe that the argument works also for $\alpha_0$ or $\beta_0$ equal to $0$.

The case $\omega^{\omega \cdot 2} \leq \beta < \omega^{\omega \cdot 3}$ is analogous to $\omega \leq \beta < \omega^\omega$. \qed

\vspace{5pt}

\begin{figure}
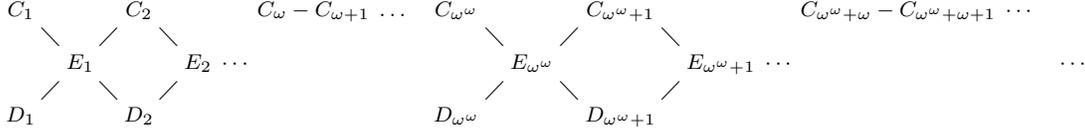

\centering
{\scriptsize {\setlength\arraycolsep{1pt}
$\begin{array}{cccccccccccccccccccccccccccccccc}
C_1 &           &     &           & C_2 &           &               & C_\omega - C_{\omega+1} \; \dots \quad  & C_{\omega^\omega} &           &                 &          & C_{\omega^\omega+1} &           &                             &  C_{\omega^\omega+\omega} - C_{\omega^\omega+\omega + 1}\; \cdots \quad &        \\
    & \diagdown &     & \diagup   &     & \diagdown &               &                                       &                 & \diagdown &                 &\diagup   &                   & \diagdown &                             &                                                                   &        \\
    &           & E_1 &           &     &           & E_2 \; \cdots &                                       &                 &           & E_{\omega^\omega} &          &                   &           & E_{\omega^\omega +1}\; \cdots &                                                                   & \cdots \\
    & \diagup   &     & \diagdown &     &  \diagup  &               &                                       &                 & \diagup   &                 &\diagdown &                   & \diagup   &                             &                                                                   &        \\
D_1 &           &     &           & D_2 &           &               &                                       & D_{\omega^\omega} &           &                 &          & D_{\omega^\omega+1} &           &                             &                                                                   &        \\
\end{array}$}}
\caption{The Wadge ordering of the canonical automata.}
\label{fig:wadgecanonical}
\end{figure}

For a~complete description of the ordering on the canonical automata (see Fig.~\ref{fig:wadgecanonical}) we need the strictness of the inequalities from the previous lemma.

\begin{thm} \label{strictorder}
Let $0 < \alpha \leq \beta < \omega^{\omega\cdot3}$. Whenever the respective automata are defined, it holds that  $D_\alpha \nleq C_\alpha$, $D_\alpha \ngeq C_\alpha$, $D_\alpha < E_\beta, \; C_\alpha < E_\beta$, and  for $\alpha <\beta$,  $C_\alpha < C_\beta$, $E_\alpha < D_\beta, \; E_\alpha < C_\beta$.
\end{thm}

\proof By Lemma \ref{order} it is enough to prove $C_{\alpha} < C_{\alpha+1}$, $D_\alpha < E_\alpha$, $C_\alpha<E_\alpha$, $D_\alpha \nleq C_\alpha$, $D_\alpha \ngeq C_\alpha$. We will only give a~proof of the first inequality; the others can be argued similarly.  We will proceed by induction on $\alpha$. If $\alpha < \omega$, the claim follows by the word languages case.

Suppose $\omega \leq \alpha < \omega^\omega$. Then  $\alpha = \omega^k + \alpha'$ with $k \geq 1$, $\alpha' < \omega^{k+1}$. Let $\alpha'\geq 1$ (the remaining case is similar). We shall describe a~winning strategy for Spoiler in $G=G(C_{\omega^k + \alpha' + 1}, C_{\omega^k + \alpha'})$. Spoiler should first follow the winning strategy for $G(C_{\alpha'+1}, C_{\alpha'})$, which exists by the induction hypothesis. Suppose that Duplicator enters the head loop of $C_{\omega^k}$. We may assume that he removes all his non-critical tokens (Corollary \ref{resetrule2}). Spoiler should remove all his non-critical tokens, move his critical token to any accepting loop in $C_{\alpha'+1}$. Let us check that such a loop is always reachable for the critical token. 

Let $C_{\alpha'+1} = A \oplus B$ with $B$ simple. If $B = C_1 \to B'$, Spoiler can move his critical token to $C_1$. If $B$ is not of this form, then by definition of canonical automata, $C_{\alpha'+1} = C_{2n+1}$ or $C_{\alpha'+1} = D_{2n}$. Recall that $C_{2n+1} \equiv WF_{(0,2n)} \equiv WF_{(0,2n-1)} \oplus C_1$ and $D_{2n} \equiv WF_{(1,2n)} \equiv  WF_{(1,2n-1)} \oplus C_1$ (see page \pageref{canonicalweakflowers}). It follows that in any play on $C_{2n+1}$ or $D_{2n}$, if one has a winning strategy, one also has a winning strategy never entering the rejecting loop of the tail component. Hence, the accepting loop in the tail component is always reachable (or has been reached already).  

 Thus, Spoiler can move his critical token to an accepting loop in the tail component of $C_{\alpha'+1}$ and loop there until Duplicator leaves the head loop. If Duplicator stays forever in the head loop of $C_{\omega^k}$, he loses. Suppose that Duplicator leaves the head loop of $C_{\omega^k}$ after producing $r$ tokens. The rest of the game is equivalent to $G'= G(C_1 \oplus C_{\omega^k}, A)$ for $A = A_1 \land \ldots \land A_r$, where $A_j$ is the part of $C_{\omega^k}$ accessible for the Duplicator's $j$th token. If $k=1$, then $A_j \leq WF_{(0,2)}$ for each $j$. Hence $A \leq WF_{(0,2r)}$ and by Corollary \ref{hardautomata} Spoiler has a winning strategy in $G'$. Let us suppose $k>1$. Then $A_j \leq C_1 \oplus C_{\omega^{k-1}}$ for $j=1, \ldots, r$ and so, by Lemma \ref{substitution},  $A \leq (C_1\oplus C_{\omega^{k-1}})^r$. Hence, by Lemma~\ref{replicators}, $A \leq C_{\omega^{k-1}r + 1}$. Since $\omega^{k-1}r+1 < \omega^{k-1}r+2 \leq \omega^k$, we may use the induction hypothesis to get a~winning strategy for Spoiler in $G'$. In either case Spoiler has a~winning strategy in $G$ as well.

 Now, assume $\omega^\omega \leq \alpha < \omega^{\omega\cdot2}$. Let $\alpha=\omega^\omega\alpha_1 + \alpha_0$ with  $\alpha_0<\omega^\omega$, $1\leq \alpha_1 < \omega^\omega$. Again, we describe a~strategy for Spoiler in $G=G(C_{\omega^\omega{\alpha_1} + \alpha_0 + 1}, C_{\omega^\omega{\alpha_1}  + \alpha_0})$ only for $\alpha_0\geq 1$, leaving the remaining case to the reader. First follow the winning strategy from $G(C_{\alpha_0+1}, C_{\alpha_0})$. If Duplicator does not leave the $C_{\alpha_0}$ component, he will lose. After leaving $C_{\alpha_0}$, Duplicator has to choose $D_{\omega^{\omega}{\alpha_1}}$ or $C_{\omega^{\omega}{\alpha_1}}$. Suppose he chooses $D_{\omega^{\omega}{\alpha_1}}$. Again, by Corollary \ref{resetrule2} we may assume that he removes all non-critical tokens. Now, Spoiler has to remove all non-critical tokens and move the critical token to the initial state of $E_{\omega^\omega{\alpha_1}}$ and use the winning strategy from $G(E_{\omega^\omega{\alpha_1}}, D_{\omega^{\omega}{\alpha_1}})$. 

For $\alpha=\omega^{\omega\cdot2+k} + \alpha'$ argue like for $\alpha=\omega^k + \alpha'$.\qed

\section{Patterns in Automata} \label{patternsinautomata}

Compare the notion of $(\iota,\kappa)$-flower defined
in Sect. \ref{sect:automata} and the canonical flower $F_{(\iota,
  \kappa)}$. It is fairly clear that if $A$ contains
a~$(\iota,\kappa)$-flower, Duplicator can win in $G(F_{(\iota,
  \kappa)}, A)$ by copying Spoiler's actions. In that case it seems plausible to
look at $A$ as if it ``contained'' a copy of $F_{(\iota,\kappa)}$. In this section we provide a~notion which captures this intuition.

Two paths $p\stackrel{\sigma'_1, d'_1}{\longrightarrow} p'_1 \stackrel{\sigma'_2, d'_2}{\longrightarrow} \ldots \stackrel{\sigma'_m, d'_m}{\longrightarrow} p'_m$ and  $p\stackrel{\sigma''_1, d''_1}{\longrightarrow} p''_1 \stackrel{\sigma''_2, d''_2}{\longrightarrow} \ldots \stackrel{\sigma''_n, d''_n}{\longrightarrow} p''_n$ in a~deterministic automaton $A$ are {\em branching} iff there exists $i<\min(m,n)$ such that for all $j<i$ it holds that $(\sigma'_j,d'_j)=(\sigma''_j,d''_j)$,  $\sigma'_i=\sigma''_i$, and $d'_i\neq d''_i$. Note that the condition implies that $p'_j=p''_j$ for $j \leq i$. 

An automaton $B$ can be {\em embedded} into an automaton $A$, if there exists a~function $e_Q: Q^B \to Q^A$ and a~function $e_\delta: Q^B\times \Sigma^B\times\{0,1\} \to \Pi^A$, where $\Pi^A$ is the set of paths in $A$, satisfying the following conditions:
\begin{enumerate}[(1)]
\item if $p \stackrel{\sigma,d}{\longrightarrow} q$ and $e_\delta(p,\sigma, d) = r_0 \stackrel{\sigma_1, d_1}{\longrightarrow} r_1 \stackrel{\sigma_2, d_2}{\longrightarrow} \ldots \stackrel{\sigma_n, d_n}{\longrightarrow} r_n$ then $r_0=e_Q(p)$, $r_n = e_Q(q)$,
\item for all $p,\sigma$ the paths $e_\delta(p,\sigma,0)$ and $e_\delta(p,\sigma,1)$ are branching,
\item for every loop $\lambda$ in $B$, the corresponding loop in $A$ (obtained by concatenating the paths assigned to the edges of $\lambda$) is accepting iff $\lambda$ is accepting. 
\end{enumerate}

For each tree automaton $A$, let $A'$ be the automaton obtained from $A$ by unravelling the DAG of strongly connected components into a~tree (for the purpose of this definition, we allow multiple copies of $\bot$). An automaton $A$ {\em admits} an automaton $B$, in symbols $B \sqsubseteq A$, if the automaton $B'$ can be embedded into $A$. Note that if $B$ can be embedded into $A$, then $A\sqsubseteq B$.

\begin{lem} \label{admittance}
For all deterministic tree automata $A$ and $B$ \[A \sqsubseteq B \implies A \leq B\,.\]
\end{lem} 

\proof Since $L(B') = L(B)$, without loss of generality we may assume that $B=B'$. We have to provide a~winning strategy for Duplicator in $G(B,A)$. Without loss of generality, we may assume that Spoiler never removes his tokens. Let $e_Q$  and $e_\delta$ be the embedding functions.  We will show that Duplicator can keep a~collection of {\em doppelg\"angers}, one for each Spoiler's token, such that if some Spoiler's token $x$ is in the state $p$, its {\em doppelg\"anger} $y$ is in the state $e_Q(p)$. 

Let us first assume that $e_Q(q_0^B)=q_0^A$. Then the invariant above holds when the play starts. As long as Spoiler does not enter $\bot$, the invariant can be maintained by means of the function $e_\delta$ as follows. Suppose that Spoiler fires a~transition  $q\stackrel{\sigma}{\longrightarrow}q',q''$ for some token $x$ obtaining new tokens $x'$ and $x''$. Let  \[e_\delta(q,\sigma,0) = p_0 \stackrel{\sigma_1, d_1}{\longrightarrow}  \ldots  \stackrel{\sigma_{l-1}, d_{l-1}} {\longrightarrow} p_{l-1}  \stackrel{\sigma_{l}, d'_{l}}{\longrightarrow} p'_{l} \stackrel{\sigma'_{l+1}, d'_{l+1}}{\longrightarrow} \ldots \stackrel{\sigma'_{m}, d'_{m}}{\longrightarrow} p'_m\,,\]
\[e_\delta(q,\sigma,1) = p_0 \stackrel{\sigma_1, d_1}{\longrightarrow}  \ldots  \stackrel{\sigma_{l-1}, d_{l-1}} {\longrightarrow} p_{l-1}  \stackrel{\sigma_{l}, d''_{l}} {\longrightarrow} p''_{l} \stackrel{\sigma''_{l+1}, d''_{l+1}}{\longrightarrow} \ldots \stackrel{\sigma''_{n}, d''_{n}}{\longrightarrow} p''_n\,,\]
with $d'_{l} = 1-d''_{l}$. 
 
Let $r_i, r'_j, r''_k$ be such that $p_{i-1} \stackrel{\sigma_i, \overline {d_i}}{\longrightarrow} r_i$, $p'_{j-1} \stackrel{\sigma'_j, \overline {d'_j}}{\longrightarrow} r'_j$, and $p''_{k-1} \stackrel{\sigma''_k, \overline{ d''_k}}{\longrightarrow} r''_k$ for $1\leq i < l$, $l+1\leq j \leq m$, $l+1 \leq k \leq n$, where $\overline d = 1-d$.

Recall that we assume that for every transition, either both target states are $\bot$ or none.  Since $q' \neq \bot$ and $q'' \neq \bot$ then, by the condition (3) of the definition of admittance, $p'_m \neq \bot$ and $p''_n \neq \bot$ and consequently all the states $p_i$, $r_i$,  $p'_j$, $r'_j$, $p''_k$, $r''_k$ are not equal to $\bot$. Hence, Duplicator can proceed as follows. Starting with the token $y$ (the {\em doppelg\"anger} of $x$), fire the transitions forming the common prefix of both paths, each time removing the token sent to $r_i$. Thus he reaches the state $p_{l-1}$ with a~descendant of the token $y$. Then he should fire the next transition producing two tokens $y'$ and $y''$, and  for each of them fire the remaining sequence of transitions (again removing the tokens in the states $r'_j$ and $r''_k$). Thus he ends up with two tokens in the states $p'_m = e_Q(q')$ and $p''_n=e_Q(q'')$. Hence, the token in $e_Q(q')$ may be the {\em doppelg\"anger} of $x'$, and the token in $e_Q(q'')$ may be the {\em doppelg\"anger} of $x''$.

Let us see that if Spoiler never enters $\bot$, Duplicator wins. Observe that the function $e_\delta$ induces a~function $e$ from the set of infinite paths in $B$ to the set of infinite paths in $A$. Owing to the condition (3), $e(\pi)$ is accepting iff $\pi$ is accepting. The strategy used by Duplicator guarantees that for each path $\pi$ in Spoiler's run, Duplicator's run contains the path $e(\pi)$. The paths in Duplicator's run that are not images of paths from Spoiler's run were all declared accepting by removing the corresponding tokens. Hence, Duplicator's run is accepting iff Spoiler's run is accepting. 

Now, if Spoiler enters $\bot$, Duplicator proceeds as before, only if some $r_i$, $r'_j$, or $r''_k$ is equal to $\bot$, instead of removing the token from there (he is not allowed to do that), he lets the token and all its descendants loop there forever.  In the end, again each path in Spoiler's run has a~counterpart in Duplicator's run. The images of the rejecting paths (which exist in Spoiler's run), will be rejecting too. Hence, Duplicator also wins in this case. 

Finally we have to consider the situation when $e_Q(q_0^B) \neq q_0^A$. In this case, Duplicator should first move his initial token to the state $e_Q(q_0^B)$,  removing the other tokens produced on the way whenever possible, and then proceed as before.  \qed

\vspace{5pt}

Another property that makes admittance similar to containment is transitivity. 

\begin{lem} \label{transitivity}
For all deterministic tree automata $A$, $B$, and $C$,
 \[ A\sqsubseteq B  \sqsubseteq C \implies A\sqsubseteq C\,.\]
\end{lem}

\proof Again, we may assume that $A'=A$. Furthermore, since the states from one SCC have to be mapped to states from one SCC, then $A$ can be embedded directly into $B'$. Hence, we may also assume that $B=B'$. Let $e_Q^{X,Y}$, $e_\delta^{X,Y}$ be functions embedding the automaton $X$ into  $Y$. The embedding of $A$ into $C$ is simply a~composition of two given embeddings: $e_Q^{A,C} = e_Q^{B,C} \circ e_Q^{A,C}$, $e_\delta^{A,C} = e_\Pi^{B,C} \circ e_\delta^{A,B}$, where $e_\Pi^{B,C}: \Pi^B \to \Pi^C$ is the function induced by $e_\delta^{B,C}$ in the natural way. It is easy to see that $e_Q^{A,C}$ and $e_\delta^{A,C}$ satisfy the conditions from the definition of admittance. \qed

\vspace{5pt}

Embedding for automata on words is defined analogously, only the function $e_\delta$ is defined on $Q\times \Sigma$ instead of $Q \times \Sigma \times \{0,1\}$, and the condition (2) is dropped. Admittance is defined identically. The two lemmas above carry over with analogous proofs. 

\section{Hard Automata} \label{wagnerhierarchyandbeyond}

In previous sections we have described an extended hierarchy of
canonical automata. As we have already mentioned there are still three
canonical automata left to define. In their definition we will make
the first use of a stronger variant of the operation $\to$.

\begin{figure}
\centering
% 6ta
\includegraphics[width=0.9\textwidth]{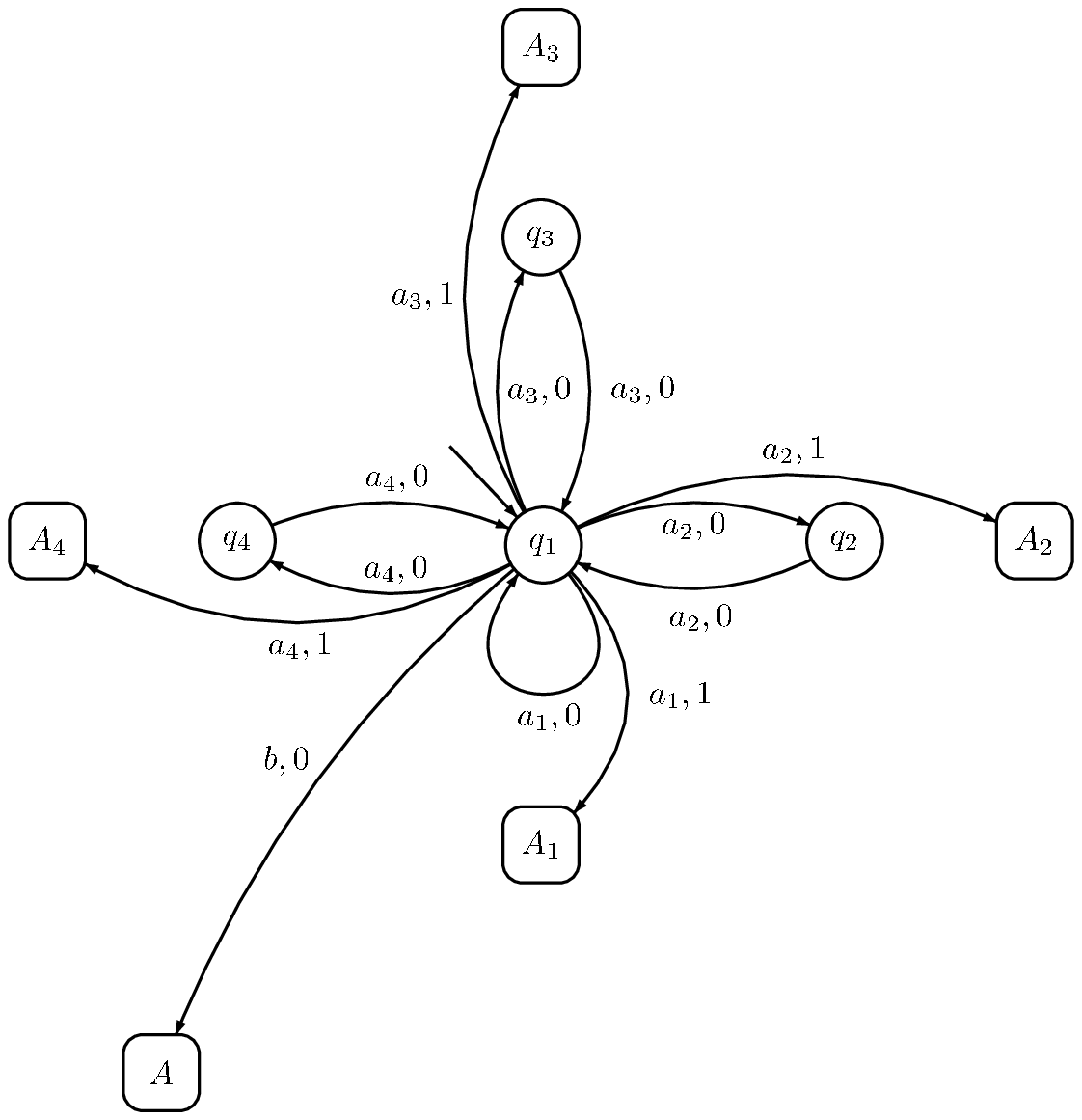}
\caption{The $(1,4)$-replication $A \stackrel{(1,4)}{\longrightarrow} A_1, A_2, A_3, A_4$.}
\label{fig:flower}
\end{figure}

In the operation $\stackrel{(\iota, \kappa)}{\longrightarrow}$,
instead of one (rejecting) loop replicating an automaton, we have a whole flower whose each
loop replicates a different automaton. Recall that $F_{(\iota, \kappa)}$ 
is an automaton whose input alphabet is $\{a_\iota, a_{\iota+1}
\ldots, a_\kappa\}$, the states are $q_\iota, q_{\iota+1},\ldots,
q_\kappa$, $\mathrm{rank}(q_i)=i$, the initial state is $q_\iota$,  and transitions  \[q_\iota \stackrel{a_\iota}{\longrightarrow} q_\iota,\top\,, \;\; q_\iota \stackrel{a_j}{\longrightarrow} q_j,\top\,, \;\;   q_j\stackrel{a_j}{\longrightarrow} q_0, \top\,, \;\; \textrm{and} \;\;  q_j\stackrel{a_k}{\longrightarrow} \top,\top \;\;\]
for $j=\iota+1, \iota+2, \ldots, \kappa$ and $k \neq j$. Let $A, A_\iota, \ldots, A_\kappa$ be deterministic tree automata over $\Sigma$. The {\em $(\iota, \kappa)$-replication}  $A \stackrel{(\iota, \kappa)}{\longrightarrow} A_\iota, \ldots, A_\kappa$   (see Fig. \ref{fig:flower}) 
is obtained as follows. Take a~copy of $F_{(\iota, \kappa)}$ over the
extended alphabet $\{a_\iota, a_{\iota+1} \ldots, a_\kappa\} \cup
\Sigma \cup \{b\}$, where $b$ is a~fresh letter. Add single disjoint
copies of  $A_\iota, \ldots, A_\kappa$ and $A$ over the extended
alphabet $\Sigma \cup \{a_\iota, a_{\iota+1} \ldots, a_\kappa\} \cup
\{b\}$. Finally, in $F_{(\iota, \kappa)}$ over the extended alphabet,
replace the transition $q_\iota\stackrel{b, 0}{\longrightarrow} r$
(where $r\in\{\bot,\top\}$) with $q_\iota\stackrel{b,
  0}{\longrightarrow} q_0^{A}$, and $q_\iota \stackrel{a_i,
  1}{\longrightarrow} \top$ with  $q_\iota \stackrel{a_i,
  1}{\longrightarrow} q_0^{A_i}$  for $i=\iota, \ldots, \kappa$. 

Using Lemma \ref{substitution} it is easy to see that the
$\equiv$-class of the defined automaton depends only on $(\iota,
\kappa)$ and the $\equiv$-classes of $A, A_\iota, \ldots,
A_\kappa$. Hence, $\equiv$ is a congruence with respect to
$\stackrel{(\iota, \kappa)}{\longrightarrow}$ for every
$(\iota,\kappa)$.  

Note also that $A \to B$ and $A\stackrel{(1,1)}{\longrightarrow}B$ 
are equal up to the names of letters and states. In particular  
$L(A \to B) \equiv_W L(A\stackrel{(1,1)}{\longrightarrow}B)$.

Let us now define the three missing automata. Let $C_{\omega^{\omega\cdot 3}} = C_1 \stackrel{(0,1)}{\longrightarrow} C_1, C_{(0,2)}$ and  $C_{\omega^{\omega\cdot3}+1} = C_1 \stackrel{(0,0)}{\longrightarrow} F_{(0,1)}$. The last automaton, $C_{\omega^{\omega\cdot3}+2}$ consists of the states $q_0$, $q_1$, $\top$ with $\mathrm{rank}(q_i)= i$ and transitions \[q_0 \stackrel{a}{\longrightarrow}q_0,q_1\,, \quad q_0 \stackrel{b}{\longrightarrow} \top,\top\,,\] \[q_1 \stackrel{a}{\longrightarrow}q_0,\top\,, \quad q_1 \stackrel{b}{\longrightarrow}\top,\top\,.\] 
Using canonical automata we can formulate results from Sect. \ref{sect:topology} in a~uniform way. In the proof we will need the following technical lemma. 

\begin{lem} \label{embeddingtop}
Let $A$ be a~deterministic tree automaton. For every productive state $p$ in $A$ there exists a~state $q$, a~path $\pi_p$ from $p$ to $q$,  and pair of branching paths $\pi_q^0$, $\pi_q^1$ from $q$ to $q$ forming accepting loops.
\end{lem}

\proof Take an accepting run starting in $p$. For  each node $v$ of
the run let $S_v$ be the set of states of the automaton that appear
below $v$. Note, that if $v'$ is a descendant of $v$, $S_{v'} \subseteq S_v$. Since all $S_v$'s
are non-empty, there exists a node $u$ such that for all descendants $u'$ of
$u$, $S_{u'} = S_u$. Pick a state $q \in
S_u$. There exists a node $w$ under $u$, labeled with $q$. Both
$w0$ and $w1$ are labeled with $S_u$, so there exist a nodes $w_0$
under $w0$ and $w_1$ under $w1$ that are also labeled with $q$. To
conclude, let $\pi_p$ be the path in $A$ induced by the path from
$\varepsilon$ to $w$, and let $\pi_q^i$ be the path induced by the path
between $w$ and $w_i$ for $i=0,1$. \footnotemark[1] \footnotetext[1]{This elegant proof was suggested by one
  of the referees in place of a clumsier inductive argument.} \qed

\begin{thm} \label{borelautomata} 
Let $A$ be a~deterministic automaton.
\begin{enumerate}[\em(1)]
\item $L(C_1\oplus D_1)$ is $\Pi^0_1$-complete; $L(A) \in \Sigma^0_1$  iff $A$ does not admit $C_1\oplus D_1$.
\item $L(D_1 \oplus C_1)$ is  $\Sigma^0_1$-complete; $L(A) \in \Pi^0_1$ iff $A$ does not admit $D_1 \oplus C_1$.
\item $L(F_{(1,2)})$ and  $L(C_1 \stackrel{(0,0)}{\longrightarrow} (D_1 \oplus C_1))$ are $ \Pi^0_2$-complete; \\ $L(A) \in \Sigma^0_2$ iff $A$ does not admit $F_{(1,2)}$ nor $C_1 \stackrel{(0,0)}{\longrightarrow} (D_1 \oplus C_1)$. 
\item $L(F_{(0,1)})$ is $\Sigma^0_2$-complete; $L(A) \in \Pi^0_2$ iff $A$ does not admit $F_{(0,1)}$.
\item $L(C_{\omega^{\omega\cdot3}+1})$ is $\Pi^0_3$-complete; $L(A) \in \Sigma^0_3$ iff $A$ does not admit $C_{\omega^{\omega\cdot3}+1}$. 
\item  $L(A) \in \Pi^0_3$ iff $A$ does not admit $C_{\omega^{\omega\cdot3}+2}$.
\item  $L(C_{\omega^{\omega\cdot3}+2})$ is $\Pi^1_1$-complete; $L(A)$ is $\Pi^1_1$-complete iff $A$ admits $C_{\omega^{\omega\cdot3}+2}$.
\end{enumerate}
\end{thm}

\proof It is enough to check that for an automaton it is the same to contain the patterns from Theorem \ref{borelch} (page \pageref{borelch}) and to admit the respective automata. It is straightforward to check that it indeed is so. The only difficulty is embedding the transitions to all-accepting states, but this is solved by Lemma \ref{embeddingtop}. Let us just see the case of $C_{\omega^{\omega\cdot3}+2}$. If $A$ admits $C_{\omega^{\omega\cdot3}+2}$, then the image of the two loops in  $C_{\omega^{\omega\cdot3}+2}$ that contain the initial state is a~split. 

Suppose that $A$ contains a~split consisting of an $i$-loop $p \stackrel{\sigma,0}{\longrightarrow} p_1 \stackrel {\sigma_1,d_1 }{\longrightarrow} \ldots \stackrel {\sigma_m,d_m}{\longrightarrow} p_{m+1} = p$ and a~$j$-loop $p \stackrel{\sigma,1}{\longrightarrow} p'_1 \stackrel {\sigma'_1,d'_1}{\longrightarrow} \ldots  \stackrel {\sigma'_n,d'_n}{\longrightarrow} p'_{n+1} = p$, such that $i$ is even, $j$ is odd, and $i<j$. Without loss of generality we may assume that $m,n \geq 1$. Let $p'_1\stackrel{\sigma'_1,\overline{d'_1}}{\longrightarrow}q'$, and let $t_p$, $t_{p'_1}$, $t_{q'}$ be the states guaranteed by Lemma \ref{embeddingtop} for $p$, ${p'_1}$, and ${q'}$ respectively. 

Let $B$ be the automaton obtained from $C_{\omega^{\omega\cdot3}+2}$ by unravelling the DAG of SCCs. The only way it differs from $C_{\omega^{\omega\cdot3}+2}$ is that instead of one state $\top$ it contains 5 all-accepting states $\top_1, \ldots, \top_5$, one for each transition from the root SCC: 
\[q_0 \stackrel{a}{\longrightarrow}q_0,q_1\,, \quad q_0 \stackrel{b}{\longrightarrow} \top_1,\top_2\,,\] \[q_1 \stackrel{a}{\longrightarrow}q_0,\top_3\,, \quad q_1 \stackrel{b}{\longrightarrow}\top_4,\top_5\,.\] 
Define $e_Q(q_0)=p$, $e_Q(q_1)=p'_1$, $e_Q(\top_1) = e_Q(\top_2) = t_p$, $e_Q(\top_3) = t_{q'}$, $e_Q(\top_4) = e_Q(\top_4) = t_{p'_1}$. The function $e_\delta$ is defined as follows:
\[\begin{array}{ll}
e_\delta(q_0, a, 0)  =  p \stackrel{\sigma,0}{\longrightarrow} p_1
\stackrel {\sigma_1,d_1 }{\longrightarrow} \ldots \stackrel
{\sigma_m,d_m}{\longrightarrow} p \,, \qquad \qquad&
e_\delta(q_1, a, 0)  =  p'_1 \stackrel
{\sigma'_1,d'_1}{\longrightarrow} \ldots  \stackrel
{\sigma'_n,d'_n}{\longrightarrow} p \,, \\

e_\delta(q_0, a, 1)  =  p \stackrel{\sigma,1}{\longrightarrow} p'_1
\,, &
e_\delta(q_1, a, 1)  = 
(p'_1\stackrel{\sigma'_1,\overline{d'_1}}{\longrightarrow}q') \pi_{q'}
\,, \\

e_\delta(q_0, b, 0)  =  \pi_p \pi_{t_p}^0 \,, &
e_\delta(q_1, b, 0)  =  \pi_{p'_1} \pi_{t_{p'_1}}^0 \\

e_\delta(q_0, b, 1)  =  \pi_p \pi_{t_p}^1 \,, &
e_\delta(q_1, b, 1)  =  \pi_{p'_1} \pi_{t_{p'_1}}^1 \,, \\

e_\delta(\top_i, *, 0)  =  \pi_{e_Q(\top_i)}^0 \,,  \\
e_\delta(\top_i, *, 1)  =  \pi_{e_Q(\top_i)}^1 \,, 
\end{array}\]
where $*$ denotes any letter and by $ \pi_1 \pi_2$ we mean the concatenation of two paths. Checking that this is an embedding is straightforward. \qed

\vspace{5pt}

For the proof of the next theorem, settling the position of the last canonical automaton,  $C_{\omega^{\omega\cdot3}}$, we will need the following property of replication. 

\begin{lem}[Replication Lemma] \label{replicationlemma} 
A state occurs in infinitely many incomparable nodes of an accepting run iff it is productive and is replicated by an accepting loop. 
\end{lem}

\proof If a~state $p$ is replicated by an accepting loop, then by productivity one may easily construct an accepting run with infinitely many incomparable  occurrences of $p$. Let us concentrate on the converse implication. 

Let $p$ occur in an infinite number of incomparable nodes $v_0, v_1, \ldots$ of an accepting run $\rho$. Let $\pi_i$ be a~path of $\rho$ going through the node $v_i$. Since $2^\omega$ is compact, we may assume, passing to a~subsequence, that the sequence $\pi_i$ converges to a~path $\pi$. Since $v_i$ are incomparable, $v_i$ is not on $\pi$. Let the word $\alpha_i$ be the sequence of states labelling the path from the last common node of $\pi$ and $\pi_i$ to $v_i$. Cutting the loops off if needed, we may assume that $|\alpha_i| \leq |Q|$ for all $i\in \omega$. Consequently, there exist a~word $\alpha$ repeating infinitely often in the sequence $\alpha_0, \alpha_1, \ldots$. Moreover, the path $\pi$ is accepting, so the starting state of $\alpha$ must lay on an accepting productive loop. This loop replicates $p$.\qed

\begin{thm} \label{hardcore}
$L(C_{\omega^{\omega\cdot3}})$ is Wadge complete for deterministic $\Delta^0_3$ tree languages. In particular, $A \leq C_{\omega^{\omega\cdot3}}$ for each $A\in {\mathcal C}$.
\end{thm}

\proof Since $C_{\omega^{\omega\cdot3}}$ admits neither $C_{\omega^{\omega\cdot3}+2}$ nor $C_{\omega^{\omega\cdot3}+1}$, $L(C_{\omega^{\omega\cdot3}})$ is a~deterministic $\Delta^0_3$ language (Theorem \ref{borelautomata}). Let us see that it is hard in that class. 

Take a~deterministic automaton $A$ recognising a~$\Delta^0_3$-language.  By Theorem \ref{borelautomata} (5),  $A$ does not admit $C_{\omega^{\omega\cdot3}+1} = C_1 \stackrel{(0,0)}{\longrightarrow}F_{(0,1)}$.  Let us divide the states of $A$ into two categories: a~state is {\em blue} if it is replicated (see page \pageref{replicated}) by an accepting loop, otherwise it is {\em red}. Note that every state reachable from a~blue state is blue.

Let $A'$ be the automaton $A$ with the ranks of red states set to $0$, and let $A''$ be $A$ with the ranks of the blue states set to $0$. Let us see that  $A \leq A' \land A''$. The strategy for Duplicator in $G(A, A'\land A'')$ is to copy Spoiler's actions in $A$, both in $A'$ and $A''$. To show that this strategy is winning it is enough to show that for each $t$ a run of $A$ on $t$ is accepting iff  the runs of $A'$ and $A''$ on $t$ are accepting. Take a path $\pi$ of the run of $A$. Let $\pi'$ and $\pi''$ be the corresponding paths of the computations of $A'$ and $A''$. If $\pi$ only visits red states, then the ranks on $\pi$ and $\pi''$ are identical, and $\pi'$ contains only $0$'s. Otherwise, $\pi$ enters a blue state at some point, and then stays in blue states forever. In such case, the blue suffixes of $\pi$ and $\pi'$ have the same ranks, and the blue suffix of $\pi''$ contains only $0$'s. Thus, $\pi$ is accepting iff $\pi'$ and $\pi''$ are accepting and the claim follows. 

Since $A$ does not admit $C_1  \stackrel{(0,0)}{\longrightarrow} F_{(0,1)}$, it follows that all $(0,1)$-flowers in $A$ are red. Consequently, $A'$ does not admit $F_{(0,1)}$, and so $L(A')$ is $\Pi^0_2$. Since  $L(F_{(1,2)})$ is $\Pi^0_2$-hard (Theorem \ref{borelautomata} (3)), $A' \leq F_{(1,2)}$.

Now consider $A''$. Once you enter a blue state, you can never move to a red state. Consequently, since in $A''$ all blue states have rank $0$, we may actually replace them all with one all-accepting state $\top$ without changing the recognised language. Recall that, by convention, instead of putting tokens into $\top$ we simply remove them. Hence, when for some token in $p$ a~transition of the form $p\stackrel{\sigma}{\longrightarrow} \top, q$ or $p\stackrel{\sigma}{\longrightarrow} q, \top$ is fired, we imagine that the token is moved to $q$ without producing any new tokens. By the Replication Lemma (Lemma \ref{replicationlemma}) the occurrences of red states in an accepting run may be covered by a~finite number of infinite paths. Hence, by our convention, only finitely many tokens may be produced in a~play if the constructed run is to be accepting.
 
Let us now show that Duplicator has a~winning strategy in  $G(A'', (C_1\stackrel{(0,1)}{\longrightarrow} C_1, F_{(\iota, \kappa)}))$, where $(\iota, \kappa)$ is the index of $A$.  Whenever Spoiler produces a~new token (including the starting token), Duplicator should loop once around the head $1$-loop producing a~{\em doppelg\"anger}  in $F_{(\iota, \kappa)}$, and keep looping around the head $0$-loop. The new token is to visit states with exactly the same ranks as the token produced by Spoiler. Let us see that this strategy works. Suppose Spoiler's run was accepting. Then, there were only finitely many red tokens produced, and hence the head $1$-loop was visited only finitely often. Furthermore, each Spoiler's token visited an accepting path. But then, so did its {\em doppelg\"anger}, and Duplicator's run was also accepting. Now suppose Spoiler's run was rejecting. If infinitely many red tokens were produced, the head $1$-loop was visited infinitely often, and Duplicator's run was also rejecting. If there were finitely many tokens produced, then one of the tokens must have gone along a~rejecting path, but so did its {\em doppelg\"anger} and Duplicator's run was also rejecting. Hence $A''\leq (C_1\stackrel{(0,1)}{\longrightarrow} C_1, F_{(\iota, \kappa)})$. 

By Lemma \ref{substitution}, $A' \land A'' \leq (C_1\stackrel{(0,1)}{\longrightarrow} C_1, F_{(\iota, \kappa)}) \land F_{(1,2)} $, so it is enough to check that  $(C_1\stackrel{(0,1)}{\longrightarrow} C_1, F_{(\iota, \kappa)}) \land F_{(1,2)} \leq C_{\omega^{\omega\cdot3}}$. Consider the following strategy for Duplicator in the game $G((C_1\stackrel{(0,1)}{\longrightarrow} C_1, F_{(\iota, \kappa)}) \land F_{(1,2)}, C_{\omega^{\omega\cdot3}})$. First, loop once around the $1$-loop and produce a~new token in $F_{(0,2)}$ and use it to mimic Spoiler's actions in $F_{(1,2)}$. Then, for each new token $x$ Spoiler produces in his $1$-loop and sends to $F_{(\iota, \kappa)}$, Duplicator should produce tokens $y_1, \ldots, y_{\lfloor \frac{\kappa+1}{2} \rfloor}$ in $F_{(0,2)}$. By Lemma \ref{flowers}, $(F_{(0,2)})^{\lfloor \frac{\kappa+1}{2} \rfloor} \equiv F_{(0,2\lfloor \frac{\kappa+1}{2} \rfloor)} \leq F_{(\iota, \kappa)}$, so Duplicator has a winning strategy in $G(F_{(\iota, \kappa)}, (F_{(0,2)})^{\lfloor \frac{\kappa+1}{2} \rfloor} )$. Adapting this strategy, Duplicator can simulate the actions of Spoiler's token $x$ in $F_{(\iota, \kappa)}$ with the tokens $y_1,\ldots,y_{\lfloor \frac{\kappa+1}{2} \rfloor}$ in $F_{(0,2)}$. If Spoiler loops the $1$-loop without producing a~new token, or loops around the $0$-loop, Duplicator should copy his actions. Clearly, this strategy is winning for Duplicator.  

Finally, let us see that $A \leq C_{\omega^{\omega\cdot3}}$ for each $A\in {\mathcal C}$. Take $n<\omega$. Observe that in $C_{\omega^{\omega\cdot2+n}}$ no state is replicated by an accepting loop. Hence, $C_{\omega^{\omega\cdot2+n}}$ may not admit $C_{\omega^{\omega\cdot3}+1}$ nor $C_{\omega^{\omega\cdot3}+2}$. By Theorem \ref{borelautomata},  $L(C_{\omega^{\omega\cdot2+n}})$ is in $\Delta^0_3$. By Lemma \ref{order}, for each $A\in {\mathcal C}$ there exists $m<\omega$ such that $A \leq C_{\omega^{\omega\cdot2+m}}$. Hence, for all $A \in {\mathcal C}$, $L(A) \in \Delta^0_3$, and $A \leq C_{\omega^{\omega\cdot3}}$. \qed

\vspace{5pt}

From Theorems \ref{borelautomata} and \ref{hardcore} we obtain the
following picture of the top of the hierarchy: \[{\mathcal C} <
C_{\omega^{\omega \cdot 3}} < C_{\omega^{\omega \cdot 3}+1} <
C_{\omega^{\omega\cdot3}+2}\,.\] Let ${\mathcal C}' = {\mathcal C}
\cup \{C_{\omega^{\omega \cdot 3}}, C_{\omega^{\omega \cdot 3}+1},
C_{\omega^{\omega\cdot3}+2}\}$. Note that it already follows
that the height of the Wadge hierarchy of deterministic tree languages
is at least
$(\omega^\omega)^3+3$. In the remaining of the paper we will
show that each deterministic automaton is Wadge equivalent to one of
the canonical automata from ${\mathcal C}'$, thus providing a matching
upper bound.

\section{Closure Properties} \label{closureproperties}

Our aim is to show that each deterministic tree language is Wadge equivalent to the language recognised by one of the canonical automata. If this is to be true, the family of canonical automata should be closed (up to Wadge equivalence) by the operations introduced in Sect. \ref{operations}. In this section we will see that it is so indeed. The closure properties carry substantial part of the technical difficulty of the main theorem, whose proof is thus made rather concise. \index{closure properties}

\begin{prop}\label{lor_closure}
For $A, B \in {\mathcal C}$ one can find in polynomial time an automaton in ${\mathcal C}$ equivalent to $A \lor B$.
\end{prop}

\proof Proceed just like for nonbranching automata (Proposition~\ref{omegaclosure}, page \pageref{omegaclosure}). Take $A, B \in {\mathcal C}$. If $A \leq B$, then $A \lor B \equiv B$ and if  $B \leq A$, then $A \lor B \equiv A$. If $A$ and $B$ are incomparable, by Lemma \ref{order} we get that they must be equal to $D_\alpha$ and $C_\alpha$. It follows easily from the definitions of the canonical automata that $D_\alpha \lor C_\alpha \equiv E_\alpha$. \qed

\begin{prop}\label{oplus_closure}
For $A, B \in {\mathcal C}$ one can find in polynomial time an automaton in ${\mathcal C}$ equivalent to $A \oplus B$.
\end{prop}

\proof Recall that simple automata are those that cannot be written as $A_1\oplus A_2$ for some canonical automata $A_1, A_2$. Let us first assume that $A$ is a~simple branching automaton.  First let us prove that for $B<A$, $A\oplus B \equiv A$. Let us consider the game $G(A\oplus B, A)$. The following is a~winning strategy for Duplicator.  While Spoiler keeps inside the head loop of $A$, mimic his actions. When he exits the head loop, let all the non-critical tokens produced so far copy the actions of their counterparts belonging to Spoiler, and for the critical token (and all new tokens to be produced) proceed as follows. If $C_1\oplus B$ is a~canonical automaton, then,  by the shape of the hierarchy, $C_1\oplus B < A$ and Duplicator may use the winning strategy from $G(C_1\oplus B, A)$. If $C_1 \oplus B$ is not canonical, then $B=F_{(\iota, \kappa)}\oplus B'$ for some $(\iota, \kappa) \neq (1,1)$. It is very easy to see that $C_1\oplus F_{(\iota, \kappa)}\oplus B' \equiv F_{(\iota, \kappa)}\oplus B'$, and again Duplicator can use the winning strategy from $G(C_1\oplus B, A)$.

Let us assume now that $B = B_1 \oplus B_2 \oplus \ldots \oplus B_n$ where $B_i$ are simple and $B_1\geq A$. Suppose $B=C_{\omega^\omega\eta}$ for some $\eta<\omega^{\omega\cdot3}$. Then $A\oplus B \equiv B$. Indeed, consider the game $G(A \oplus B, B)$. While Spoiler keeps inside $A$, Duplicator should keep in $B_1$ and apply the strategy from $G(A,B_1)$. Suppose Duplicator enters $B$. Since $B_1 \geq A$, it holds that $B$ dominates $A$ and we may assume that Spoiler has removed his non-critical tokens before entering $B$. From now on Duplicator may simply mimic Spoiler's behaviour. 

An analogous argument shows that for $B_1 = D_{\omega^\omega \eta}$, we get $A\oplus B\equiv B$. For $B_1=E_{\omega^\omega \eta}$, $A \oplus B$ is a canonical automaton (up to a~permutation of the input alphabet). 

Now, consider $B=B_1 \oplus \ldots \oplus B_n \geq A$, $B_i$ simple and $B_1<A$. By the definition of canonical automata, $B_1 \leq B_2 \leq \ldots \leq B_n$, and since $B \geq A$, $B_n \geq A$. Let $k$ be the least number for which $B_k \geq A$. Let $B' = B_1 \oplus \ldots \oplus B_{k-1}$ and $B'' = B_k \oplus \ldots \oplus B_n$. In order to reduce this case to the previous one it is enough to check that $A \oplus B \leq A\oplus B''$ (the converse inequality is obvious). Consider $G(A \oplus B, A\oplus B'')$. While Spoiler's critical token stays inside $A\oplus B'$, Duplicator follows the strategy from $G(A \oplus B', A)$. If Spoiler does not leave $A\oplus B'$, he loses. Suppose that Spoiler finally enters $B''$. Note that $B''$ dominates $A$ and $B_1, \ldots, B_{k-1}$. Hence, by Lemma \ref{resetrule}, we may assume that Spoiler removes all his non-critical tokens on entering $B''$. Duplicator should simply move his critical token to the initial state of $B''$ and mimic Spoiler's actions. 

Suppose now that $A = F_{(\iota, \kappa)}$  or $A = F_{(\iota, \kappa)} \lor F_{\overline{(\iota, \kappa)}}$. Let $B=B_1\oplus \ldots \oplus B_n$ with $B_i$ simple. For $\iota < \kappa$ proceeding like in Proposition \ref{omegaclosure} (page \pageref{omegaclosure}) one proves that 
\begin{enumerate}[(1)]
\item $B<A \implies A\oplus B \equiv A$,
\item $B_1 = F_{\overline{(\iota,\kappa)}} \implies A\oplus B \equiv F_{(\iota, \kappa)} \oplus (F_{(\iota, \kappa)} \lor F_{\overline{(\iota,\kappa)}}) \oplus B_2 \oplus \ldots \oplus B_n \in {\mathcal C}$, 
\item $A \leq B_1 =(F_{(\iota', \kappa')} \lor F_{\overline{(\iota', \kappa')}}) \implies A\oplus B \in {\mathcal C}$,

\item $A \leq B_1=F_{(\iota', \kappa')} \implies A\oplus B \equiv B$,
\item $A \leq B_1=(C_1 \to B'_1) \implies A\oplus B \equiv B$.
\end{enumerate}
In the remaining case,  $B_1<A \leq B$, argue like for branching $A$.

For $\iota=\kappa$, the implications (2), (3), and (4) also hold, and give a canonical form if $B_1$ is non-branching.  If $B_1$ is branching, $A\oplus B \equiv B$ for $A=F_{(1,1)}$, and $A\oplus B \equiv F_{(0,0)} \oplus B \in {\mathcal C}$ for $A\in\{F_{(0,0)}, F_{(0,0)} \lor F_{(1,1)}\}$.

Finally let $A = A_1\oplus A_2 \oplus \ldots \oplus A_r$, where $A_i$ are simple. Using the fact that $\oplus$ is associative up to $\equiv$, and Lemma \ref{substitution} (page \pageref{substitution}), we get $(A_1\oplus A_2 \oplus \ldots \oplus A_r) \oplus B \equiv (A_1\oplus A_2 \oplus \ldots \oplus A_{r-1})  \oplus (A_r \oplus B) \equiv (A_1\oplus A_2 \oplus \ldots \oplus A_{r-1})  \oplus B'$ where $B'$ is a~canonical automaton equivalent to $(A_r \oplus B)$. Repeating this $r-1$ times more we obtain a~canonical automaton equivalent to $A\oplus B$.\qed

\vspace{5pt}

In the following proofs we will need the following property. For simple branching automata $B = (C_1 \to C_\alpha)$, let $B^- = D_1 \to C_\alpha$. \index{$B^-$}

\begin{lem} \label{auxiliary_oplus} 
For every $A\in {\mathcal C}$ and every simple branching $B$  one can find in polynomial time a~canonical automaton equivalent to $B^- \oplus A$.
\end{lem}

\proof $B$ is simple branching, so $B = C_\alpha$ where $\alpha=\omega^k$ or $\alpha = \omega^{\omega\cdot 2 + k}$.  Let $A = S \oplus A'$, where $A' \in {\mathcal C}$ and  $S$ is a~simple automaton. Suppose first that $S$ is a~branching automaton. Then $S \equiv C^-_\beta \oplus C_1$ and $A \equiv C^-_\beta \oplus C_1 \oplus A'$ with $\beta=\omega^j$ or $\beta = \omega^{\omega\cdot 2 + j}$. Let us check that $C^-_\alpha \oplus C^-_\beta \oplus C_1 \oplus A' \equiv C^-_{\max(\alpha, \beta)} \oplus C_1 \oplus A'$. Consider the following strategy for Duplicator in $G(C^-_\alpha \oplus C^-_\beta \oplus C_1 \oplus A', C^-_{\max(\alpha,\beta)} \oplus  C_1 \oplus A')$. While Spoiler's critical token $x$ can reach the head loop of $C^-_{\alpha}$ or $C^-_{\beta}$ , Duplicator may keep his critical token $y$ looping in the head loop of his automaton $C^-_{\max(\alpha,\beta)}$. For every new token produced by Spoiler in the head loop of $C^-_\alpha$ or $C^-_\beta$, Duplicator produces a~{\em doppelg\"anger} in  the head loop of $C^-_{\max(\alpha,\beta)}$. When Spoiler moves his critical token $x$ to $C_1 \oplus A'$, Duplicator does the same with $y$ and lets it copy $x$'s actions. As the converse inequality is obvious, $C^-_{\max(\alpha, \beta)} \oplus C_1 \oplus A' \equiv C_{\max(\alpha, \beta)} \oplus A'$ gives the canonical form for $C^-_{\alpha} \oplus A$.

Now, let $S$ be non-branching. Suppose first that $S$ is one of the automata $D_{\omega^{\omega+k}}$, $C_{\omega^{\omega+k}}$, $E_{\omega^{\omega+k}}$ for $k \geq 0$. If $\alpha=\omega^k$, $C^-_\alpha \oplus S \oplus A' \leq C_\alpha \oplus S \oplus A' \equiv S \oplus A'$ by the proof of the closure by $\oplus$. The converse inequality is obvious. Similarly, if $\alpha=\omega^{\omega\cdot2+k}$, $C^-_\alpha \oplus S \oplus A' \leq C_\alpha \oplus S \oplus A' \equiv C_\alpha \oplus A'$. The converse inequality is obvious again.

The remaining possible values for $S$ are $C_1$, $D_1$ and $E_1$. If $S = C_1$, $C^-_\alpha \oplus C_1 \oplus A\equiv C_\alpha \oplus A$, and the canonical automaton is obtained via closure by $\oplus$. For $S = E_1$, observe that  $C^-_\alpha \oplus E_1 \oplus A' \leq C^-_{\omega^k} \oplus C_2 \oplus A' = C_{\omega^k} \oplus D_1 \oplus A$. By the proof of the closure by $\oplus$ we get $C_{\omega^k} \oplus D_1 \oplus A\equiv C_{\omega^k} \oplus A$. Hence $C^-_{\omega^k} \oplus E_1 \oplus A\leq C_{\omega^k} \oplus A$. The converse inequality is obvious. Finally, if $S = D_1$, we get $C^-_{\alpha} \oplus D_1 \oplus A' \equiv C^-_{\alpha} \oplus A'$. By the structure of canonical automata, $A'$ must start with $E_1$ or $C_{\omega^{\omega+k}}$. In both cases we can use one of the previous cases to get an equivalent canonical automaton.

If $A = S$ the whole argument is analogous, only in the last case, for $S=D_1$,  we have $C^-_{\alpha} \oplus D_1 \equiv D_1$. \qed

\begin{prop}\label{land_closure}
For $A, B \in {\mathcal C}$ one can find in polynomial time an automaton in ${\mathcal C}$ equivalent to $A \land B$.
\end{prop}

\proof We will proceed by induction on $(A,B)$ with the product order induced by $\leq$. Let $A=A_1 \oplus A_2 \oplus \ldots \oplus A_m$, $B=B_1 \oplus B_2 \oplus \ldots \oplus B_n$ with $A_i$, $B_j$ simple. Let $A'=A_2 \oplus \ldots \oplus A_m$ for $m>1$ and $B'=B_2\oplus \ldots \oplus B_n$ for $n>1$.

First, assume that $B_1=C_1 \to C_\beta$, and either $A_1 = F_{\iota,\kappa}$ for some $(\iota,\kappa)$, or $A_1 = C_1 \to C_\alpha$ for $\alpha \leq \beta$. Let $m,n>1$. Let us see that  $A \land B \equiv  B_1^- \oplus (A' \land B \lor A \land C_1\oplus B')$. In the first move Spoiler produces token $x^A$ in $A$ and $x^B$ in $B$. While $x^A$ stays in $A_1$ and $x^B$ stays in the head loop of $B_1$,  Duplicator should keep his critical token in the head loop of $B_1^-$ and for each $x$, a~child of $x^B$ or $x^A$, produce a~token $y$ whose task is to play against $x$. The token $x$ after being produced is put in the head loop of $C_\beta$ or, if $A_1 = C_1\to C_\alpha$, in the head loop of $C_\alpha$. The token $y$ is put in the head loop of $C_\beta$. Since $\alpha\leq\beta$, $y$ can adapt the strategy from $G(C_\alpha, C_\beta)$ if $x$ is in $C_\alpha$, or simply copy $x$'s actions if $x$ is in $C_\beta$. Now, two things may happen. If $x^A$ enters $A'$ while $x^B$ stays in the head loop of $B_1$, Duplicator should move his critical token to $A' \land B$ and split it into $y^A$ sent to $A'$ and $y^B$ sent to $B$. Then $y^A$ should mimic $x^A$, and $y^B$ should mimic $x^B$. If $x^B$ exits the head loop of $B_1$, Duplicator should move to $A \land C_1\oplus B'$, produce two tokens, and mimic Spoiler's actions. The converse inequality is even simpler. In a~similar way we prove $A \land B \equiv  B_1^- \oplus ( A\land C_1\oplus B')$ for $n>m=1$, $A \land B \equiv  B_1^- \oplus (A' \land B \lor A\land C_1)$ for $m>n=1$, and $A \land B \equiv  B_1^- \oplus (A \land C_1)$ for $m=n=1$. In all four cases using the induction hypothesis, the closure by $\lor$, $\oplus$, and the Substitution Lemma (Lemma \ref{substitution}, page \pageref{substitution}) we obtain an automaton of the form $B^-_1 \oplus C$, where $C$ is canonical. Lemma \ref{auxiliary_oplus} gives an equivalent canonical automaton. 

Next, suppose that $A_1=F_{(\iota,\kappa)}$, $B_1 = F_{(\iota',\kappa')}$. Assume $m,n>1$. Using Lemma \ref{flowers} one proves easily that $A \land B \equiv F_{(\iota, \kappa)\land(\iota',\kappa')} \oplus ((A \land B') \lor (A' \land B))$.  Similarly, for $m>1$, $n=1$, we have $A \land B \equiv F_{(\iota, \kappa)\land(\iota',\kappa')} \oplus  (A' \land B)$ and the canonical form follows from the induction hypothesis. For $m=1$, $n>1$ proceed symmetrically. For $m=n=1$, $A \land B \equiv F_{(\iota, \kappa)\land(\iota',\kappa')}$. Again, using the induction hypothesis, the closure by $\lor$, $\oplus$, and the Substitution Lemma, we get an equivalent canonical automaton. 

The general case may be reduced to one of the special cases above, because  $E_\alpha \land A\equiv (C_\alpha \land A) \lor (D_\alpha \land A)$. \qed

\vspace{5pt}

Since $(\iota, \kappa)$-replication requires a~rather involved analysis, let us first consider $\to$. 

\begin{prop}\label{to_closure}
For $A, B \in {\mathcal C}$ one can find in polynomial time an automaton in ${\mathcal C}$ equivalent to $A \to B$.
\end{prop}

\proof First, let us deal with two special cases for which the general method does not work. For $B \ngeq C_3$ simple calculations give the following equivalences: $A\to B\; \equiv\; (D_1 \oplus A) \land B$ for $B \in \{C_1, E_1, C_2,D_2, D_3\}$, $A\to D_1 \;\equiv\; A\lor D_1$, $A\to E_2 \;\equiv\; A\to D_3$. By the Substitution Lemma, the equivalent canonical forms follow from the closure by $\oplus$, $\lor$, and $\land$.

The second special case is when $B$ contains non-trivial flowers but $B \ngeq F_{(0,2)}$. First, let us see that  $A\to F_{(0,1)} \; \equiv \;(D_1 \oplus A) \land F_{(0,1)}$. The inequality $A\to F_{(0,1)} \; \geq \;(D_1 \oplus A) \land F_{(0,1)}$ follows easily from Lemma \ref{replication}. For the converse it remains to observe that the following strategy is winning for Duplicator in $G(A\to F_{(0,1)}, (D_1 \oplus A) \land F_{(0,1)})$: in $D_1 \oplus A$ mimic Spoiler and  in $F_{(0,1)}$ apply the strategy from $G(C_1 \to F_{(0,1)}, F_{(0,1)})$ given by Theorem \ref{borelautomata} (3 and 4). An analogous argument shows that $A\to F_{(1,2)} \; \equiv \;(D_1 \oplus A) \land F_{(1,2)}$. For the remaining possible values of $B$ we will show  $A\to B \; \equiv (D_1 \oplus A) \land F_{(0,1)} \land F_{(1,2)}$.  Again,  $A\to B \; \geq \;  (D_1 \oplus A) \land B \land B \geq (D_1 \oplus A) \land F_{(0,1)} \land F_{(1,2)}$ is easy. For the converse, observe that $B$ only uses ranks $1,2,3$. Consider the following strategy for Duplicator in $G(A \to B, (D_1 \oplus A) \land F_{(0,1)} \land F_{(1,2)})$. In the component $D_1 \oplus A$ simply mimic the behaviour of Spoiler's critical token. In $F_{(0,1)}$ use the strategy from $G(C_1 \to B', F_{(0,1)})$, where $B'$ denotes $B$ with ranks $1$ and $2$ replaced by $0$ and rank $3$ replaced by $1$. In $F_{(1,2)}$ use the strategy from $G(C_1 \to B'', F_{(1,2)})$, where $B''$ denotes $B$ with all $3$'s replaced by $1$'s. The combination of these three strategies is winning for Duplicator.

For the remaining automata, we will show that what really matters is the maximal simple branching automaton contained in $C_1 \to B$. There are two main cases: either $C_{\omega^{k-1}} < B \leq C_{\omega^{k}}$ ($C_3 \leq B \leq C_\omega$ for $k=1$),  or $C_{\omega^{\omega\cdot 2+(k-1)}} < B \leq C_{\omega^{\omega\cdot2+k}}$ ($ F_{(0,2)} \leq B \leq C_{\omega^{\omega\cdot2}}$ for $k=1$).  In the first case $A \to B \equiv C^-_{\omega^k}\oplus A$, in the second case $A \to B \equiv C^-_{\omega^{\omega\cdot2+k}}\oplus A$. Since the proofs are entirely analogous, we will only consider the first case. We only need to argue that $A \to B \leq C^-_{\omega^k}\oplus A$, since the converse inequality is obvious.

Let us start with $B = C_{\omega^k}$. Denote the head loop of $C_{\omega^k}$ by $\lambda_0$. It is enough to show a~winning strategy in $G(A\to B, C^-_{\omega^k} \oplus A)$. Since no path from the head loop of $A\to B$ to $\lambda_0$ goes through an accepting loop, Duplicator may keep his critical token in the head loop of $C^-_{\omega^k}$ as long as at least one of Spoiler's tokens can reach $\lambda_0$. Hence, for every token produced by Spoiler in $\lambda_0$, Duplicator can produce a~{\em doppelg\"anger}. When none of Spoiler's tokens can reach $\lambda_0$ any more, Duplicator moves his critical token to $A$ and mimics Spoiler.

Let us now suppose that $C_{\omega^{k-1}} < B < C_{\omega^{k}}$, $k \geq 2$ (for $k=1$ the proof is very similar). The strategy for Duplicator in $G(A \to B, C^-_{\omega^k} \oplus A)$ is as follows. Let $m$ be such that $B \leq C_{\omega^{k-1} m}$. For every token $x_i$ produced by Spoiler using the head loop of $A \to B$, Duplicator produces $m$ tokens $y^1_i, \ldots, y^m_i$ using the head loop of $C^-_{\omega^k}$. Then the tokens $y^1_i, \ldots, y^m_i$ play against $x_i$ simulating Duplicator's winning strategy from $G(B, (C_1 \oplus C_{\omega^{k-1}})^m)$. When Spoiler moves his critical token to $A$, Duplicator does the same and keeps mimicking Spoiler in $A$.

Thus we managed to simplify $A \to B$ to $C^-_{\alpha} \oplus A$ where $\alpha=\omega^k$ or $\alpha = \omega^{\omega\cdot 2 + k}$. An equivalent canonical automaton is provided by Lemma~\ref{auxiliary_oplus}.  \qed

\vspace{5pt}

Now we are ready to deal with $(\iota,\kappa)$-replication. Since $C_{\omega^{\omega\cdot3}} = C_1 \stackrel{(0,1)}{\longrightarrow} C_{\omega^{\omega+1}}$, the class ${\mathcal C}$ is not closed by $\stackrel{(\iota,\kappa)}{\longrightarrow}$. However, adding the three top canonical automata is enough to get the closure property.

\begin{prop}\label{longrightarrow_closure} 
For $ A,  A_\iota, \ldots,  A_\kappa \in {\mathcal C}$, $\iota, \kappa < \omega$, one can find in polynomial time  an automaton in ${\mathcal C}'$ equivalent to $A \stackrel{(\iota, \kappa)}{\longrightarrow}  A_\iota, \ldots,  A_\kappa$. 
\end{prop}

\proof Let $ B= A\stackrel{(\iota, \kappa)}{\longrightarrow}  A_\iota, \ldots,  A_\kappa$. If $ B$ admits any of the automata $C_{\omega^{\omega\cdot3}}$, $C_{\omega^{\omega\cdot3}+1}$, $C_{\omega^{\omega\cdot3}+2}$, then it is equivalent to the maximal one it admits (see Theorems \ref{borelautomata} and \ref{hardcore}). Let us assume $ B$ admits none of the three automata above. Let us also assume that $\iota<\kappa$.

\paragraph{{\em  1. If some $ A_i$ contains a~$(0,1)$-flower  and some $A_j$ contains a $(1,2)$-flower, then $ B \equiv (F_{(\iota,\kappa)} \oplus  A)\land F_{(1,2)} \land F_{(0,1)}$.}} It is easy to show that $(F_{(\iota,\kappa)} \oplus  A)\land F_{(1,2)} \land F_{(0,1)} \leq  B$ (c.~f. Lemma \ref{replication}). We shall concentrate on the converse inequality. From the hypothesis that $ B$ does not admit $C_{\omega^{\omega\cdot3}+1}$, it follows easily that $\kappa$ must be odd and $ A_\iota, \ldots,  A_{\kappa-1}$ must be $(1,2)$ automata. Furthermore, since $ B$ does not admit $C_{\omega^{\omega\cdot3}}$, $ A_k$ uses only ranks $1,2,3$. The strategy for Duplicator in $G( B, (F_{(\iota,\kappa)} \oplus  A)\land F_{(1,2)} \land F_{(0,1)})$ is analogous to the one used in the proof of the previous proposition. In the component $F_{(\iota,\kappa)} \oplus  A$ simply mimic the behaviour of Spoiler's critical token. In $F_{(0,1)}$, loop around the $1$-loop whenever Spoiler loops around the $1$-loop of a $(0,1)$-flower in $ A_\kappa$ (again, if the run is to be accepting, this may happen only finitely many times), otherwise loop around $0$-loop. For the strategy in $F_{(1,2)}$, treat all the ranks appearing in Spoiler's $F_{(\iota,\kappa)}$ or $ A$ as $2$'s, and the $3$'s in $ A_\kappa$ as $1$'s. Seen this way, $ B$ is a~$(1,2)$-automaton, and by Theorem \ref{borelautomata} Spoiler's actions can be simulated in $F_{(1,2)}$.

\paragraph{{\em 2. If $ A_i$ contain only $(1,2)$-flowers, then  $ B \equiv (F_{(\iota,\kappa)} \oplus  A)\land F_{(1,2)}$.}} This is proved just like the first case.

\paragraph{{\em 3. If $ A_i$ contain only $(0,1)$-flowers, then $ B \equiv  (A \stackrel{(\iota, \kappa)}{\longrightarrow}  A_\iota, \ldots,  A_{\kappa-1}, C_1)\land F_{(0,1)}$  (use case 4 or 5 to get a~canonical form).}} Like in the first case,  $\kappa$ must be odd, $A_\iota, \ldots,  A_{\kappa-1}$ must be $(1,2)$-automata. Consequently, it must be $A_\kappa$ that contains a $(0,1)$-flower. Since $A_\kappa$ contain no $F_{(0,2)}$ (by the hypothesis no $A_i$ does), $A_\kappa = F_{(0,1)}$.  Again $B \geq ( (A \stackrel{(\iota, \kappa)}{\longrightarrow}  A_\iota, \ldots,  A_{\kappa-1}, C_1)\land F_{(0,1)})$ is easy. The strategy for Duplicator in $G( B,  (A \stackrel{(\iota, \kappa)}{\longrightarrow}  A_\iota, \ldots,  A_{\kappa-1}, C_1)\land F_{(0,1)})$ is to copy Spoiler's actions in $A \stackrel{(\iota, \kappa)}{\longrightarrow}  A_\iota, \ldots,  A_{\kappa-1}, C_1$ and in $F_{(0,1)}$ keep record of all $1$'s appearing in $ A_\kappa$ (if the run is to be accepting, there may be only finitely many altogether).

\paragraph{{\em 4. If $ A_i$ contain no non-trivial flowers, $\iota=0$, and $ A_\iota$ contains a~$D_2$, then $ B \equiv (F_{(\iota,\kappa)} \oplus  A)\land F_{(1,2)}$.}} The inequality $ B \leq (F_{(\iota,\kappa)} \oplus  A)\land F_{(1,2)}$ is proved just like in the first case. Let us see that the converse holds. Consider the game $G((F_{(\iota,\kappa)} \oplus  A)\land F_{(1,2)},  B)$ and the following strategy for Duplicator. Copy Spoiler's actions in $F_{(\iota, \kappa)}\oplus  A$, but whenever Spoiler enters the $1$-loop in $(1,2)$, loop once around $0$-loop, move the extra token to the head loop of $D_2$, and keep looping around until Spoiler leaves his $1$-loop. Then remove your extra token, and so on. It is easy to see that the strategy is winning for Duplicator. 

\paragraph{{\em 5. If $ A_i$ contain no non-trivial flowers and either $\iota\neq 0$ or $ A_\iota$ contains no $D_2$, then $ B \equiv F_{(\iota,\kappa)} \oplus  A$.}} To prove it, we have to describe the strategy for Duplicator in $G( B, F_{(\iota,\kappa)} \oplus  A)$. During the whole play keep numbering the new tokens produced by Spoiler according to their birth time. (As usual, the left token is considered a~parent, the right token is born, transitions of the form $p \longrightarrow \top,q$ or $p \longrightarrow q, \top$ do not produce new tokens.) The strategy is as follows. While there are no new tokens in rejecting loops in $ A_\iota, \ldots,  A_\kappa$, keep copying Spoiler's moves in his $F_{(\iota, \kappa)}$. When the first new token, say $x_{i_1}$, enters a $1$-loop, start looping around the $1$-loop of your $F_{(\iota, \kappa)}$ (the loop exists since $\iota<\kappa$), and keep doing it until $x_{i_1}$ leaves the $1$-loop. If it does not happen, Spoiler will lose. When it does happen, stop looping around $1$-loop. Investigate all the ranks used by Spoiler in $(\iota, \kappa)$-flower while you were simulating $x_{i_1}$, choose the highest one, say $k$, and loop once a~$k$-loop. Afterwords, if there are no tokens in rejecting loops in $ A_i$, copy Spoiler's moves. Otherwise, choose the token with the smallest number, say $x_{i_2}$, start looping around the loop with the highest rank $1$ in your $(\iota, \kappa)$-flower, and so on.

Let us see that if Spoiler does not enter $ A$, he loses the game. If the run constructed by Spoiler is to be rejecting, either the highest rank used infinitely often in $F_{(\iota, \kappa)}$ is odd, or some token stays forever in a~rejecting loop in one of $ A_\iota, \ldots,  A_\kappa$. In any case Duplicator's strategy guarantees a~rejecting run for him as well. Let us suppose that Spoiler's run is accepting. If only finitely many new tokens entered rejecting loops in $ A_\iota, \ldots,  A_\kappa$, then there was a round such that from this round on Duplicator was simply mimicking Spoiler's actions in $F_{(\iota, \kappa)}$ and so Duplicator's run is also accepting. Suppose that infinitely many new tokens visited rejecting loops in $ A_\iota, \ldots,  A_\kappa$. We have assumed that either $\iota\neq 0$ or $ A_\iota$ contains no $D_2$. In either case the ranks greater then $0$ must have been used infinitely many times in $F_{(\iota, \kappa)}$. Consequently, the highest rank used in $F_{(\iota, \kappa)}$ is greater then $1$, and Duplicator's run is accepting despite infinitely many $1$'s used in $F_{(\iota, \kappa)}$.

Suppose now that Spoiler leaves $F_{(\iota, \kappa)}$. Following the argument used in the proof of the closure by $\oplus$, we may suppose that the simple automaton containing the head loop of $A$ is at least a~$\overline{(\iota, \kappa)}$. When Spoiler enters $ A$, he may produce no more tokens in $ A_\iota, \ldots,  A_\kappa$.  From now on Duplicator should mimic Spoiler's behaviour in his copy of $ A$, handling rejecting loops in $ A_\iota, \ldots,  A_\kappa$ in the usual way.

\vspace{10pt}

What is left is the case $\iota=\kappa$. If $\kappa$ is odd, $ B= A \to  A_1$. If $\kappa$ is even, $ A_0$ must be a~$(1,2)$-automaton. In the cases 2 and 4 proceed just like before. In the case 5, the automaton $A_0$ cannot contain $D_2$. If $ A_0 \in \{C_1, D_1, C_2\}$, then   $ B \equiv (C_1 \oplus  A) \land  A_0$. If $ A_0 = E_1$, then  $ B \equiv C_1 \oplus ( A\lor C_1)$. \qed

\vspace{5pt}

 The following corollary sums up the closure results. 

\begin{cor} \label{closure}
The class of canonical automata ${\mathcal C}'$ is closed by $\lor$, $\oplus$, $\land$, $\stackrel{(\iota, \kappa)}{\longrightarrow}$, and the equivalent automaton can be found in polynomial time. 
\end{cor} 

\proof The claim is an almost immediate consequence of the preceding propositions. Only the automata $C_{\omega^{\omega\cdot3}}$, $C_{\omega^{\omega\cdot3}+1}$, $C_{\omega^{\omega\cdot3}+2}$ need special care: if the result of the operation in question admits any of these automata, it is equivalent to the hardest one it admits (Theorems \ref{borelautomata} and \ref{hardcore}). \qed

\section{Completeness} \label{sect:completeness}

In this section we show that the canonical automata represent the $\equiv_W$-classes of all deterministically recognisable tree languages. We will implicitly use Corollary \ref{closure} and the Substitution Lemma (Lemma \ref{substitution}, page \pageref{substitution}) on several occasions. 

We will say that a~transition is {\em positive} \index{positive transition} if one of its branches lies on an accepting loop, and {\em negative} \index{negative transition} if one of its branches lies on a~rejecting loop. Note that a~transition may be positive and negative at the same time. Recall the notion of replication (see page \pageref{replicated}). We say that a~state is $j$-replicated if it is replicated by a~$j$-loop. An automaton is $j$-replicated if its initial state is $j$-replicated.

Finally, let us recall the {\em lifting operation} \index{lifting operation} invented by Niwi\'nski and Walu\-kiewicz and used to prove the decidability of the deterministic index hierarchy (Theorem \ref{omegaindch}, page \pageref{omegaindch}). 

\begin{lem}[Niwi\'nski and Walukiewicz \cite{kwiatek}] \label{lifting} 
For each deterministic automaton $A$ one can compute (in polynomial time if the productive states are given) an automaton $A\uparrow^0\uparrow^1\ldots\uparrow^n$ such that $L(A)=L(A\uparrow^0\uparrow^1\ldots\uparrow^n)$ and if a~state $q$ has the rank $j\leq n$ than $q$ lies on a~$j$-loop of a~$(j,n)$-flower. \qed
\end{lem}

\begin{thm} \label{completeness}
For every deterministic tree automaton there exists an equivalent canonical automaton.
\end{thm}

\proof Let $A$ be a~deterministic tree automaton. From Theorem \ref{borelautomata} (7) it follows that if $A$ admits $C_{\omega^{\omega\cdot3}+2}$, $A\equiv C_{\omega^{\omega\cdot3}+2}$. If $A$ does not admit $C_{\omega^{\omega\cdot3}+2}$, then by Theorem \ref{borelautomata} (5 and 6) if $A$ admits $C_{\omega^{\omega\cdot3}+1}$, $A \equiv C_{\omega^{\omega\cdot3}+1}$. Otherwise $L(A) \in \Delta_3$ and if $A$ admits $C_{\omega^{\omega\cdot3}}$, then $A \equiv C_{\omega^{\omega\cdot3}}$ (Theorem \ref{hardcore}). In the remaining of the proof we will assume that $A$ admits none of these three automata. We will proceed by induction on the height of the DAG of SCCs of $A$. Let $X$ denote the root SCC of $A$. We will say that $X$ contains a~transition $p \longrightarrow p',p''$, if $X$ contains all three states,  $p$, $p'$, and $p''$. We consider four separate cases.

\paragraph{{\em 1.  $X$ contains a~positive transition.}} Observe that each state of $A$ is replicated by an accepting loop. Therefore, if $A$ admits $F_{(0,1)}$, it must also admit $C_1\stackrel{(0,0)}{\longrightarrow} F_{(0,1)} =  C_{\omega^{\omega\cdot3}+1}$, which is excluded by our initial assumption. Consequently, $A$ is a~$(1,2)$-automaton (Theorem \ref{indch}). Without loss of generality we may assume that $A$ uses only ranks $1$ and $2$.

By Theorem \ref{borelautomata} (3 and 4), $L(F_{(1,2)})$ is $\Pi^0_2$-complete and $L(A) \in \Pi^0_2$, which implies that $A \leq F_{(1,2)}$. If $A$ admits $D_1 \oplus C_1$, then it also admits $C_1\stackrel{(0,0)}{\longrightarrow} D_1 \oplus C_1$, and so $A \geq C_1\stackrel{(0,0)}{\longrightarrow} D_1 \oplus C_1$. From Theorem \ref{borelautomata} (3) it follows that $C_1\stackrel{(0,0)}{\longrightarrow} D_1 \oplus C_1 \equiv F_{(1,2)}$. Consequently, $A \equiv  F_{(1,2)}$.

Suppose that $A$ does not admit $D_1 \oplus C_1$, but $X$ contains a rejecting loop $\lambda_1$. Let $p_1$ be a state on that loop. Since $X$ contains a positive transition, it must contain an accepting loop and, in particular, a state with rank 2, say $p_2$. Since $X$ is strongly connected, we may find a loop $\lambda_2$ going from $p_1$ to $p_1$ via $p_2$. Since $X$ only uses ranks $1$ and $2$, and $\mathrm{rank}(p_2)=2$, $\lambda_2$ is accepting. Hence, $\lambda_1$ and $\lambda_2$ form a $(1,2)$-flower. In consequence, $A \geq F_{(1,2)}$.  Hence, $A\equiv F_{(1,2)}$. 

Finally, suppose that $X$ contains no rejecting loops and $A$ does not admit $D_1 \oplus C_1$. By Theorem \ref{borelautomata} (2), $L(A) \in \Pi^0_1$ and since $L(C_1 \oplus D_1)$ is $\Pi^0_1$-complete, $A \leq C_1 \oplus D_1$. If $A$ admits $D_1$ it also admits $C_1 \oplus D_1$, and so $A \equiv C_2$. If $A$ does not admit $D_1$ it means that it contains no rejecting loop. Hence, $A$ accepts every tree and $A\equiv C_1$.

\paragraph{{\em 2. $X$ contains an accepting loop and a~negative transition, but no positive transitions. }} Let $\lambda_+$ be an accepting loop in $X$ and $\lambda_X$ be a~loop visiting all $X$'s nodes and containing a~branch of the (negative) transition contained in $X$. Since $X$ does not contain positive transitions, $\lambda_X$ is rejecting. The loops  $\lambda_+$ and $\lambda_X$ form a~$(0,1)$-flower. Hence, $A$ admits $F_{(0,1)}$.  Furthermore, should $A$ contain a~$(0,2)$-flower, it would obviously be replicated by $\lambda_X$ and $A$ would admit $C_1 \to F_{(0,2)} = C_{\omega^{\omega \cdot 3}}$, which contradicts our general hypothesis. Hence, $A$ does not admit $F_{(0,2)}$, which means $A$ is a $(1,3)$-automaton (Theorem \ref{indch}, page \pageref{indch}). Without loss of generality we may assume that it uses only ranks $1,2,3$. 

By Theorem \ref{borelautomata} (3 and 4), if $A$ admits neither $F_{(1,2)}$ nor  $C_1\stackrel{(0,0)}{\longrightarrow} D_1 \oplus C_1$, then $A \equiv F_{(0,1)}$. Suppose that $A$ admits one of these two automata. Consider the game $G(F_{(0,1)} \land F_{(1,2)}, A)$. Let $x^1$ and $x^2$ be Spoiler's tokens in $F_{(0,1)}$ and $F_{(1,2)}$, respectively. Since $X$ contains a~(negative) transition, Duplicator can split his critical token into $y^1$ and $y^2$ within $X$, and move $y^1$ to the $(0,1)$-flower in $X$, and $y^2$ to the $(1,2)$-flower, or to the accepting loop replicating a~weak $(1,2)$-flower (if $A$ admits $C_1\stackrel{(0,0)}{\longrightarrow} D_1 \oplus C_1$). Then $y^1$ should mimic $x^1$, and $y^2$ should mimic $x^2$ --- either directly, or adapting the strategy from $G(F_{(1,2)}, C_1\stackrel{(0,0)}{\longrightarrow} D_1 \oplus C_1)$. Hence, Duplicator has a~strategy to win the game. It follows that $F_{(0,1)} \land F_{(1,2)}\leq A$. 

For the converse inequality, let us call the states with rank 3 contained in a~$(0,1)$-flower {\em red}, and the remaining {\em blue}. Since $A$ does not admit $C_1\stackrel{(0,0)}{\longrightarrow} F_{(0,1)}$, no red state is replicated by an accepting loop. Consider the game $G(A, F_{(0,1)} \land F_{(1,2)})$. For a strategy in  $F_{(1,2)}$ Duplicator should treat all the red states as if they had rank 1; the automaton  $A$ modified this way does not admit $F_{(0,1)}$, so Duplicator may use the strategy given by Theorem \ref{borelautomata} (3 and 4). In $F_{(0,1)}$ Duplicator should loop a~$1$-loop whenever some Spoiler's token is in a~red state. Otherwise, Duplicator should loop a~$0$-loop. Let us see that this strategy is winning. 

Suppose that Spoiler's run is accepting. After changing the ranks of red states from $3$ to $1$ it is still accepting, so Duplicator's token in $F_{(1,2)}$ visited an accepting path. By the Replication Lemma (Lemma \ref{replication}, page \pageref{replication}), the occurrences of red states in Spoiler's run may be covered by a~finite number of paths. Furthermore, each of these paths is accepting, so it may only contain a~finite number of red states. Hence, there may be only finitely many red states in Spoiler's run and the path visited by Duplicator's token in $F_{(0,1)}$ is also accepting. 

Suppose now, that Spoiler's run is rejecting. If red states occurred only finitely often, Spoiler's run is still rejecting after changing their ranks to $1$, so Duplicator's token in $F_{(1,2)}$ visited a~rejecting path. If there were infinitely many red states in Spoiler's run, Duplicator's token in $F_{(0,1)}$ visited a~rejecting path. 

Hence, $A \equiv F_{(0,1)} \land F_{(1,2)}$ and by Lemma \ref{flowers}, $A \equiv F_{(1,3)}$.

\paragraph{{\em 3. $X$ contains some transitions but no accepting loops.}} Let $q_i \stackrel{\sigma_i}{\longrightarrow} q'_i, q''_i$, $i=1,\ldots, n$ be all the transitions such that $q_i\in X$ and $q'_i, q''_i \notin X$. Let $p_j \stackrel{\sigma_i, d}{\longrightarrow} p'_j$ $j=1,\ldots, m$ be all the remaining transitions such that  $p_j\in X$ and $p'_j \notin X$. We will call the automata $(A)_{q'_i}$, $(A)_{q''_i}$ and $(A)_{p'_j}$ the {\em child automata of $X$}. By the induction hypothesis we may assume that they are in canonical forms. Let $B = ((A)_{q'_1} \land (A)_{q''_1}) \lor \ldots \lor ((A)_{q'_n} \land (A)_{q''_n}) \lor (A)_{p'_1} \lor \ldots \lor (A)_{p'_m}$. It is not difficult to see that $A$ is equivalent to $C_1 \to B$. 

\paragraph{{\em 4.  $X$ contains no transitions. }} Recall that this means exactly  that at most one branch of every transition stays in $X$. First replace subtrees rooted in the target states of transitions whose all branches leave $X$ with one canonical automaton $B$ just like above. Let $(\iota,\kappa)$ denote the highest index of a~flower contained in $X$. It is well defined, because a~strongly connected component admitting  $F_{(0,j)}$ and $F_{(1,j+1)}$ must also admit $F_{(0,j+1)}$. We may assume that $X$ uses only ranks $\iota, \ldots, \kappa$, and that each $j$-loop is indeed a~$j$-loop in a $(j,\kappa)$-flower (Lemma \ref{lifting}).  For each $j=\iota, \ldots, \kappa$, let $B_j$ be the alternative of all the child automata replicated by a~$j$-loop in $X$. By induction hypothesis, we may assume that $B_\iota, \ldots, B_\kappa$ and $B$ are canonical automata. Let $A' = B \stackrel{(\iota, \kappa)}{\longrightarrow} B_\iota, \ldots, B_\kappa$. We will show that $A \equiv A'$.

If $\iota=\kappa$, the assertion is clear. Suppose that $\iota < \kappa$. Obviously, $A' \geq A$. Let us see that  $A' \leq A$. Let $A''$ denote the result of the following simplifications performed on $A'$.
\begin{enumerate}[$\bullet$]
\item If some $B_i$ contains a~$(0,1)$-flower and some $B_j$ contains a $(1,2)$-flower, replace $B_\kappa$ with a~$(1,3)$-flower.
\item If some $B_i$ contains a~$(0,1)$-flower and no $B_j$ contains a $(1,2)$-flower, replace $B_\kappa$ with a~$(0,1)$-flower.
\item If some $B_i$ contains a~$(1,2)$-flower and no $B_j$ contains a $(0,1)$-flower, replace $B_\kappa$ with a~$(1,2)$-flower.
\item If $B_\iota, \ldots, B_\kappa$ admit no $F_{(\iota,\kappa)}$ with $\iota<\kappa$, remove $B_\kappa$.
\item If $\iota=0$ and $B_\iota$ admits $D_2$, replace $B_\iota$ with $D_2$. Otherwise, remove $B_\iota$.
\item Remove all $B_{\iota+1}, \ldots, B_{\kappa-1}$.
\end{enumerate}
  Examination of the five cases considered in the proof of Proposition
  \ref{longrightarrow_closure} reveals that $A'$ and $A''$ have
  identical canonical forms. Consequently, $A' \equiv A''$, and it is
  enough to show that $A'' \leq A$. Consider all $(\iota,
  \kappa)$-flowers in $X$. Choose any one whose $\iota$-loop
  replicates $D_2$, if there is one, or take any $(\iota,
  \kappa)$-flower otherwise. Then, extend the $\kappa$-loop to a~loop
  using all the transitions in $X$. Denote this flower, together with
  the subtrees replicated by $\iota$-loop or $\kappa$-loop, by
  $F$. One can prove easily that $A'' \leq F \oplus B$, and obviously
  $F \oplus B \leq A$. \qed

\begin{algorithm}
\caption{The canonical form of deterministic tree automata}
\begin{algorithmic}[1]
\IF {$A$ admits $C_{\omega^{\omega\cdot3}+2}$} \STATE return $C_{\omega^{\omega\cdot3}+2}$
\ELSIF {$A$ admits $C_{\omega^{\omega\cdot3}+1}$} \STATE return $C_{\omega^{\omega\cdot3}+1}$
\ELSIF {$A$ admits $C_{\omega^{\omega\cdot3}}$} \STATE return $C_{\omega^{\omega\cdot3}}$

\ELSE

\STATE $X:=$ the root SCC of $A$ 
\IF {$X$ contains a~positive transition}
\IF {$A$ admits $F_{(1,2)}$ or $A$ admits $\emptyset \stackrel{(0,0)}{\longrightarrow} D_2$}
\STATE return $F_{(1,2)}$
\ELSIF {$A$ admits $D_1$}
\STATE return $C_2$
\ELSE 
\STATE return $C_1$
\ENDIF
\ELSIF {$X$ contains a~negative transition}
\IF {$X$ admits $C_1$}
\IF {$A$ admits $F_{(1,2)}$ or  $A$ admits $\emptyset\stackrel{(0,0)}{\longrightarrow} D_2$}
\STATE return $F_{(1,3)}$
\ELSE
\STATE return $F_{(0,1)}$ 
\ENDIF
\ELSE
\STATE $B:=$ the alternative of the canonical forms of $X$'s children
\STATE return $C_1 \to B$
\ENDIF
\ELSE[$X$ contains no transitions]
\STATE $B:=$ the alternative of  the canonical forms of $X$'s non-replicated children 
\STATE lift ranks in $X$
\STATE $(\iota, \kappa):=$ the index of the maximal flower
\FOR {$j:=\iota$ to $\kappa$} 
\STATE $B_j:=$ the alternative of the canonical forms of $X$'s $j$-replicated children
\ENDFOR
\STATE return $B\stackrel{(\iota, \kappa)}{\longrightarrow} B_\iota, \ldots, B_\kappa$
\ENDIF
\ENDIF
\end{algorithmic}
\end{algorithm}

From the proof of the Completeness Theorem one easily extracts an algorithm to calculate the canonical form of a~given deterministic automaton (Algorithm 1). 

\begin{cor} \label{canonical}
For a~deterministic tree automaton, a~Wadge equivalent canonical automaton can be calculated within the time of finding the productive states of the automaton.
\end{cor}

\proof It is easy to see that the size of the canonical forms returned by the recursive calls of each depth is bounded by the size of $A$ (up to a~uniform constant factor). To prove the time complexity of the algorithm assume that the productive states of $A$ are given. Checking if $A$ admits any of the automata mentioned in the algorithm can be easily done in polynomial time. The operations on the automata returned by the recursive calls of the procedure (lines 25, 26, 29, 33, and 35) are polynomial in the size of those automata, and by the initial remark also in the size of the automaton. By Lemma \ref{lifting} the lifting operation is also polynomial. Therefore, when implemented dynamically, this procedure takes polynomial time for each SCC. Processing the entire automaton increases this polynomial by a~linear factor. \qed

\vspace{5pt}

Instead of a~canonical automaton, the algorithm above can return its
``name'', i.~e., a~letter $C$, $D$, or $E$, and an ordinal
$\alpha\leq\omega^{\omega\cdot3}+2$ presented as a~polynomial in
$\omega^\omega$, with the coefficients presented as polynomials in
$\omega$. Since for such presentation it is decidable in linear time
if $\alpha\leq\beta$, as an immediate consequence of Corollary
\ref{canonical} and Theorem \ref{strictorder} we get an algorithm for Wadge reducibility.

\begin{cor} 
For deterministic tree automata $A$, $B$ it is decidable if $L(A) \leq_W L(B)$ (within the time of finding the productive states of the automata). \qed
\end{cor}

\section*{Acknowledgements}

The author thanks Damian Niwi\'nski for drawing his attention to the Wadge hierarchy problems, and for reading carefully a preliminary version of this paper. The author is also grateful to the anonymous referees for their sharp yet extraordinarily useful comments that had a great impact on the present version of the paper.

\end{document}